\documentclass[a4paper,fleqn,usenatbib]{mnras}

\usepackage{txfonts}

\usepackage[T1]{fontenc}
\usepackage{ae,aecompl}

\usepackage{psfig}
\usepackage{graphicx}
\usepackage{amssymb}
\usepackage{subfigure}

\title[Evolution of Friends of Friends]{Evolution of spatio-kinematic structures in star-forming regions: are Friends of Friends worth knowing?}

\author[R. J. Parker \& N. J. Wright]{Richard  J. Parker$^{1}$\thanks{E-mail: R.Parker@sheffield.ac.uk}\thanks{Royal Society Dorothy Hodgkin Fellow} and Nicholas J. Wright$^{2}$ \vspace*{0.1cm}\\
$^{1}$Department of Physics and Astronomy, The University of Sheffield, Hicks Building, Hounsfield Road, Sheffield, S3 7RH, UK\\
$^{2}$Astrophysics Group, Keele University, Keele, Staffordshire ST5 5BG, UK}

\begin{document}

\date{}
                             
\pagerange{\pageref{firstpage}--\pageref{lastpage}} \pubyear{2018}

\maketitle

\label{firstpage}

\begin{abstract}
The Friends of Friends algorithm identifies groups of objects with similar spatial and kinematic properties, and has recently been used extensively to quantify the distributions of gas and stars in young star-forming regions. We apply the Friends of Friends algorithm to $N$-body simulations of the dynamical evolution of subvirial (collapsing) and supervirial (expanding) star-forming regions. We find that the algorithm picks out a wide range of groups (1 -- 25) for statistically identical initial conditions, and cannot distinguish between subvirial and supervirial regions in that we obtain similar mode and median values for the number of groups it identifies. We find no correlation between the number of groups identified initially and either the initial or subsequent spatial and kinematic tracers of the regions' evolution, such as the amount of spatial substructure, dynamical mass segregation, or velocity dispersion. We therefore urge caution in using the Friends of Friends algorithm to quantify the initial conditions of star formation. 
\end{abstract}


\section{Introduction}

The majority of stars form in regions where the median stellar density exceeds the density in the Galactic field by several orders of magnitude \citep[e.g.][]{Lada03,Bressert10}. These young stars often follow a hierarchical or substructured spatial distribution \citep{Cartwright04,Gutermuth09,Gouliermis14,Kuhn14,Wright14,Jaehnig15} and also display correlated velocities (i.e.\,\,low velocity dispersions) and kinematic substructure on local scales \citep[e.g.][]{Larson81,Jeffries14,Foster15,Hacar16,DaRio17}.

There is also mounting evidence that the gas from which stars form also exhibits significant spatial \citep{Cartwright06,Henshaw16,Williams18} and kinematic \citep{Peretto06,Hacar13,Hacar17} substructure, although analysis of simulations that directly follow the conversion of gas to stars suggests that the link between their respective spatial and kinematic properties is highly non-trivial \citep{Parker15c,Vazquez17,Kuznetsova18} and may not be a direct one-to-one mapping.

Quantifying how these spatial and kinematic structures form and evolve is crucial in order to understand the typical environment where most stars form, and the implications for planet formation and stability \citep[e.g.][]{Scally01,Adams04,Parker12a,Vincke15,Zwart16,Cai17a} as well as their collective evolution in the context of Galaxy-scale astrophysical processes \citep[e.g.][]{Keres09}.

A significant amount of effort has been invested in quantifying both the spatial distributions in young star-forming regions \citep{Larson95,Cartwright04,Kuhn14,Jaffa18} and the kinematic distributions \citep{Foster15,Wright16,Wright18}. Recently, the Friends of Friends algorithm, originally used to quantify clusters of Galaxies \citep{Huchra82}, has been used to quantify the initial stages of star formation by simultaneously incorporating both the spatial \emph{and} kinematic information. This is usually realised by using the $x$- and $y$- position and the radial velocity measurement of either gas parcels, or individual stars.

The most notable result from these Friends of Friends analyses has been the discovery of ``bundles of fibres'' within filaments in star-forming regions \citep{Hacar13,Hacar17,Hacar18} as well as significant substructure in both the distribution of gas \citep{Henshaw16a} and of the protostars \citep{Hacar16,DaRio17}. These results potentially indicate a universality in the spatial and kinematic substructure in star-forming regions, but crucially, the Friends of Friends method has not been extensively tested on either synthetic data \citep{Cartwright04,Lomax11,Parker15b,Jaffa18}, though a notable recent exception is the work by \citet{Clarke18}, or on simulation data.

An exception to the latter is the work by \citet{Kuznetsova18}, who show that the kinematic structures are likely influenced by the accretion histories of the groups of stars. However, even star-forming regions with a moderate stellar density ($\sim$100\,M$_\odot$\,pc$^{-3}$) evolve significantly in the first few Myr after star formation, and an obvious avenue of investigation is the longevity of groups identified by the Friends of Friends algorithm during their subsequent dynamical evolution.

Furthermore, no studies have tested the Friends of Friends algorithm on multiple random realisations of the same distribution. Tests of spatial distribution algorithms \citep[such as the $\mathcal{Q}$-parameter, or mass segregation alogorithms][]{Cartwright09b,Parker15b} are essential to understand the significance of any single observed (or simulated) result.

In this paper we test the Friends of Friends algorithm on $N$-body simulations of the dynamical evolution of star-forming regions. The paper is organised as follows. In Section 2 we describe the Friends of Friends algorithm, and the $N$-body simulations we utilise. We present our results in Section 3 and provide a discussion in Section 4. We conclude in Section 5.

\section{Method}

In this section we describe the set-up of our $N$-body simulations and the method used to define groups based on the Friends of Friends method.

\subsection{$N$-body simulations}

Observations and simulations of the early stages of star formation both suggest that stars form with a spatially and kinematically substructured distribution. (In part, this has inspired the proliferation of the use of Friends of Friends algorithms in star formation studies \citep{Hacar13,DaRio17}, as star-forming regions are clearly inhomogeneous in terms of their spatial and kinematic properties.) To mimic this substructure, our simulated star-forming regions are set up using the box fractal method described in \citet{Goodwin04a} and \citet{Cartwright04}. We direct the reader to \citet{Goodwin04a,Allison10,Parker14b} for full details of this method, but we briefly summarise it here.

A box fractal is constructed by defining a `parent' in the centre of a cube of side $N_{\rm div}$ (we adopt $N_{\rm div} = 2$), which spawns $N_{\rm div}^3$ subcubes. Each of these subcubes contains a first generation `child' at its centre. The construction of the fractal distribution proceeds by determining which of the children then go on to become parents. The probability that a child becomes a parent is given by $N^{D-3}_{\rm div}$, where $D$ is  the fractal dimension. This process is repeated recursively and the final generation of children become stars, which are positioned randomly within the fractal distribution. When the fractal dimension is low, fewer children  spawn their own offspring and the resultant fractal distribution contains more substructure.

In two sets of simulations we present in this paper, the fractal dimension is $D = 1.6$, which gives the highest degree of substructure in a three-dimensional distribution. This is in order to facilitate the detection of multiple groups of stars by the Friends of Friends algorithm. A higher fractal dimension, e.g. $D = 2.0$ or $D = 2.6$, would lead to fewer distinct groups of stars in the resultant distribution. However, as we discuss below, the fractal dimension also governs the velocity structure in our box fractal method and may influence the way spatio-kinematic groups disperse, so we ran a further two sets of simulations with the fractal dimension set to $D = 2.0$ and $D = 2.6$, respectively.

The velocities of the parents in the fractal are drawn from a Gaussian with mean zero, and the children inherit the velocities of their parents plus an extra random component, the size of which scales as $N^{D-3}_{\rm div}$ (i.e. in a similar fashion to the spatial distribution) and decreases through each successive generation. This results in a kinematic distribution where stars on local scales have very similar velocities, but on larger scales the velocities can be quite different. 

In the box fractal method, on local scales of size $L$ the velocities scale as $v(L) \propto L^{3-D}$. We expect the timescale for the erasure of substructure to be of the order $t(L) \sim L/v(L)$, so for a fractal with $D = 1.6$ the timescale for the erasure of structure is $t(L) \propto L^{-0.4}$, for $D = 2.0$  $t(L) \propto L$ and for $D = 2.6$ $t(L) \propto L^{0.6}$. This implies that structure is erased faster on large scales in the case of $D = 1.6$, but erased faster on small scales in the cases of $D = 2.0$ and $D = 2.6$.

Interestingly, the $D=2.0$ and $D=2.6$ fractals are more consistent with the \citet{Larson81} observed line-width relations\footnote{Note that recent work by \citet{Traficante18} has shown that massive star-forming clumps deviate quite strongly from the \citet{Larson81} relations.}, where $v(L) \propto L^{0.38}$ (and therefore $t(L) \propto L^{0.6}$), so we might expect that spatio-kinematic structure in observed star-forming regions would be erased faster on smaller scales. 

We scale the velocities of the stars to a virial ratio $\alpha_{\rm vir} = T/|\Omega|$, where $T$ and $|\Omega|$ are the total kinetic and potential energies, respectively. In one set of simulations $\alpha_{\rm vir} = 0.3$, i.e. subvirial, where the global motion of the stars is to fall towards the centre of the potential. In the second set of simulations  $\alpha_{\rm vir} = 1.5$, i.e. supervirial, where the global motion is for the star-forming region to expand. However, due to the (local) correlated velocities in the fractal distributions, a significant degree of violent relaxation \citep{LyndenBell67} occurs \citep[see][for examples of the dynamical evolution of these type of systems]{Allison10,Parker14b,Parker16b}. In our simulations, violent relaxation occurs within the substructure, whilst the global bulk motion of the simulation  is either to collapse (in the $\alpha_{\rm vir} = 0.3$, subvirial case) or rapidly expand (in the $\alpha_{\rm vir} = 1.5$, supervirial case).

Each simulation contains $N = 1500$ single stars, drawn from the \citet{Maschberger13} formulation of the Initial Mass Function (IMF) with a probability distribution
\begin{equation}
p(m) \propto \left(\frac{m}{\mu}\right)^{-\alpha}\left(1 + \left(\frac{m}{\mu}\right)^{1 - \alpha}\right)^{-\beta},
\end{equation}
where $\mu = 0.2$\,M$_\odot$ is the average stellar mass, $\alpha = 2.3$ is the \citet{Salpeter55} power-law exponent for higher mass stars, and $\beta = 1.4$ describes the slope of the slope of the IMF for low-mass objects \citep*[which also deviates from the log-normal form;][]{Bastian10}. We sample this distribution in the mass range 0.01 -- 50\,M$_\odot$. 

The radii of our fractal star-forming regions are set to 1\,pc. This radius, and the degree of spatial substructure as set by the fractal dimension, gives high to moderate \emph{local} stellar densities ($\tilde{\rho} \sim 10^4$\,M$_\odot$\,pc$^{-3}$ for $D = 1.6$, $\tilde{\rho} \sim 10^3$\,M$_\odot$\,pc$^{-3}$ for $D = 2.0$ and $\tilde{\rho} \sim 10^2$\,M$_\odot$\,pc$^{-3}$ for $D = 2.6$), which means that the initial substructure will dynamically evolve. It is unclear if there is a typical initial density for star formation \citep[and if there is one, what it is,][]{Marks12,Parker14e,Parker17a}, but $\tilde{\rho} \sim 10^3 - 10^4$\,M$_\odot$\,pc$^{-3}$ is consistent with models of the formation and evolution of the Orion Nebula Cluster \citep{Allison10,Allison11}. 

In summary, we evolve four sets of $N$-body simulations, each with the same number of stars and initial radius, but we vary the initial degree of substructure, the initial local density, and the initial virial ratio $\alpha_{\rm vir}$. We use the \texttt{kira} integrator in the \texttt{Starlab} environment \citep{Zwart99,Zwart01} to evolve the star-forming regions for 10\,Myr. We do not include stellar evolution. A summary of the simulation initial conditions is given in Table~\ref{initials}. 

\begin{table}
  \caption[bf]{Summary of the variables in the initial conditions of our $N$-body simulations. The columns are the fractal dimension, $D$, the  virial ratio,  $\alpha_{\rm vir}$, and the local density $\tilde{\rho}$.}
  \begin{center}
    \begin{tabular}{|c|c|c|}
      \hline
fractal dimension & virial ratio & local density  \\   
      \hline
 $D = 1.6$ & $\alpha_{\rm vir} = 0.3$ & $\tilde{\rho} \sim 10^4$\,M$_\odot$\,pc$^{-3}$ \\
 $D = 1.6$ & $\alpha_{\rm vir} = 1.5$ & $\tilde{\rho} \sim 10^4$\,M$_\odot$\,pc$^{-3}$ \\
 $D = 2.0$ & $\alpha_{\rm vir} = 1.5$ & $\tilde{\rho} \sim 10^3$\,M$_\odot$\,pc$^{-3}$ \\
 $D = 2.6$ & $\alpha_{\rm vir} = 1.5$ & $\tilde{\rho} \sim 10^2$\,M$_\odot$\,pc$^{-3}$ \\
      \hline
    \end{tabular}
  \end{center}
  \label{initials}
\end{table}

\subsection{Friends of Friends group detection}

As with other recent work \citep{Hacar13,Hacar16,DaRio17,Kuznetsova18}, we base our Friends of Friends detection algorithm on the original method for classifying clusters of galaxies in redshift space by \citet{Huchra82}. We perform our analysis using the full six-dimensional information available ($x$, $y$ and $z$, as well as the corresponding velocity components $v_x$, $v_y$ and $v_z$). Most observational studies are done in three dimensions (usually $x$, $y$ and $v_z$), but we have verified that our results do not significantly differ if done in fewer dimensions. However, by using the full 6D information we would expect the Friends of Friends algorithm to identify real group structures without being hampered by projection effects.

The algorithm proceeds as follows. If a star is not yet assigned a group, we search for companions to that star that are less than a distance $d_L$ and a velocity difference less than $v_L$. If the nearby star fulfills both criteria then we start a new group and add companions that are less than $d_L$ and $v_L$ from any star in the group. If no further stars fulfill the criteria another unassigned star is chosen randomly and we repeat the process. 

Arguably, the most challenging aspect of the Friends of Friends analysis is to define the distance and velocity thresholds, $d_L$ and $v_L$, above which stars are assigned into different groups. Because we are analysing multiple $N$-body simulations, each containing multiple snapshots of data, we have automated the process of defining $d_L$ and $v_L$. We create an ordered list of all possible separations between stars, and all possible differences in velocity.  We set $d_L$ to be the median separation divided by three ($\tilde{\Delta r}/3$) and $v_L$ to be the median velocity difference ($\tilde{\Delta v}$). There is no real physical basis behind these choices, other than they divide the star-forming region into a reasonable number of groups (i.e. 1--20). These thresholds vary from simulation to simulation (and over time), but the initial values are typically $\tilde{\Delta r}/3 \sim 0.3$pc and $\tilde{\Delta v} \sim 1$km\,s$^{-1}$, which are similar to the threshold lengths adopted in observational studies \citep{Hacar16,DaRio17}.  However, we note that dividing a hierarchical fractal distribution into constituent groups is somewhat artificial, and we discuss this issue further in Section 4. 

Finally, in order to mimic observational studies, and to avoid the potentially artificial imposition of boundaries, we set an automatic stellar density threshold where any star that resides in the lowest quartile of an ordered list of stellar densities is not assigned to a group. This is intended to reduce the prospect of ``bridges'' of only one or two stars between groups causing the algorithm to merge two otherwise distinct groups. However, as we will see, this conservative threshold does not alleviate confusion in the Friends of Friends group detection.

\subsection{Other kinematic and spatial measures}

In Section 3 we will also look for a dependence of the number of groups identified by the Friends of Friends algorithm on other kinematic and spatial measures. These techniques have been presented in previous papers so we direct the interested reader to the relevant literature, however we briefly summarise them here.

We determine the $\mathcal{Q}$-parameter \citep{Cartwright04,Cartwright09b,Lomax11,Jaffa17}, which compares the average length of the edges on the minimum spanning tree of all the stars in a region, $\bar{m}$, to the average separation between stars, $\bar{s}$:
\begin{equation}
Q = \frac{\bar{m}}{\bar{s}},
\end{equation}
where $\mathcal{Q} < 0.7$ indicates a substructured distribution and $\mathcal{Q} > 0.9$ indicates a smooth, centrally concentrated distribution. In Section 3 we will plot $\mathcal{Q}$ against three other measures. 

First, we will take the ratio of the statistical radial velocity dispersion (the velocity measured along the $z$-axis), $\sigma$, to the interquartile range (IQR) of the radial velocities ($\sigma$/IQR). \citet{Parker16b} show that this ratio exceeds unity for clusters that have formed via violent relaxation and merging of substructure. Secondly, we determine the relative surface density of the most massive stars, $\Sigma_{\rm LDR}$, by comparing the median surface density of the ten most massive stars $\tilde{\Sigma}_{10}$ to the median surface density of all stars, $\tilde{\Sigma}_{\rm all}$ \citep{Maschberger11,Kupper11,Parker14b}:
\begin{equation}
\Sigma_{\rm LDR} = \frac{\tilde{\Sigma}_{10}}{\tilde{\Sigma}_{\rm all}},
\end{equation}
where $\Sigma >> 1$ indicates that the most massive stars are in areas of significantly higher than average stellar density. Finally, we will follow the evolution of the mass segregation ratio, $\Lambda_{\rm MSR}$, which compares the length of the minimum spanning tree of the $N_{\rm MST}$ most massive stars, $l_{\rm subset}$, to the average length $\langle l_{\rm average} \rangle$ of $N_{\rm MST}$ randomly chosen stars \citep{Allison09a}. There is a dispersion associated with the average length of random MSTs, which is roughly Gaussian and can be quantified as the standard deviation of the lengths $\langle l_{\rm average} \rangle \pm \sigma_{\rm average}$. Instead of using $\sigma_{\rm average}$, we conservatively estimate the lower (upper) uncertainty as the MST length which lies 1/6 (5/6) of the way through an ordered list of all the random lengths (corresponding to a 66 per cent deviation from the median value, $\langle l_{\rm average} \rangle$). This determination prevents a single outlying object from heavily influencing the uncertainty. The mass segregation ratio is then
\begin{equation}
\Lambda_{\rm MSR} = {\frac{\langle l_{\rm average} \rangle}{l_{\rm subset}}} ^{+ {\sigma_{\rm 5/6}}/{l_{\rm subset}}}_{-{\sigma_{\rm 1/6}}/{l_{\rm subset}}},
\end{equation}
where $\Lambda_{\rm MSR} >> 1$ indicates that a star-forming region is significantly mass segregated. 

\section{Results}

\subsection{Subvirial (collapsing) star-forming regions}

\begin{figure*}
  \begin{center}
\setlength{\subfigcapskip}{10pt}

\hspace*{-1.5cm}\subfigure[0\,Myr]{\label{subvirial-a}\rotatebox{270}{\includegraphics[scale=0.20]{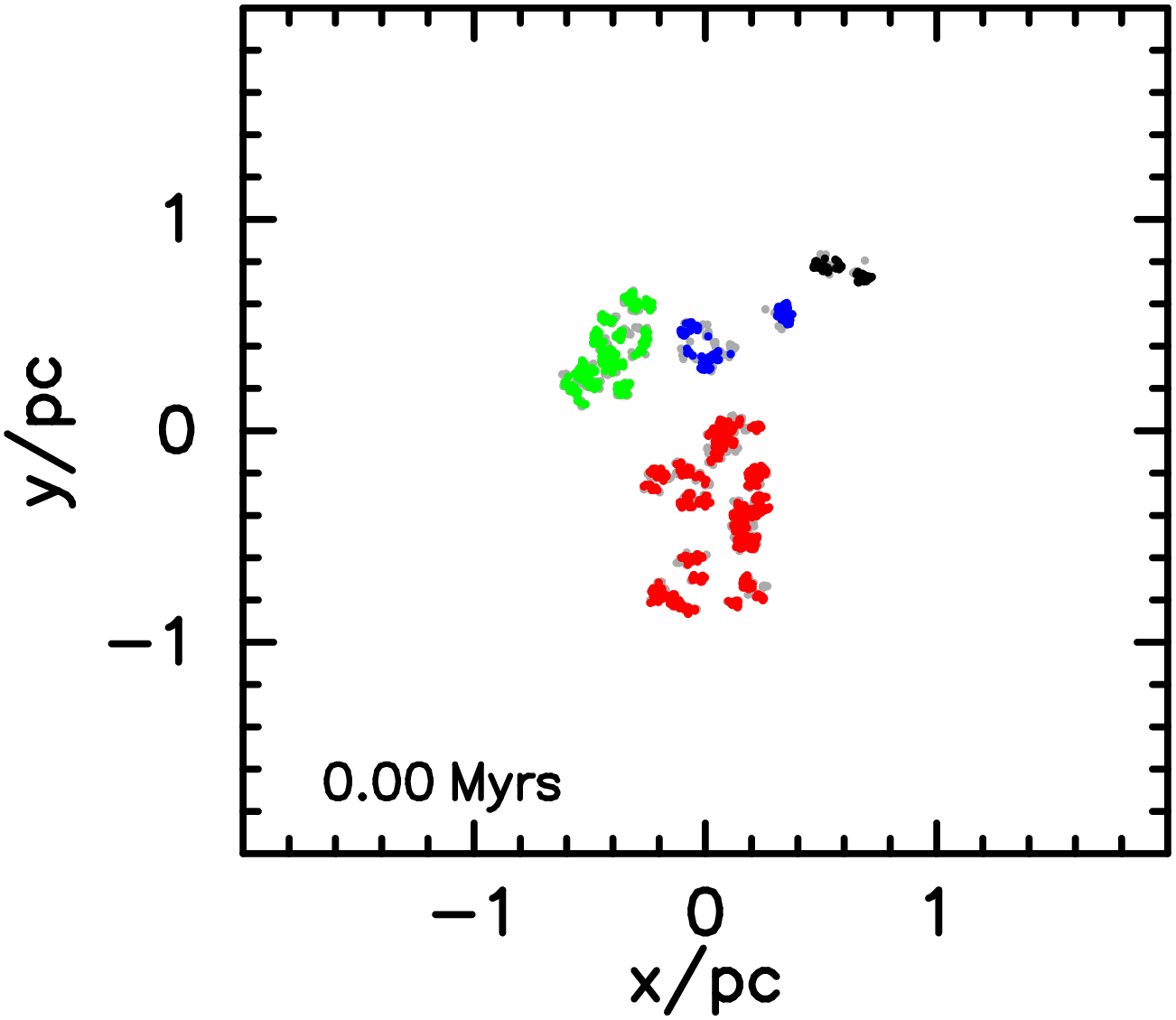}}}
\hspace*{0.3cm} 
\subfigure[0\,Myr]{\label{subvirial-b}\rotatebox{270}{\includegraphics[scale=0.20]{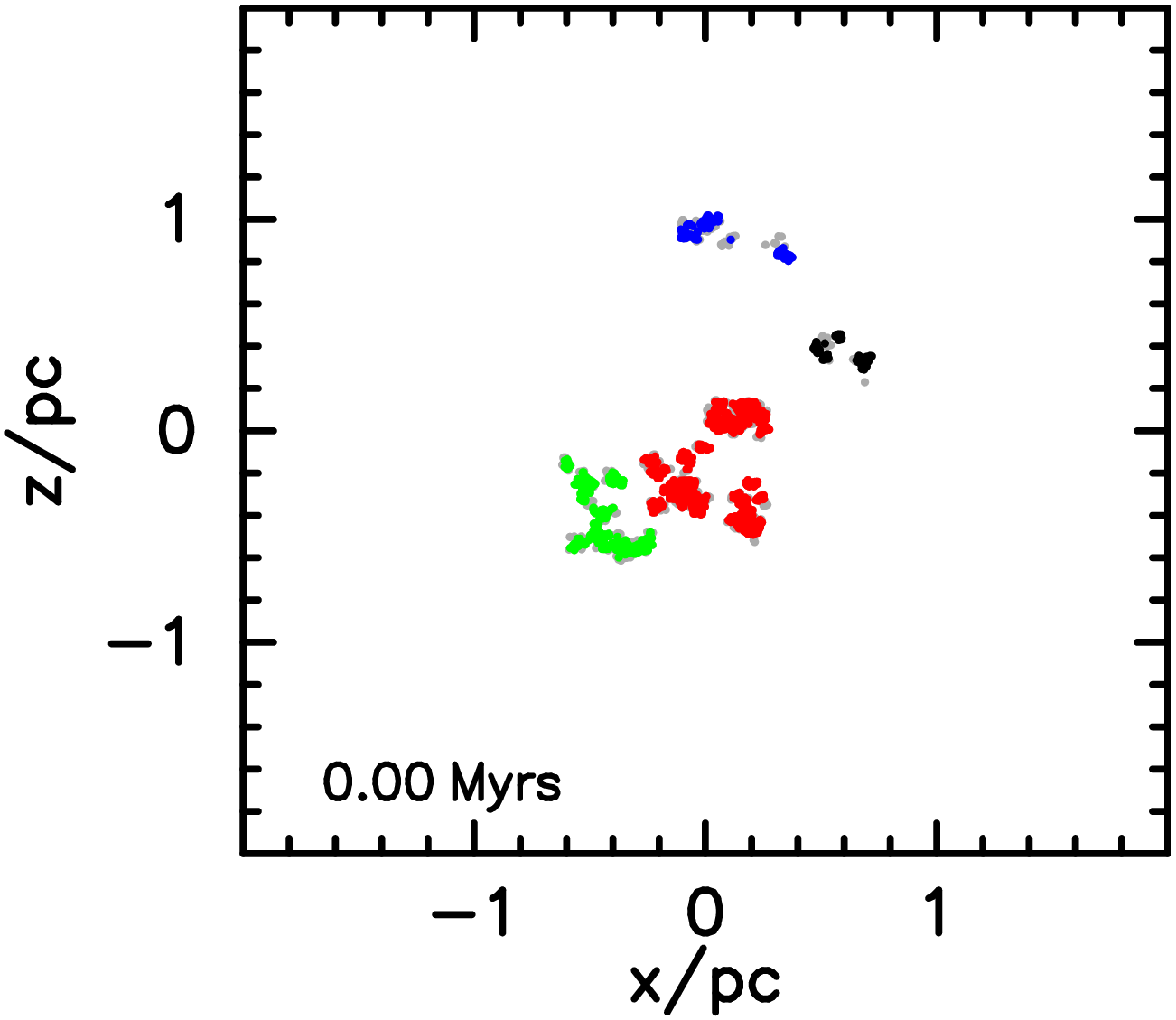}}} 
\hspace*{0.3cm}\subfigure[0\,Myr]{\label{subvirial-c}\rotatebox{270}{\includegraphics[scale=0.20]{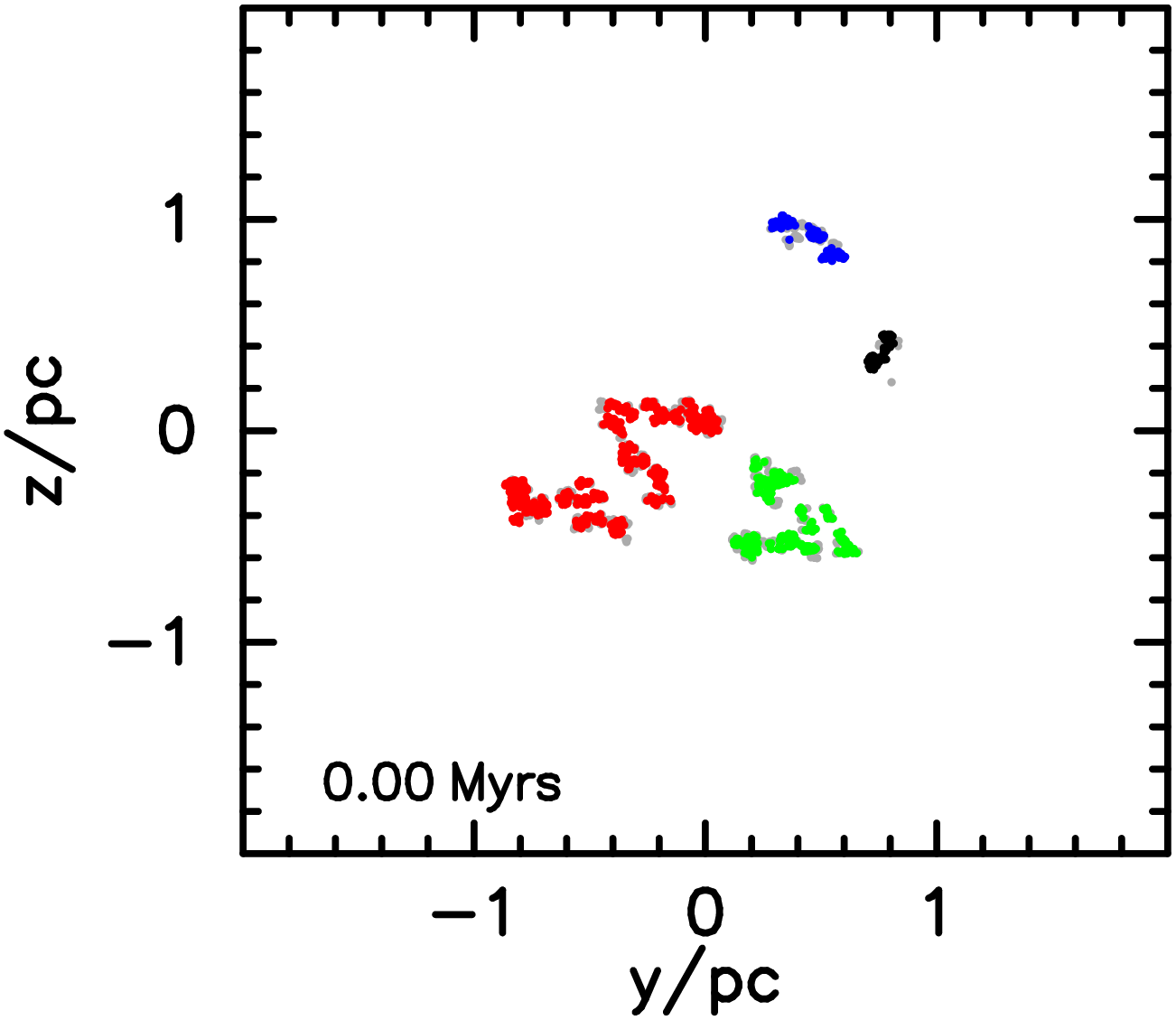}}}

\hspace*{-1.5cm}\subfigure[0.1\,Myr]{\label{subvirial-d}\rotatebox{270}{\includegraphics[scale=0.20]{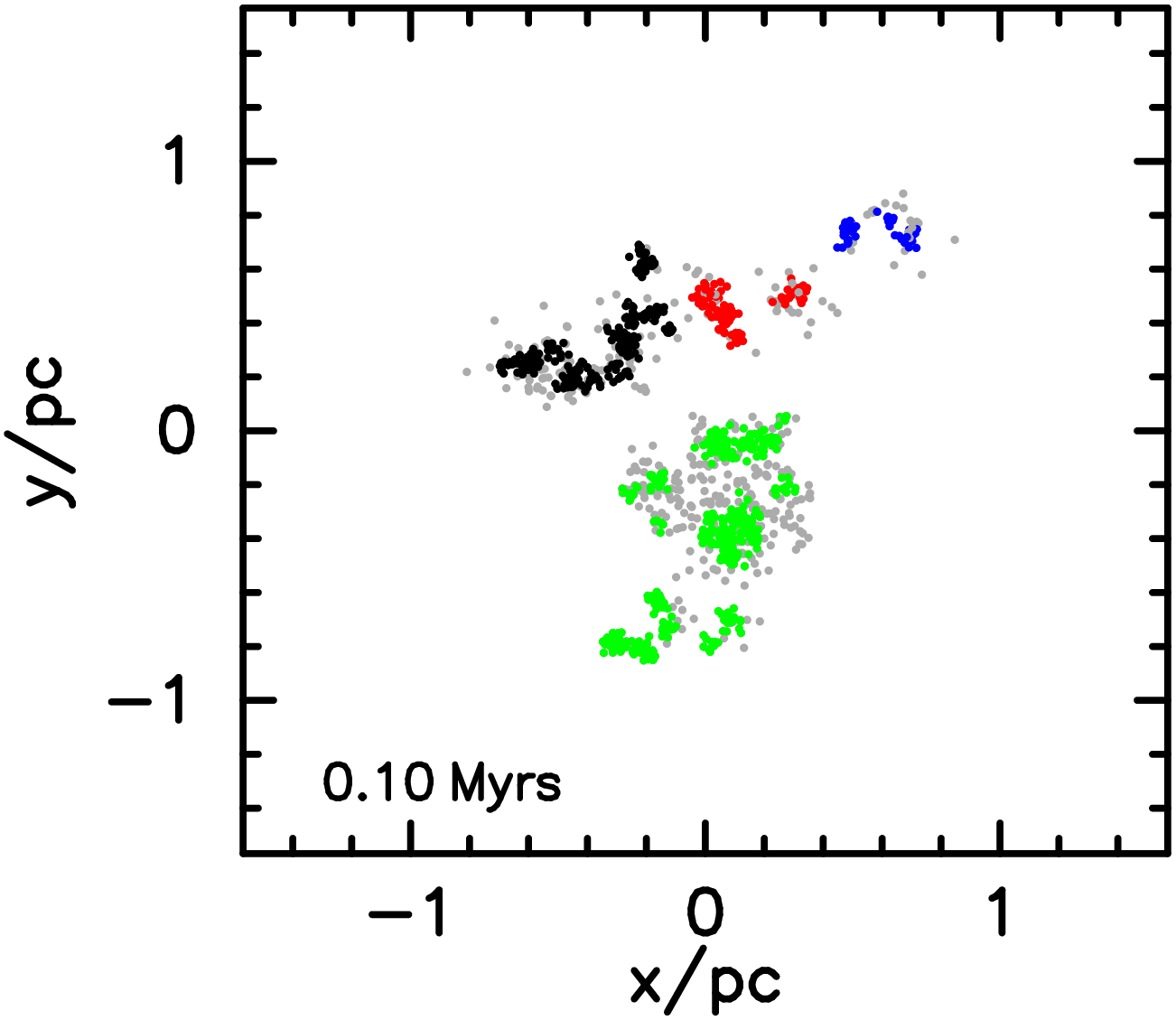}}}
\hspace*{0.3cm} 
\subfigure[0.1\,Myr]{\label{subvirial-e}\rotatebox{270}{\includegraphics[scale=0.20]{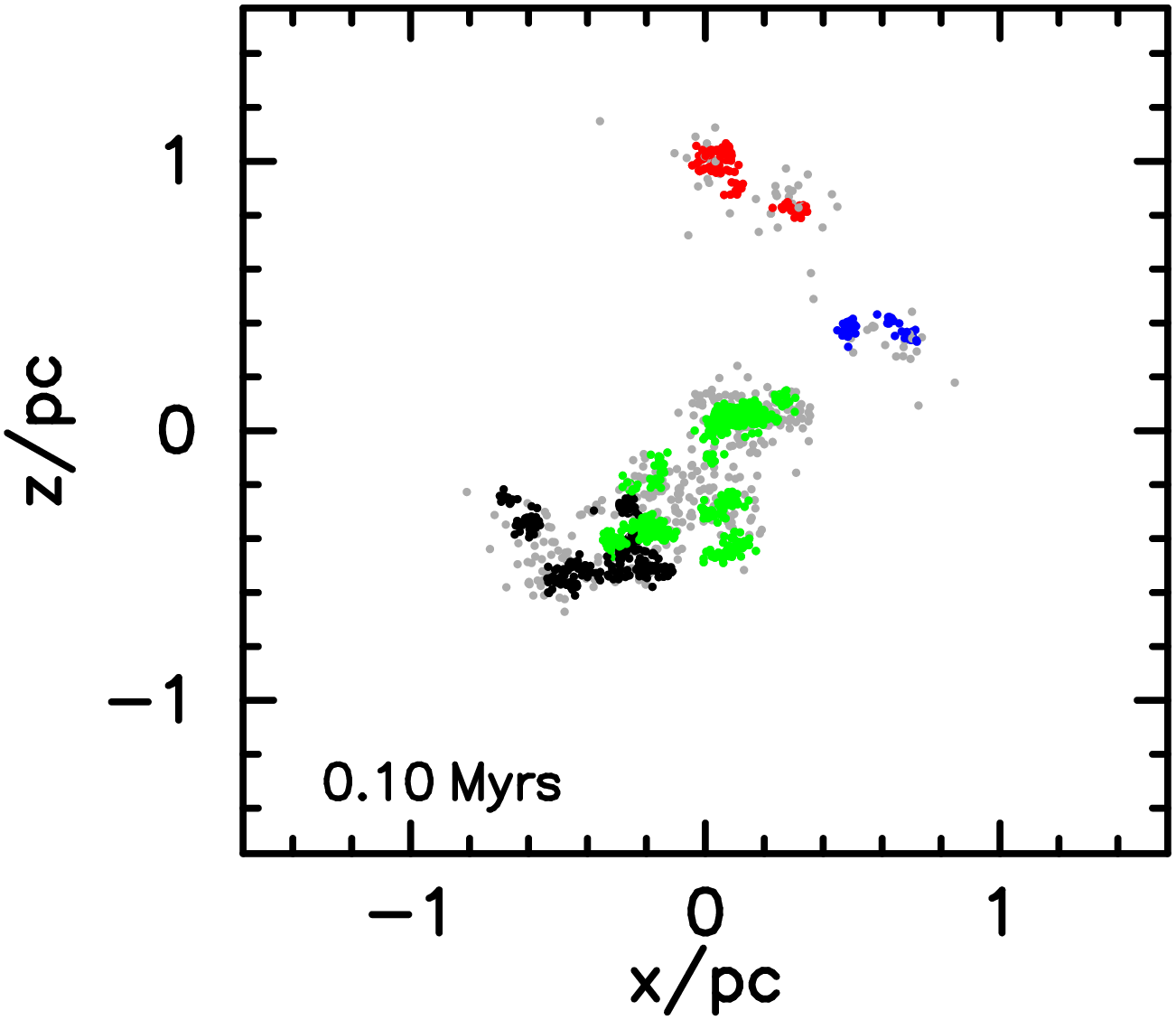}}} 
\hspace*{0.3cm}\subfigure[0.1\,Myr]{\label{subvirial-f}\rotatebox{270}{\includegraphics[scale=0.20]{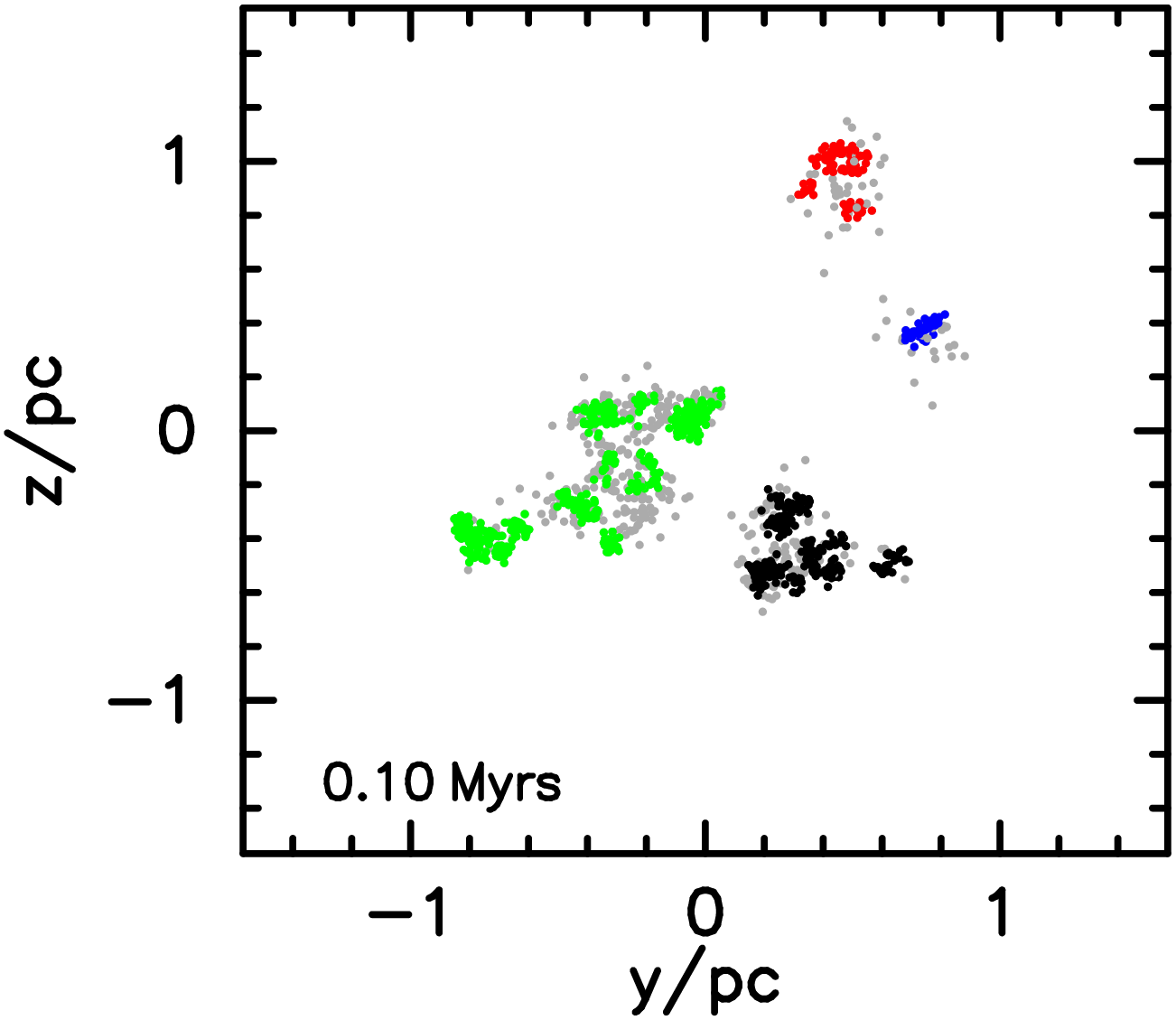}}}

\hspace*{-1.5cm}\subfigure[0.4\,Myr]{\label{subvirial-g}\rotatebox{270}{\includegraphics[scale=0.20]{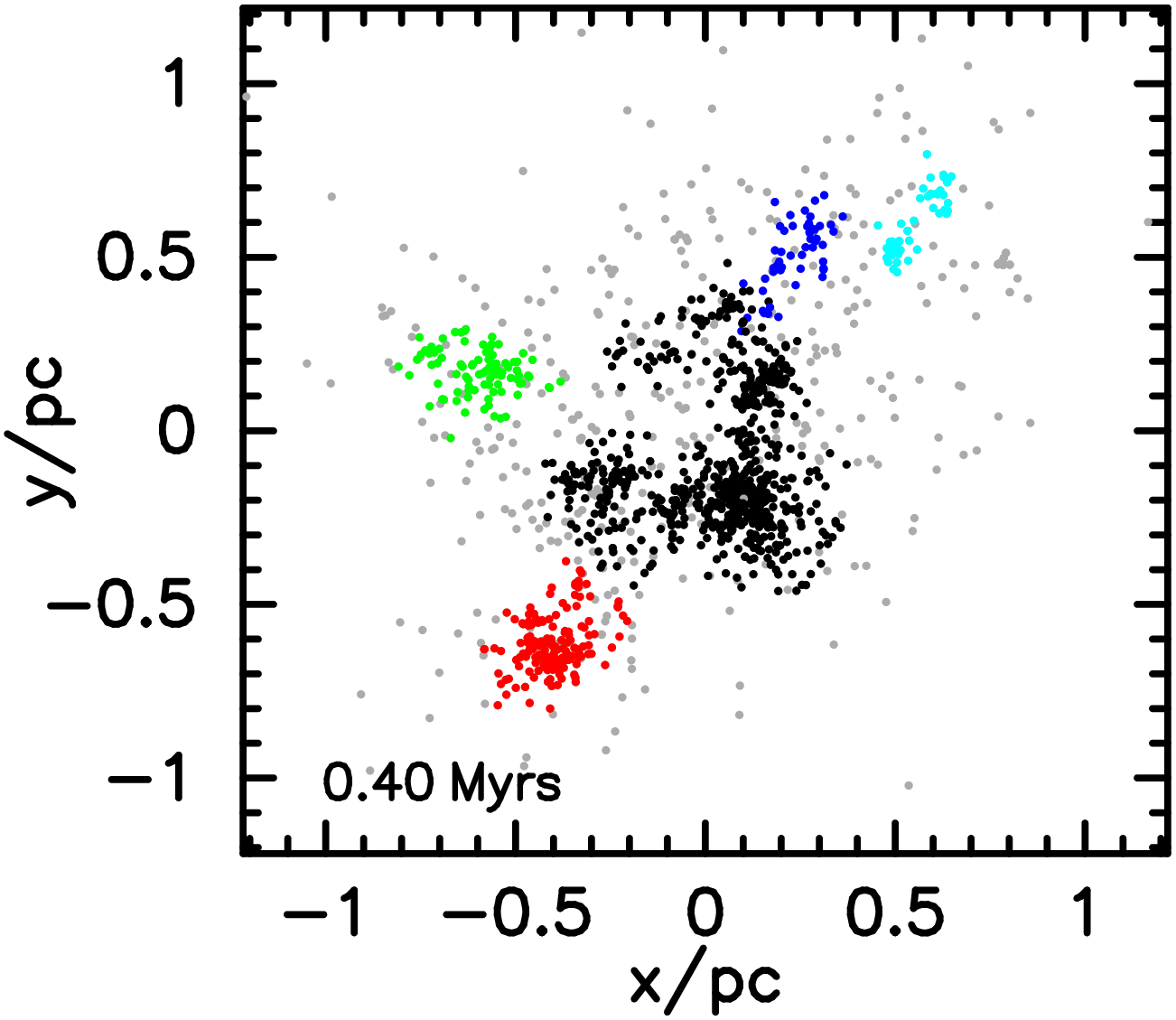}}}
\hspace*{0.3cm} 
\subfigure[0.4\,Myr]{\label{subvirial-h}\rotatebox{270}{\includegraphics[scale=0.20]{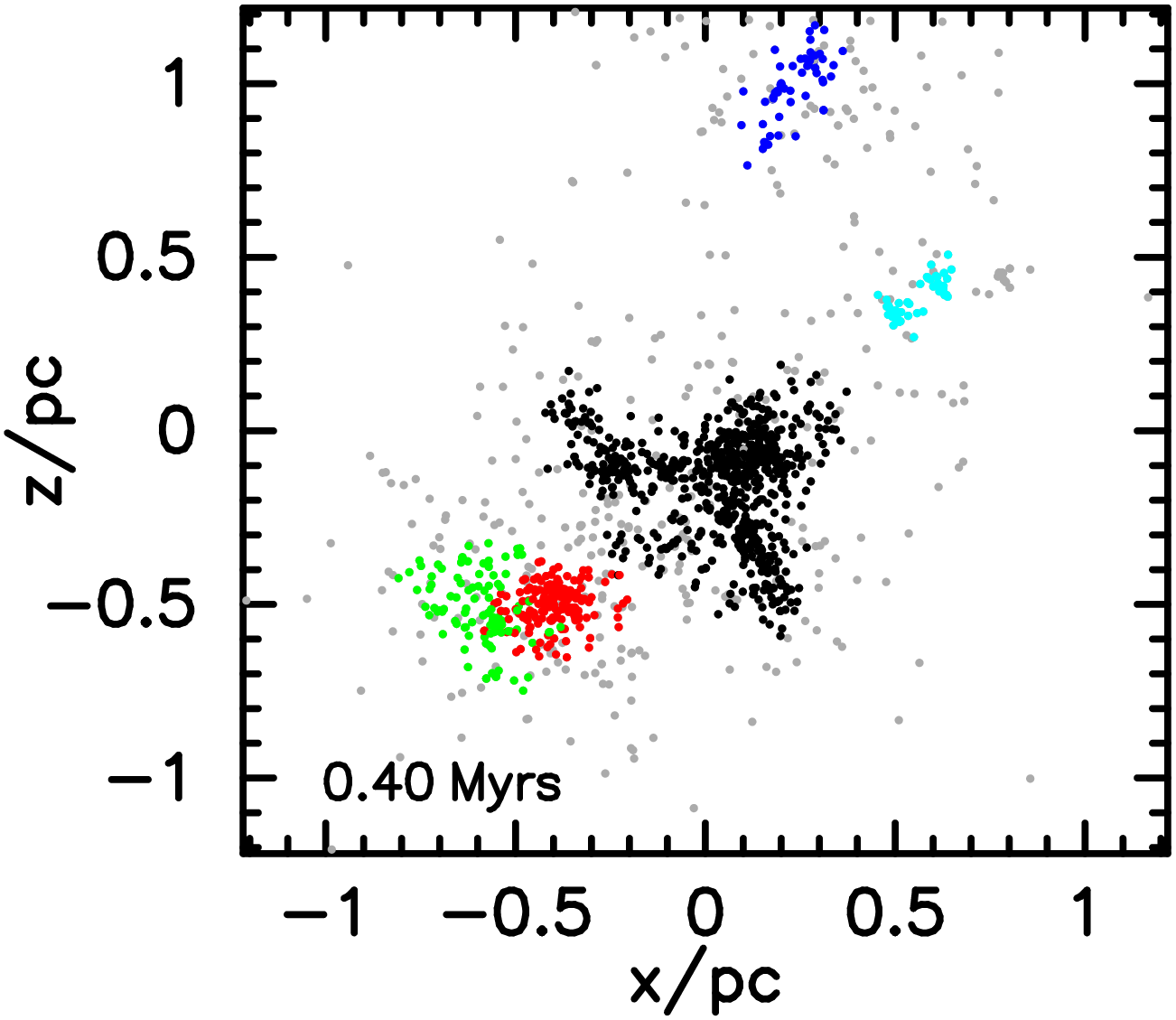}}} 
\hspace*{0.3cm}\subfigure[0.4\,Myr]{\label{subvirial-i}\rotatebox{270}{\includegraphics[scale=0.20]{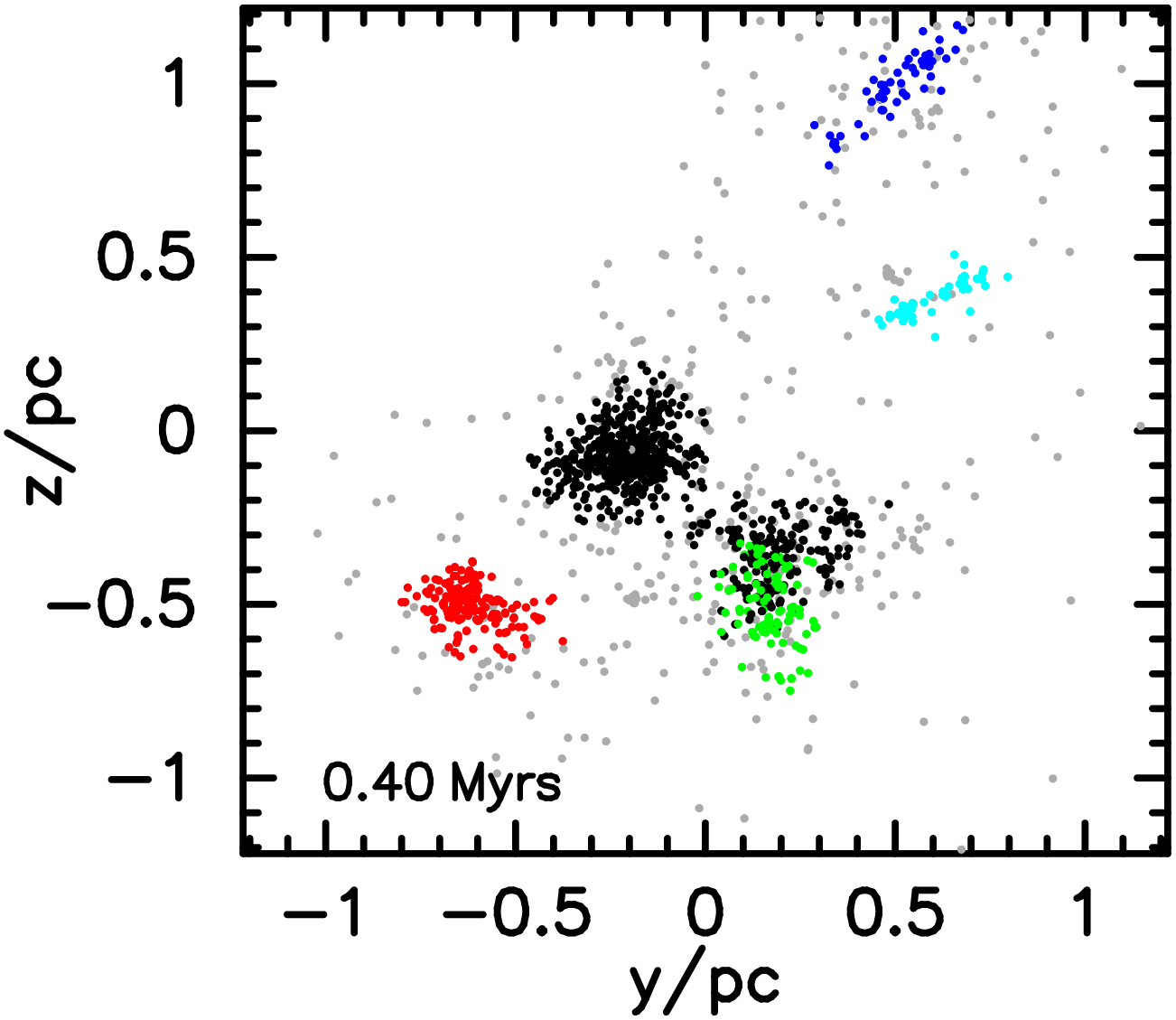}}}

\hspace*{-1.5cm}\subfigure[0.7\,Myr]{\label{subvirial-j}\rotatebox{270}{\includegraphics[scale=0.20]{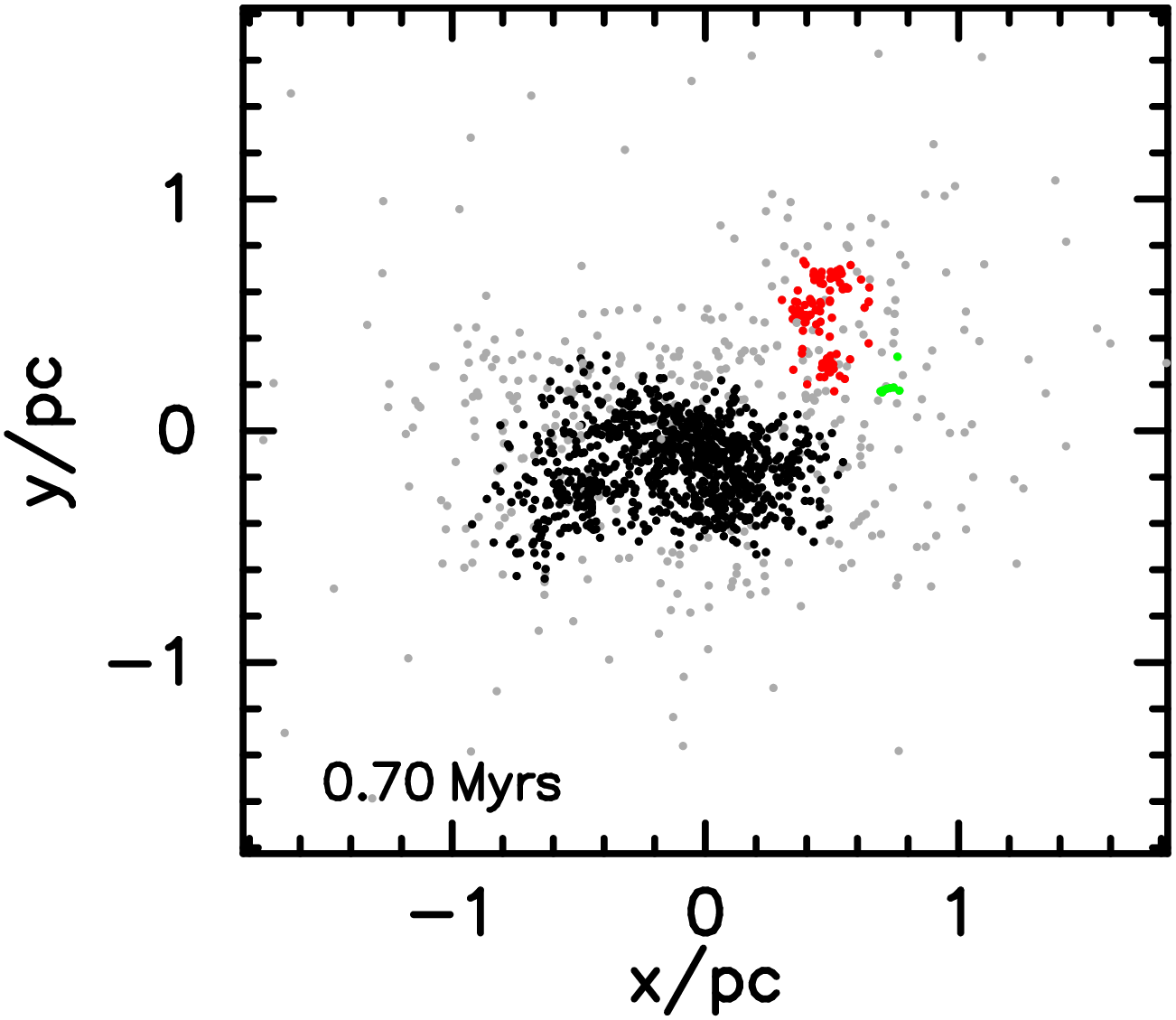}}}
\hspace*{0.3cm} 
\subfigure[0.7\,Myr]{\label{subvirial-k}\rotatebox{270}{\includegraphics[scale=0.20]{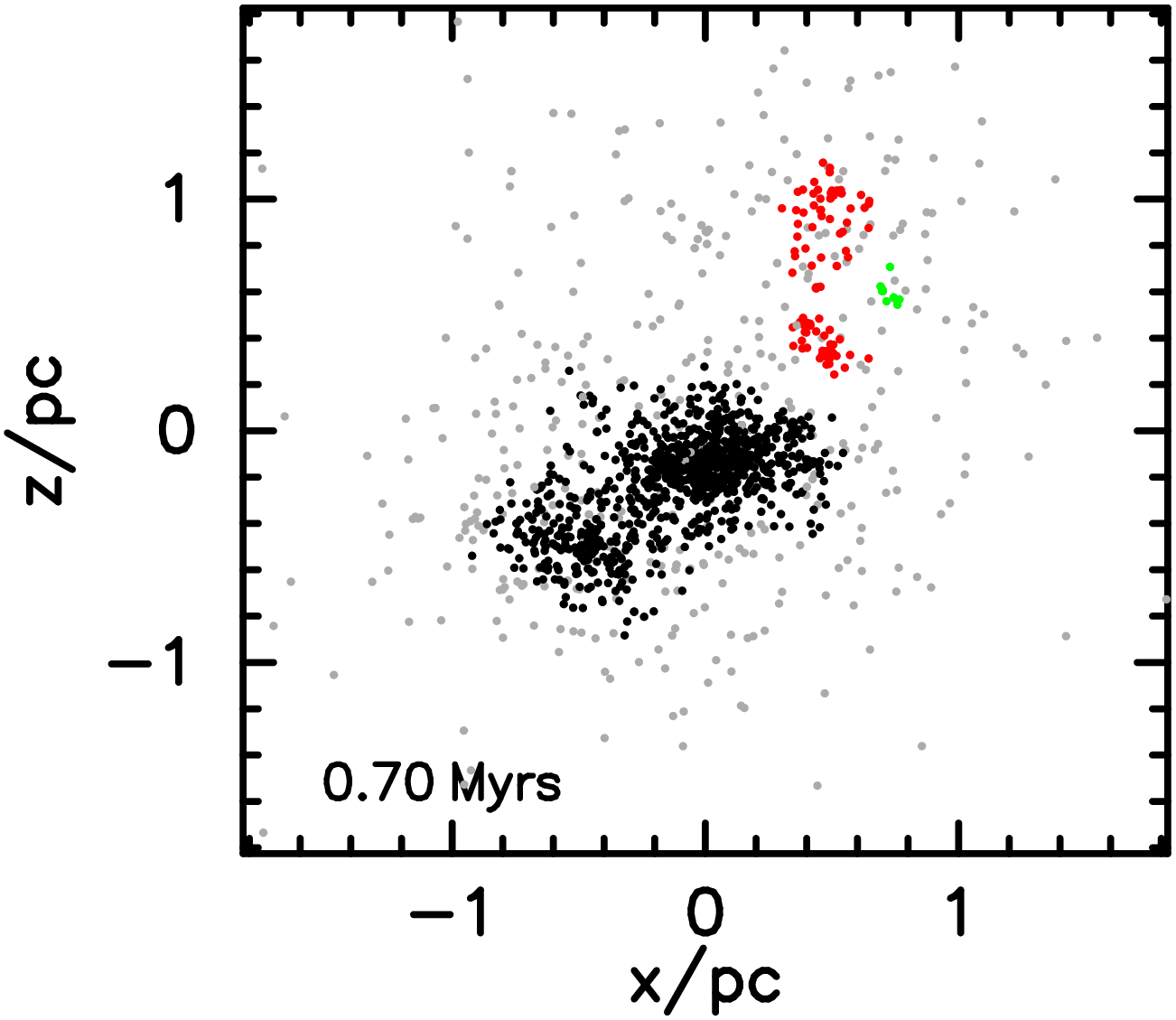}}} 
\hspace*{0.3cm}\subfigure[0.7\,Myr]{\label{subvirial-l}\rotatebox{270}{\includegraphics[scale=0.20]{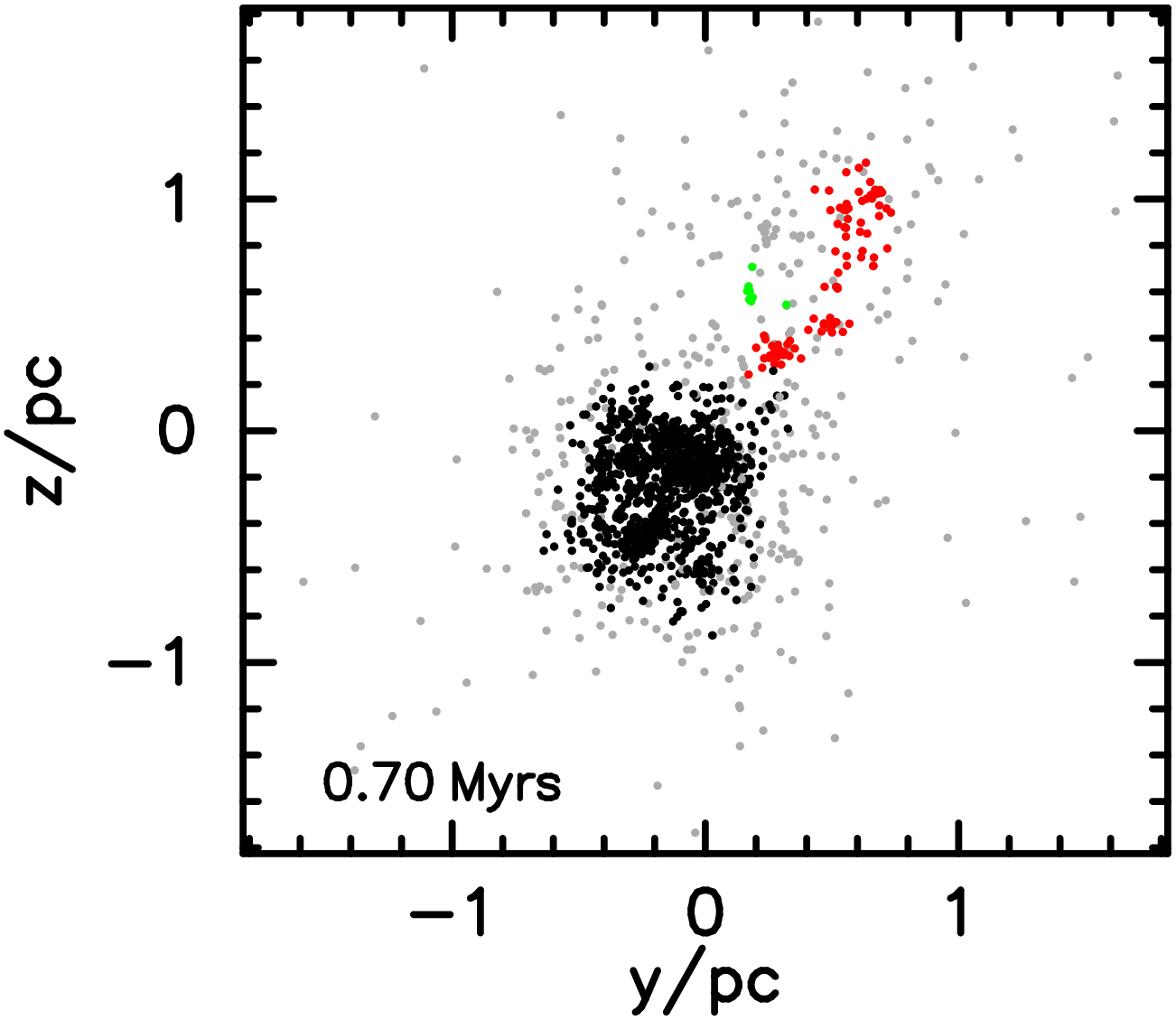}}}

\hspace*{-1.5cm}\subfigure[3\,Myr]{\label{subvirial-m}\rotatebox{270}{\includegraphics[scale=0.20]{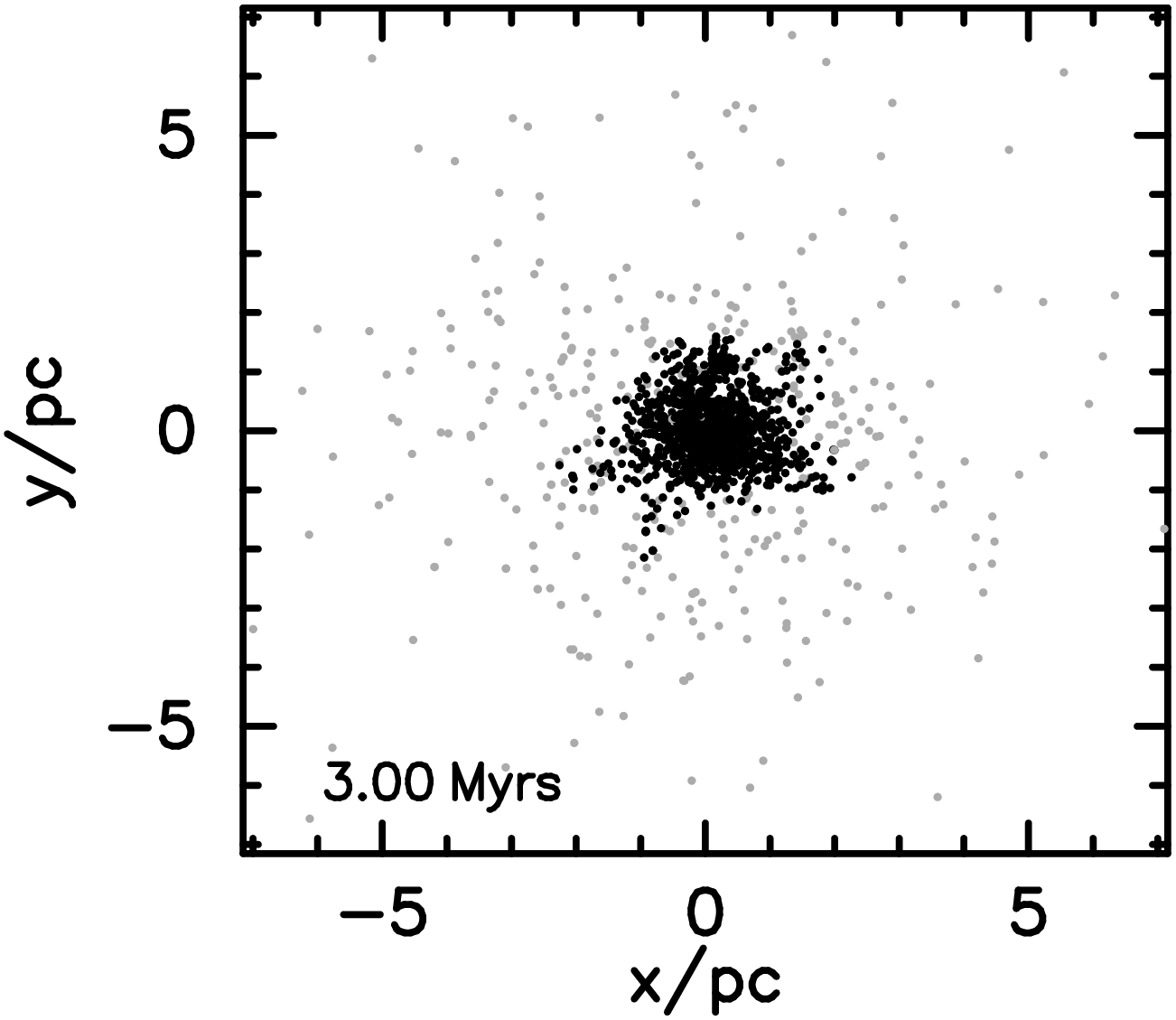}}}
\hspace*{0.3cm} 
\subfigure[3\,Myr]{\label{subvirial-n}\rotatebox{270}{\includegraphics[scale=0.20]{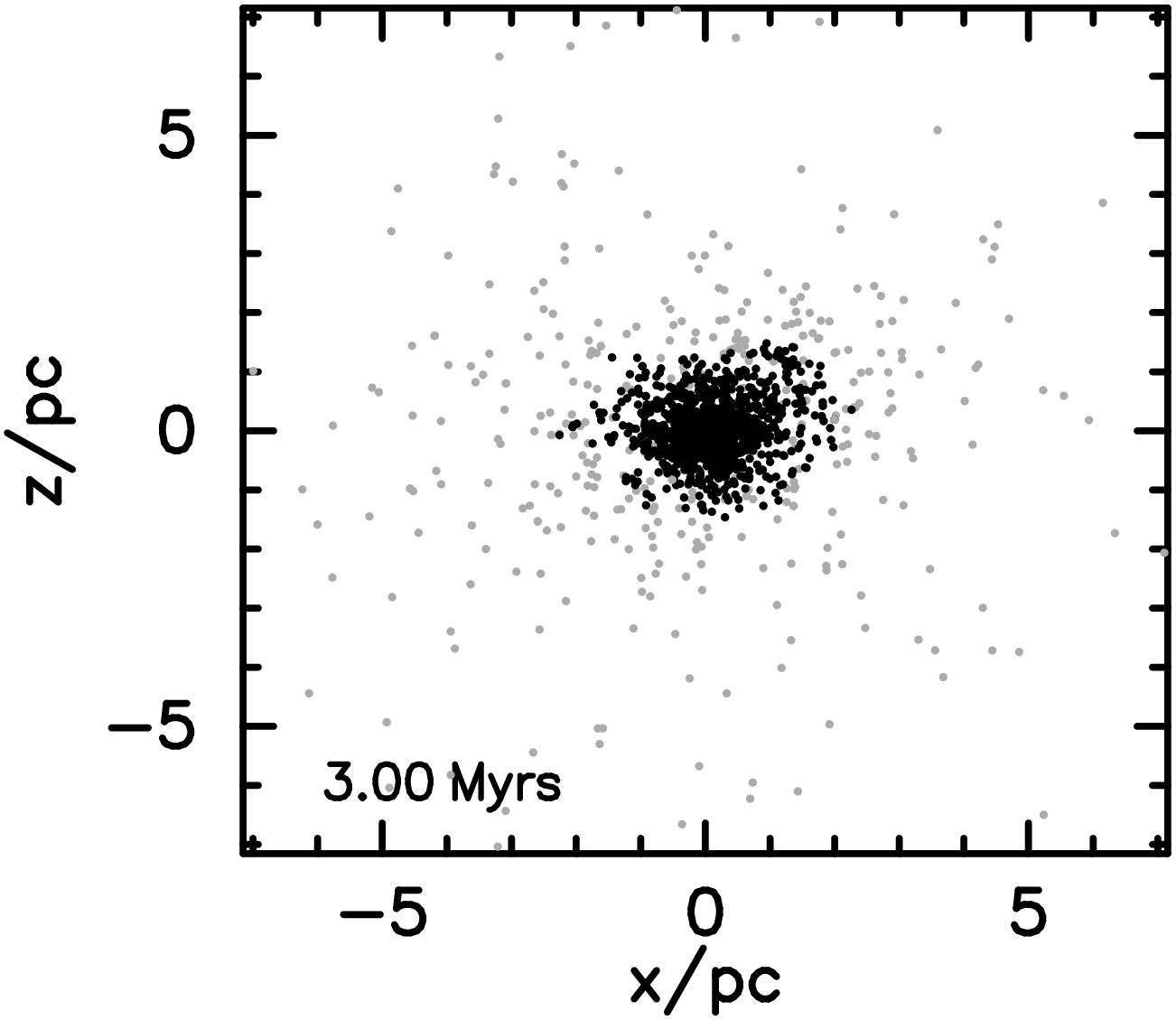}}} 
\hspace*{0.3cm}\subfigure[3\,Myr]{\label{subvirial-o}\rotatebox{270}{\includegraphics[scale=0.20]{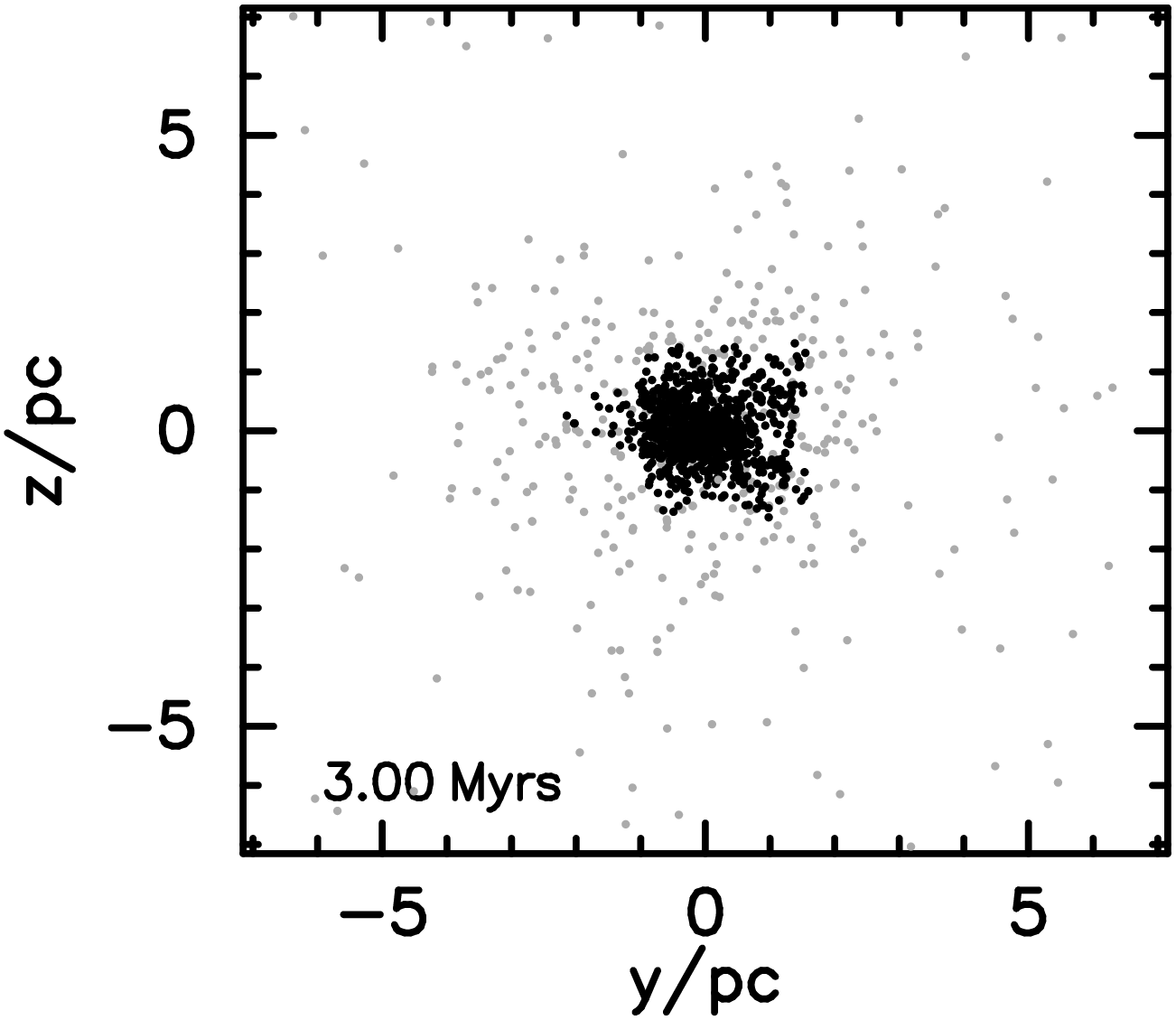}}}

\vspace*{0.25cm}
\caption[bf]{Evolution of groups defined by the Friends of Friends algorithm in a simulated subvirial ($\alpha_{\rm vir} = 0.3$) star-forming region with initial fractal dimension $D = 1.6$. Stars that have a local stellar density below the first quartile in the distribution are not assigned to a group and are coloured grey. The colours in the subsequent snapshots do not correspond to the colours in the first snapshot ($t = 0$\,Myr). }
  \label{subvirial}
  \end{center}
\end{figure*}

We show five snapshots of the evolution of a typical subvirial simulation in Fig.~\ref{subvirial}, with three different viewing angles (along the $z$, $y$, and $x$ axes, respectively) for each snapshot. Each group detected by the Friends of Friends algorithm is shown by a different colour, with stars sitting below the density threshold shown by the grey points. We emphasise that due to the disappearance (or sometimes formation) of groups, long-lived groups are not necessarily shown by the same colour in different snapshots. The Friends of Friends algorithm has detected four distinct groups at $t = 0$\,Myr in this particular simulation (panels a -- c). As this simulation has a subvirial bulk motion, we would expect the individual groups to evolve and merge as the star-forming region coalesces to a cluster, which occurs during the first 1\,Myr. Interestingly, however, the number of groups briefly increases to 5 by 0.4\,Myr (panels g -- i) before reducing to one main group after 0.7\,Myr (panels j - o).

All of the subvirial simulations lose their substructure within the first 1\,Myr and form a bound cluster \citep{Parker12d,Parker14b}. Interestingly, the number of distinct groups that the Friends of Friends algorithm identifies at $t = 0$\,Myr varies significantly. In our suite of twenty simulations, identical apart from the random number seed used to set the positions, velocities and stellar masses, the number of groups identified varies between 1 and 25, where the mode is 4 and the median number of groups is 8.

To check whether this is due to the automatically calculated threshold lengths, we plot the evolution of both the distance length, $\tilde{\Delta r}/3$ (solid lines) and the velocity length, $\tilde{\Delta v}$ (dashed lines) for simulations where the Friends of Friends algorithm identifies 4 groups (black lines), 8 groups (green lines) and 25 groups (red lines). Whilst the distance threshold is smaller when more groups are identified, this trend is not fulfilled for the velocity thresholds. We also note that the differences in both thresholds between the three simulations are very small (all are less than 0.4\,pc and $\sim$\,1\,km\,s$^{-1}$). 

\begin{figure*}
  \begin{center}
\setlength{\subfigcapskip}{10pt}
\hspace*{-1.5cm}\subfigure[0\,Myr]{\label{subvirial_other-a}\rotatebox{270}{\includegraphics[scale=0.20]{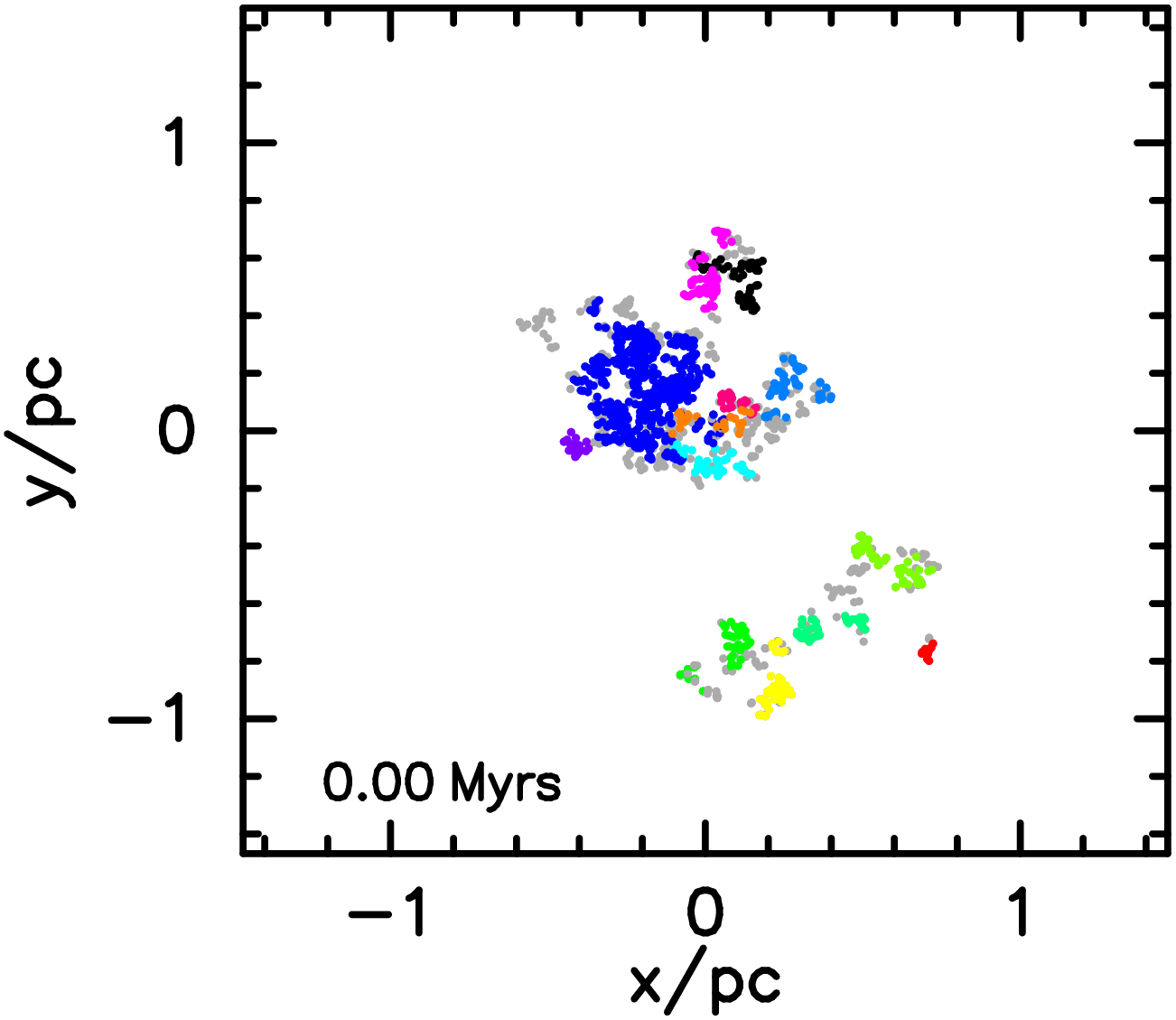}}}
\hspace*{0.3cm} 
\subfigure[0\,Myr]{\label{subvirial_other-b}\rotatebox{270}{\includegraphics[scale=0.20]{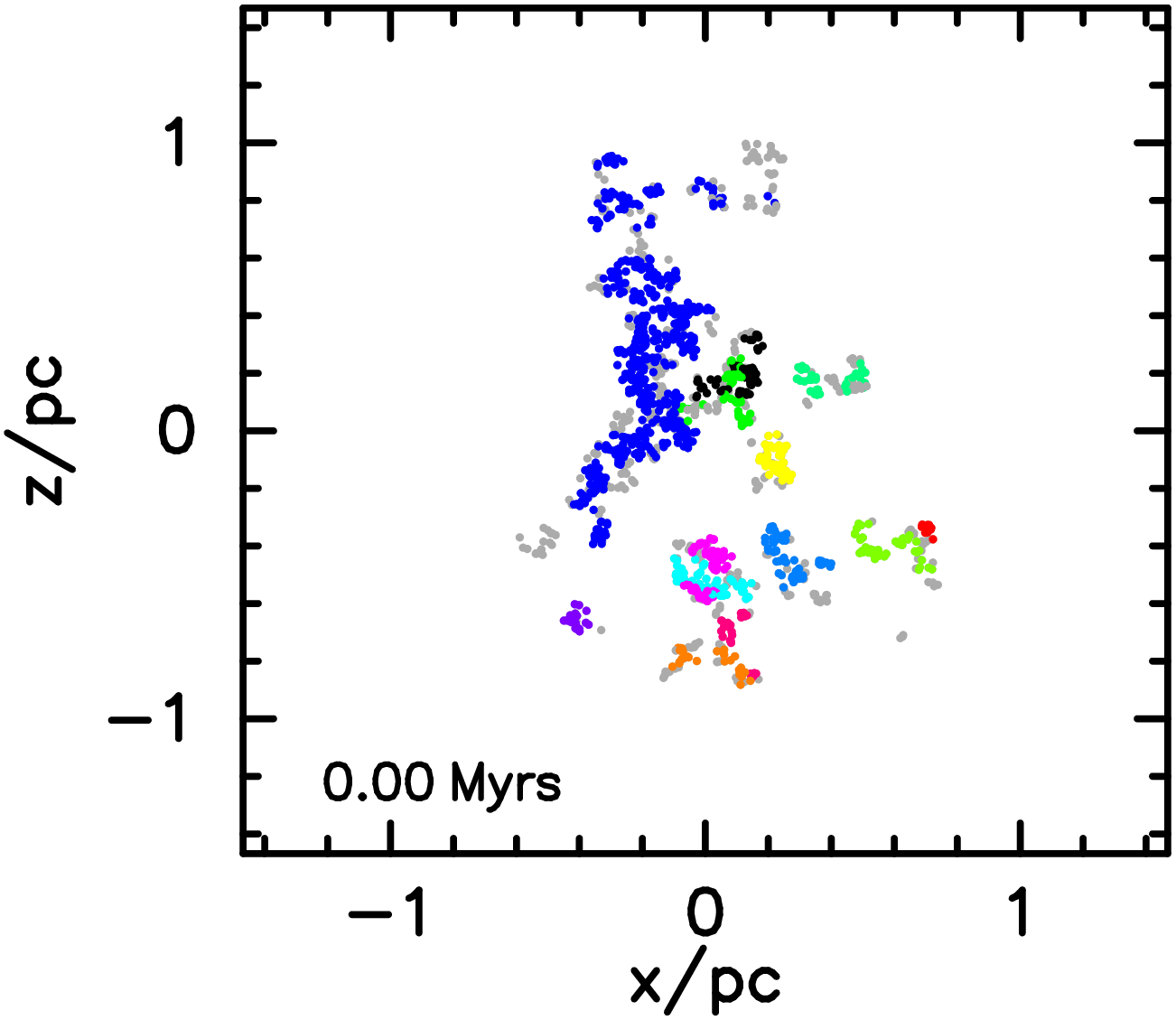}}} 
\hspace*{0.3cm}\subfigure[0\,Myr]{\label{subvirial_other-c}\rotatebox{270}{\includegraphics[scale=0.20]{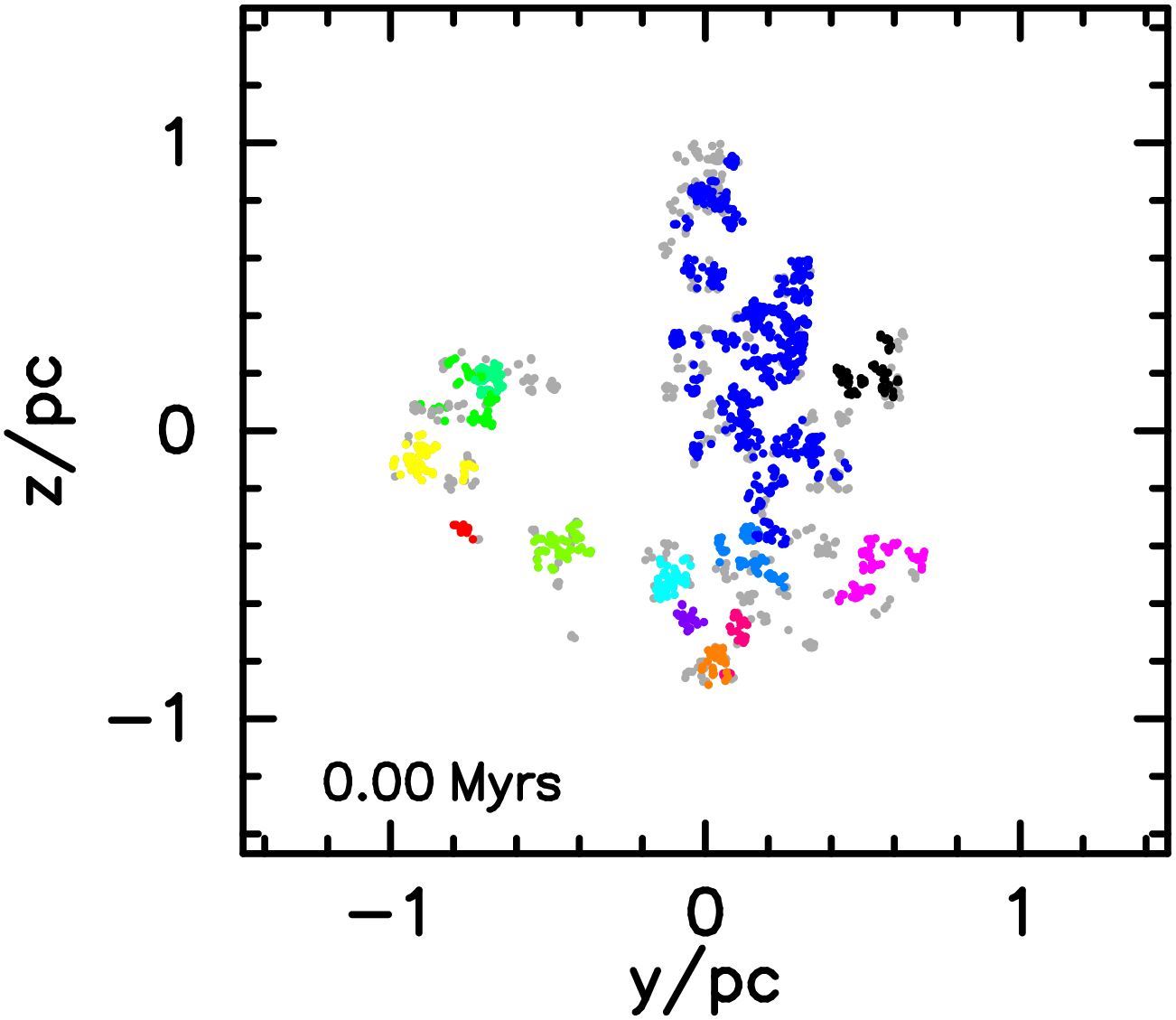}}}

\hspace*{-1.5cm}\subfigure[3\,Myr]{\label{subvirial_other-d}\rotatebox{270}{\includegraphics[scale=0.20]{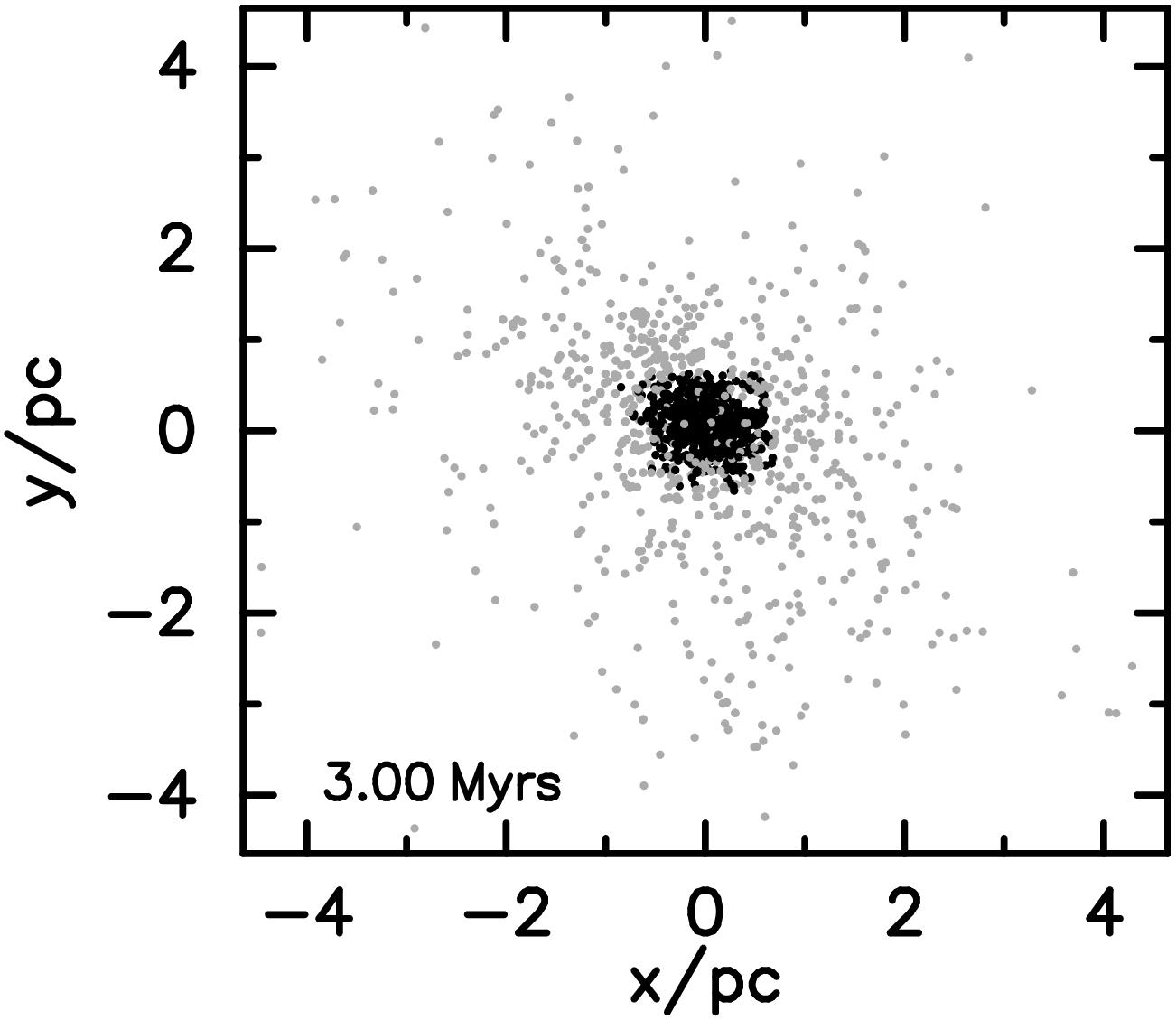}}}
\hspace*{0.3cm} 
\subfigure[3\,Myr]{\label{subvirial_other-e}\rotatebox{270}{\includegraphics[scale=0.20]{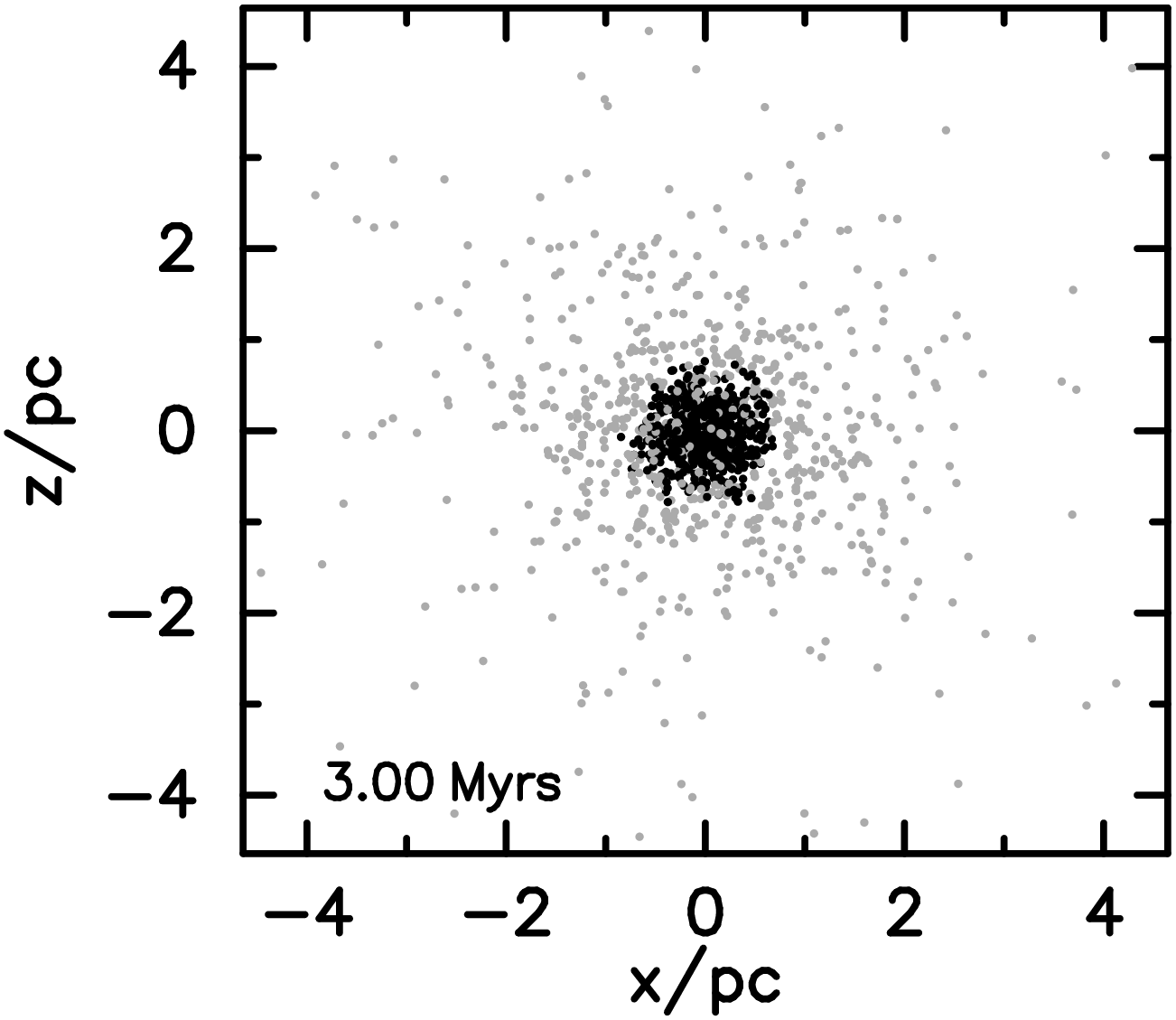}}} 
\hspace*{0.3cm}\subfigure[3\,Myr]{\label{subvirial_other-f}\rotatebox{270}{\includegraphics[scale=0.20]{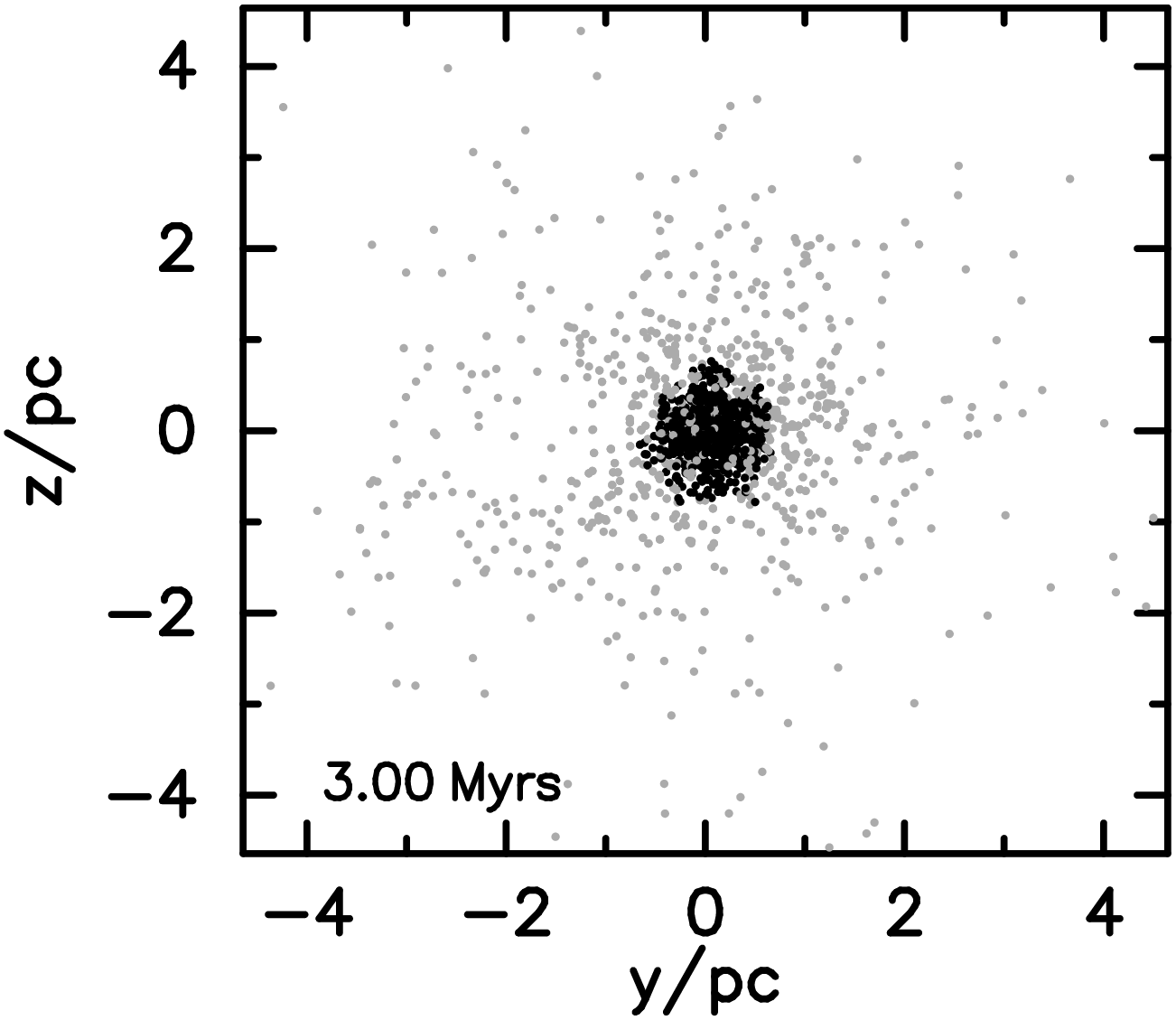}}}
\caption[bf]{As Fig.~\ref{subvirial}, but showing the evolution of groups defined by the Friends of Friends algorithm for a different realisation of a simulated subvirial ($\alpha_{\rm vir} = 0.3$) star-forming region with $D = 1.6$. This simulation is identical to that in Fig.~\ref{subvirial}, save for the random number seed used to assign positions, velocities and masses to the stars. In this simulation, 13 distinct groups are found by the Friends of Friends algorithm, compared to 4 in the simulation shown in Fig.~\ref{subvirial}. Stars that have a local stellar density below the first quartile in the distribution are not assigned to a group and are coloured grey. The colours in the subsequent snapshots do not correspond to the colours in the first snapshot ($t = 0$\,Myr). }
  \label{subvirial_other}
  \end{center}
\end{figure*}

Fig.~\ref{subvirial_other} shows a different subvirial simulation, with 13 groups identified initially, which evolves to a similar-looking single cluster after 1\,Myr, which is indistinguishable from the cluster shown in Fig.~\ref{subvirial}.

\begin{figure*}
  \begin{center}
\setlength{\subfigcapskip}{10pt}
\hspace*{-1.5cm}\subfigure[$\mathcal{Q}$ -- $\sigma$/IQR]{\label{subvirial_structure-a}\rotatebox{270}{\includegraphics[scale=0.25]{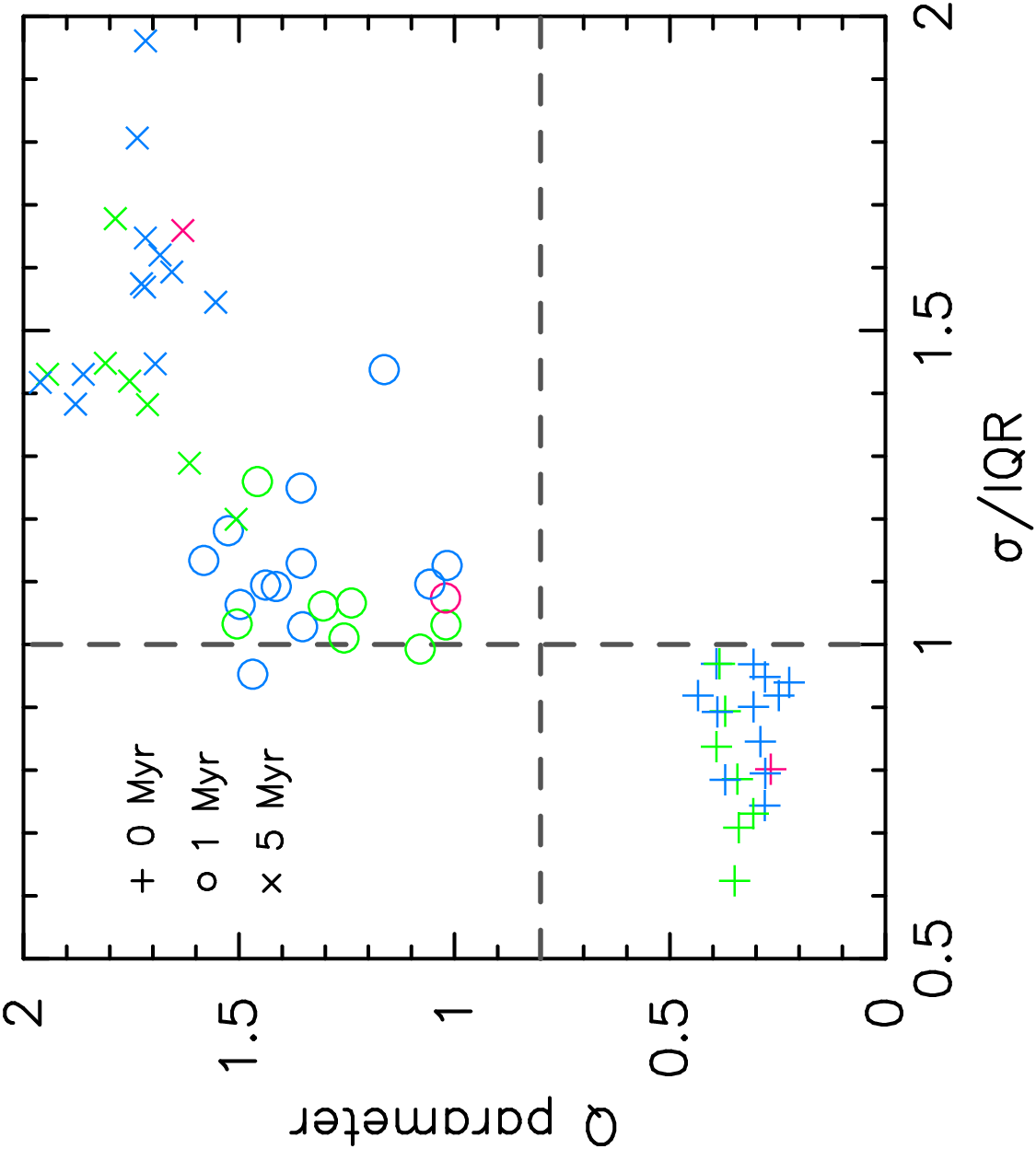}}}
\hspace*{0.3cm} 
\subfigure[$\mathcal{Q}$ -- $\Sigma_{\rm LDR}$]{\label{subvirial_structure-b}\rotatebox{270}{\includegraphics[scale=0.25]{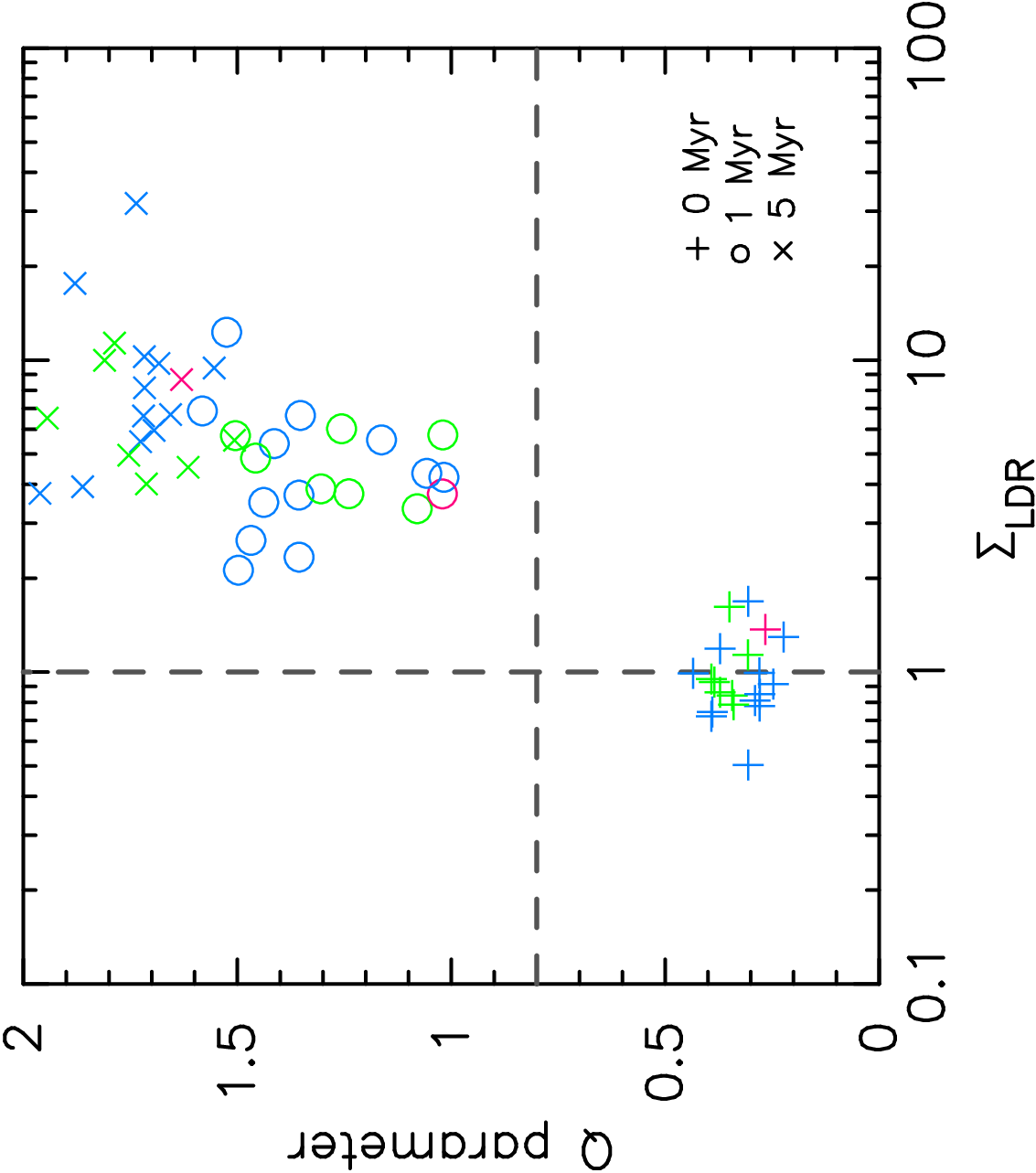}}} 
\hspace*{0.3cm}\subfigure[$\mathcal{Q}$ -- $\Lambda_{\rm MSR}$]{\label{subvirial_structure-c}\rotatebox{270}{\includegraphics[scale=0.25]{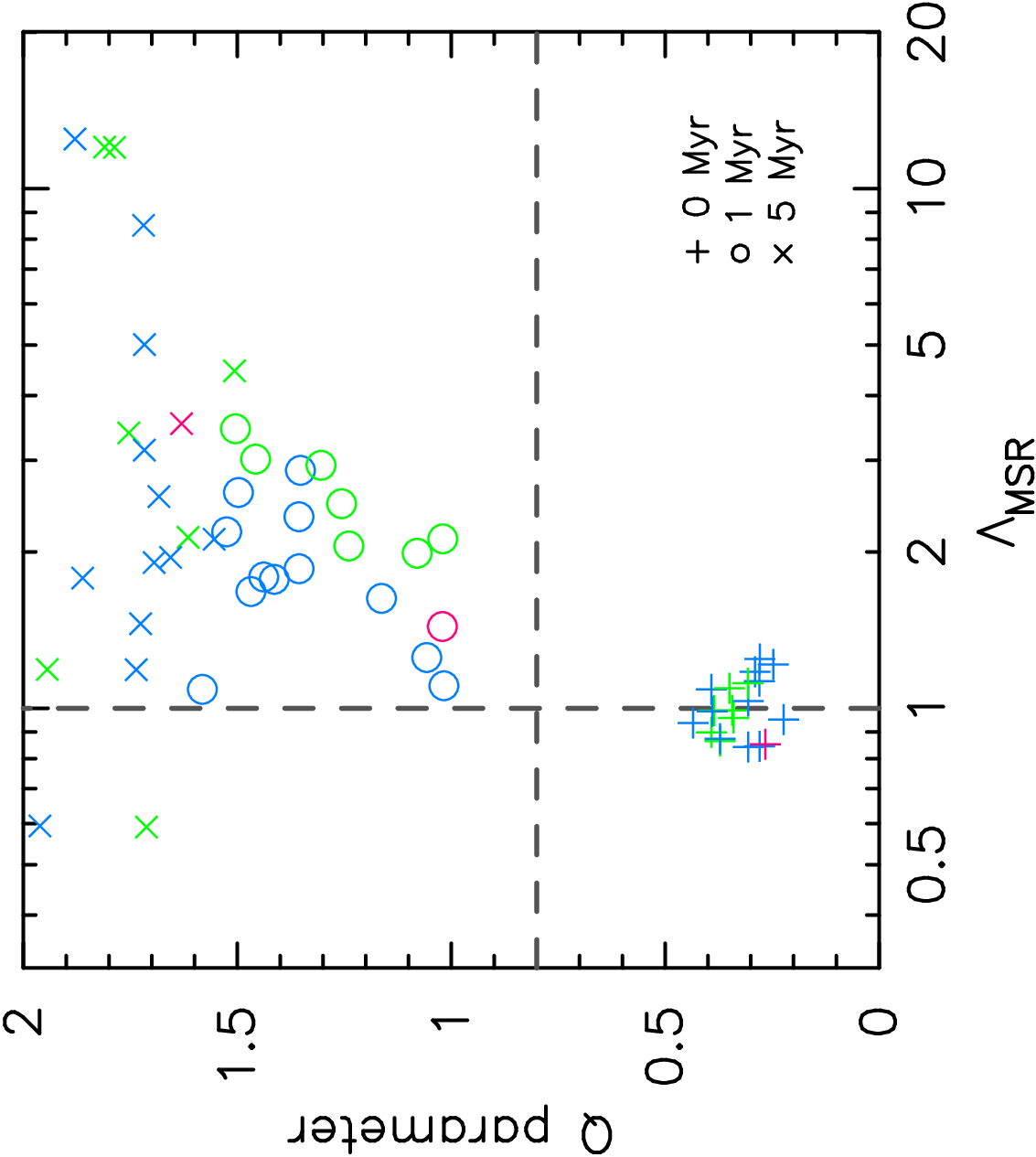}}}
\caption[bf]{Different measures of spatial and kinematic evolution in our subvirial ($\alpha_{\rm vir} = 0.3$) simulations. In panel (a) we show the \citet{Cartwright04} $\mathcal{Q}$-parameter against the radial velocity dispersion divided by the interquartile range of radial velocities \citep{Parker16b}. In panel (b) we show the $\mathcal{Q}$-parameter against the relative local surface density ratio of the ten most massive stars, $\Sigma_{\rm LDR}$ \citep{Kupper11,Parker14b}. In panel (c) we show the $\mathcal{Q}$-parameter against the mass segregation ratio $\Lambda_{\rm MSR}$ \citep{Allison09a}. The boundary between hierarchically substructured and centrally concentrated distributions is shown by the horizontal dashed lines, and the vertical dashed lines correspond to unity for the other measures, indicating no special configuration for the massive stars. In all panels, the green symbols indicate simulations where the Friends of Friends algorithm picks out 5 groups or fewer at  $t = 0$\,Myr; the blue symbols indicate between 5 and 15 groups, and the magenta symbols indicate that the Friends of Friends algorithm has picked out more than 15 groups. There is no strong dependence of the dynamical evolution on the initial number of groups.}
  \label{subvirial_structure}
  \end{center}
\end{figure*}

To investigate whether the long-term evolution of the cluster depends on the initial number of groups identified by the Friends of Friends algorithm, in Fig.~\ref{subvirial_structure} we plot the $\mathcal{Q}$-parameter against the kinematic and spatial diagnostics of cluster evolution, $\sigma$/IQR, $\Sigma_{\rm LDR}$ and $\Lambda_{\rm MSR}$. (We refer the interested reader to \citet{Parker16b} and \citet{Parker14b} for detailed descriptions of how these diagnostics evolve over time due to dynamical relaxation.) For this paper, we colour code these plots according to the number of groups the Friends of Friends algorithm picks out. The orange coloured points indicate simulations where the Friends of Friends algorithm picks out 5 groups or fewer at  $t = 0$\,Myr; the blue symbols indicate between 5 and 15 groups, and the magenta symbols indicate that the Friends of Friends algorithm has picked out more than 15 groups. It is clear that the initial number of groups picked out by the algorithm is not related to the magnitude of the $\sigma$/IQR ratio, or the spatial structure, $\mathcal{Q}$, the relative local surface density, $\Sigma_{\rm LDR}$, or the occurrence and amount of mass segregation as measured by  $\Lambda_{\rm MSR}$.

 \begin{figure}
 \begin{center}
\rotatebox{270}{\includegraphics[scale=0.39]{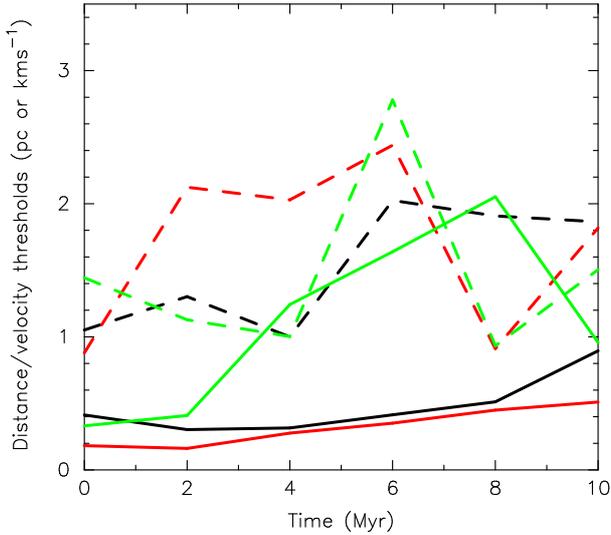}} 
\caption[bf]{The evolution of distance and velocity threshold lengths automatically calculated at every snapshot in the simulation. The solid lines are the distance thresholds, $\tilde{\Delta r}/3$ and dashed lines are the velocity thresholds $\tilde{\Delta v}$ for three simulations with 4 (the mode -- black lines), 8 (median -- green lines) and 25 (extreme -- red lines) groups initially identified by the Friends of Friends algorithm.}
\label{subvirial_thresholds}
\end{center}
 \end{figure}

\subsection{Supervirial (expanding) star-forming regions}

\begin{figure*}
  \begin{center}
\setlength{\subfigcapskip}{10pt}

\hspace*{-1.5cm}\subfigure[0\,Myr]{\label{supervirial-a}\rotatebox{270}{\includegraphics[scale=0.20]{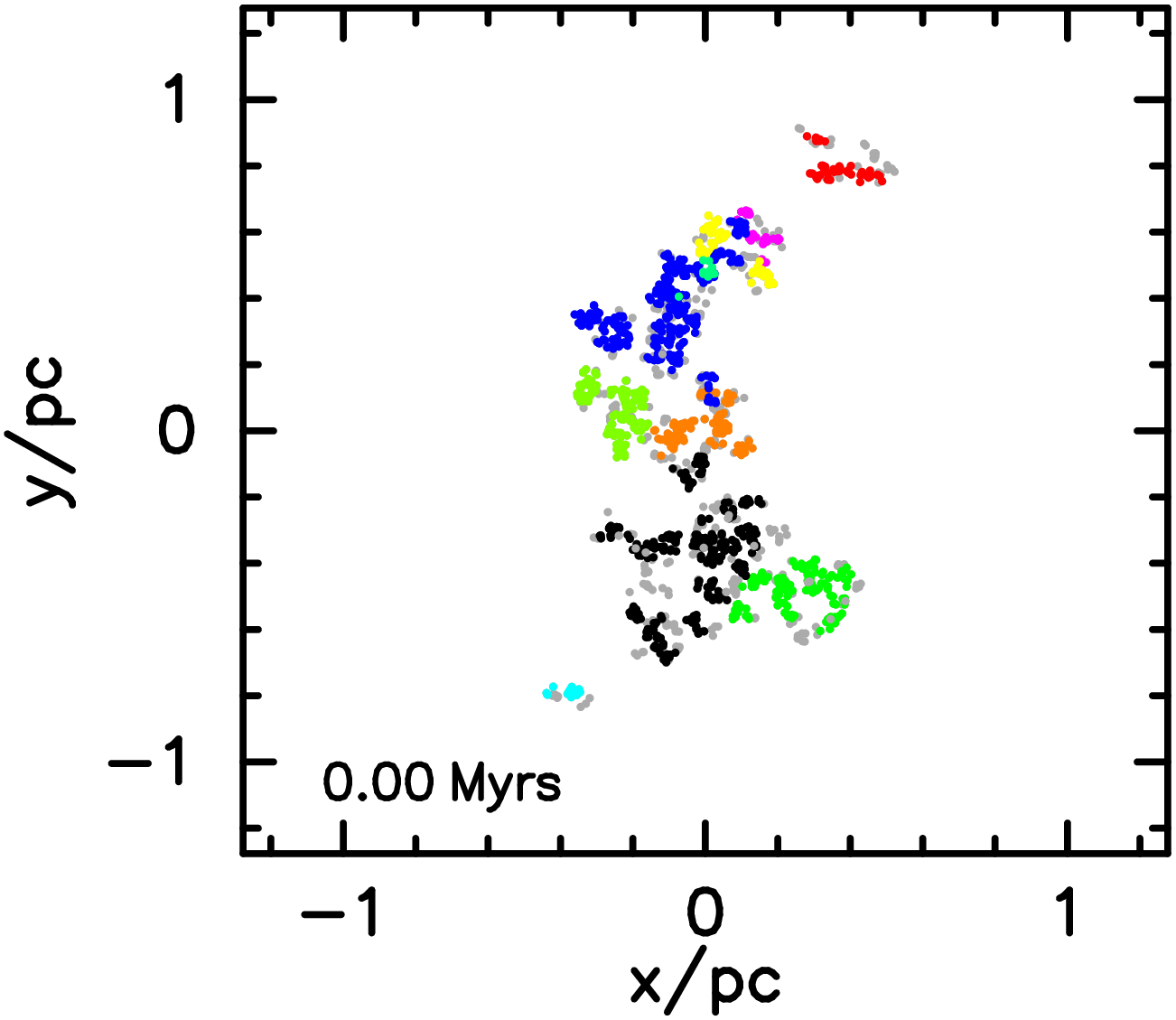}}}
\hspace*{0.3cm} 
\subfigure[0\,Myr]{\label{supervirial-b}\rotatebox{270}{\includegraphics[scale=0.20]{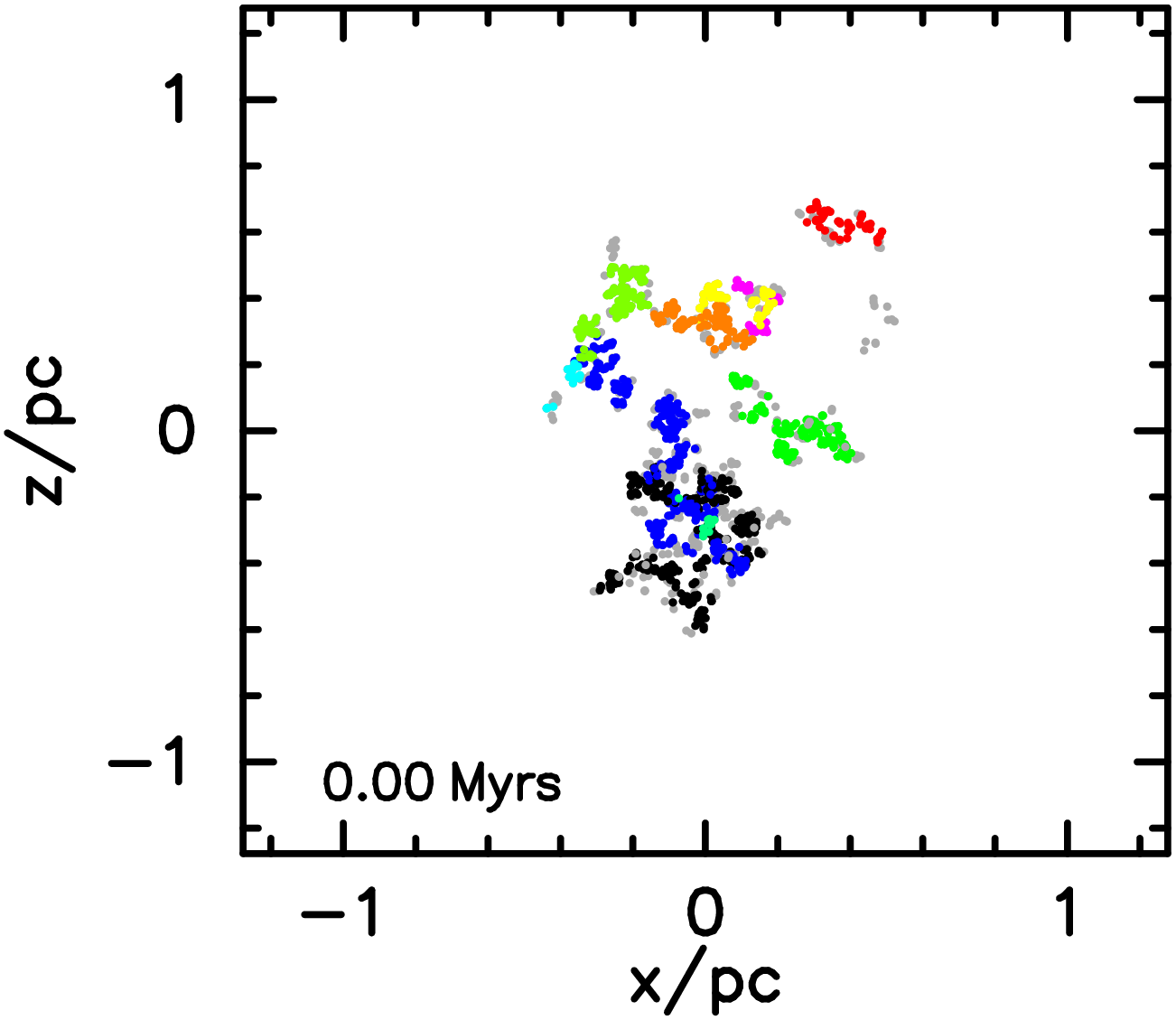}}} 
\hspace*{0.3cm}\subfigure[0\,Myr]{\label{supervirial-c}\rotatebox{270}{\includegraphics[scale=0.20]{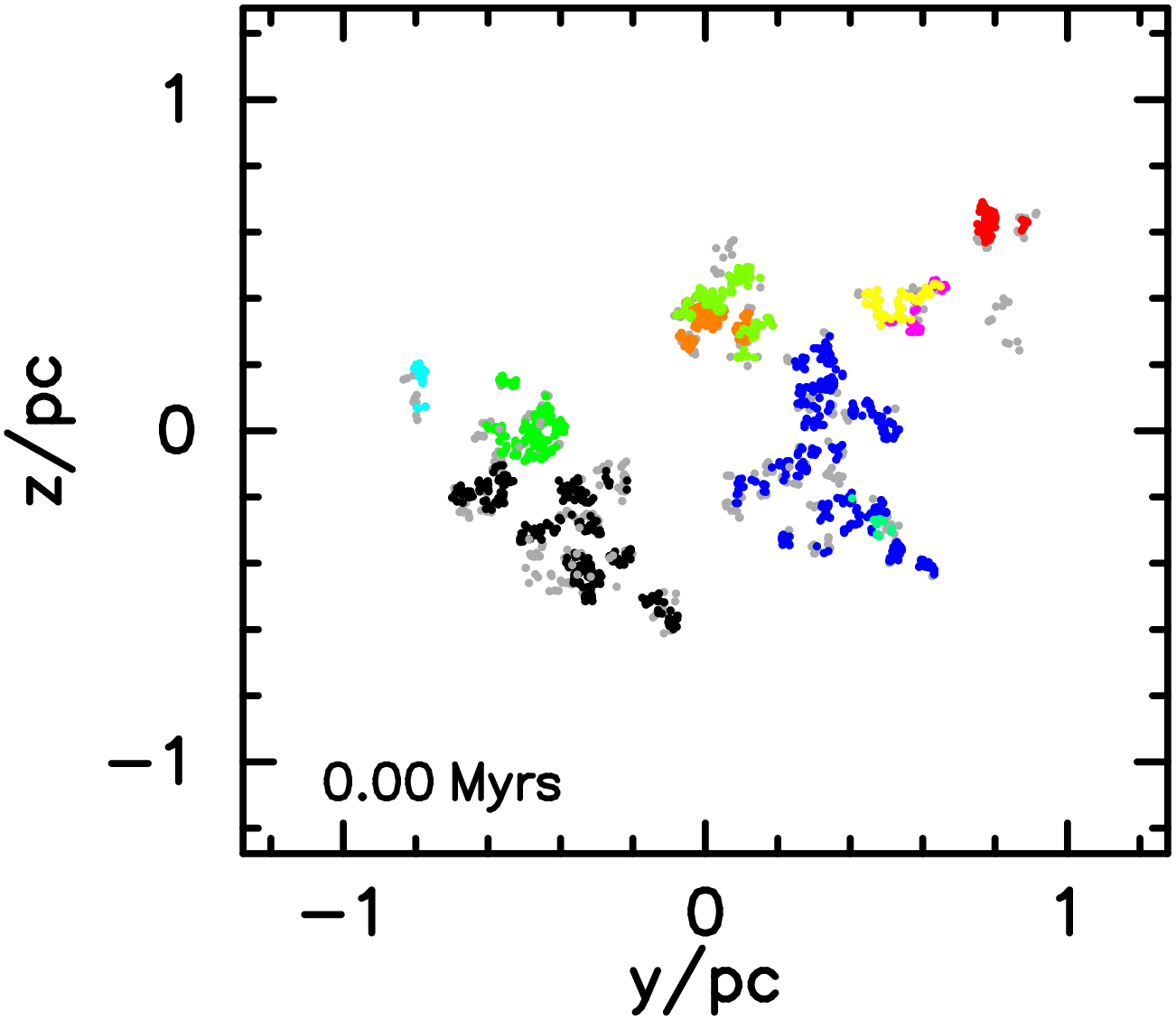}}}

\hspace*{-1.5cm}\subfigure[0.1\,Myr]{\label{supervirial-d}\rotatebox{270}{\includegraphics[scale=0.20]{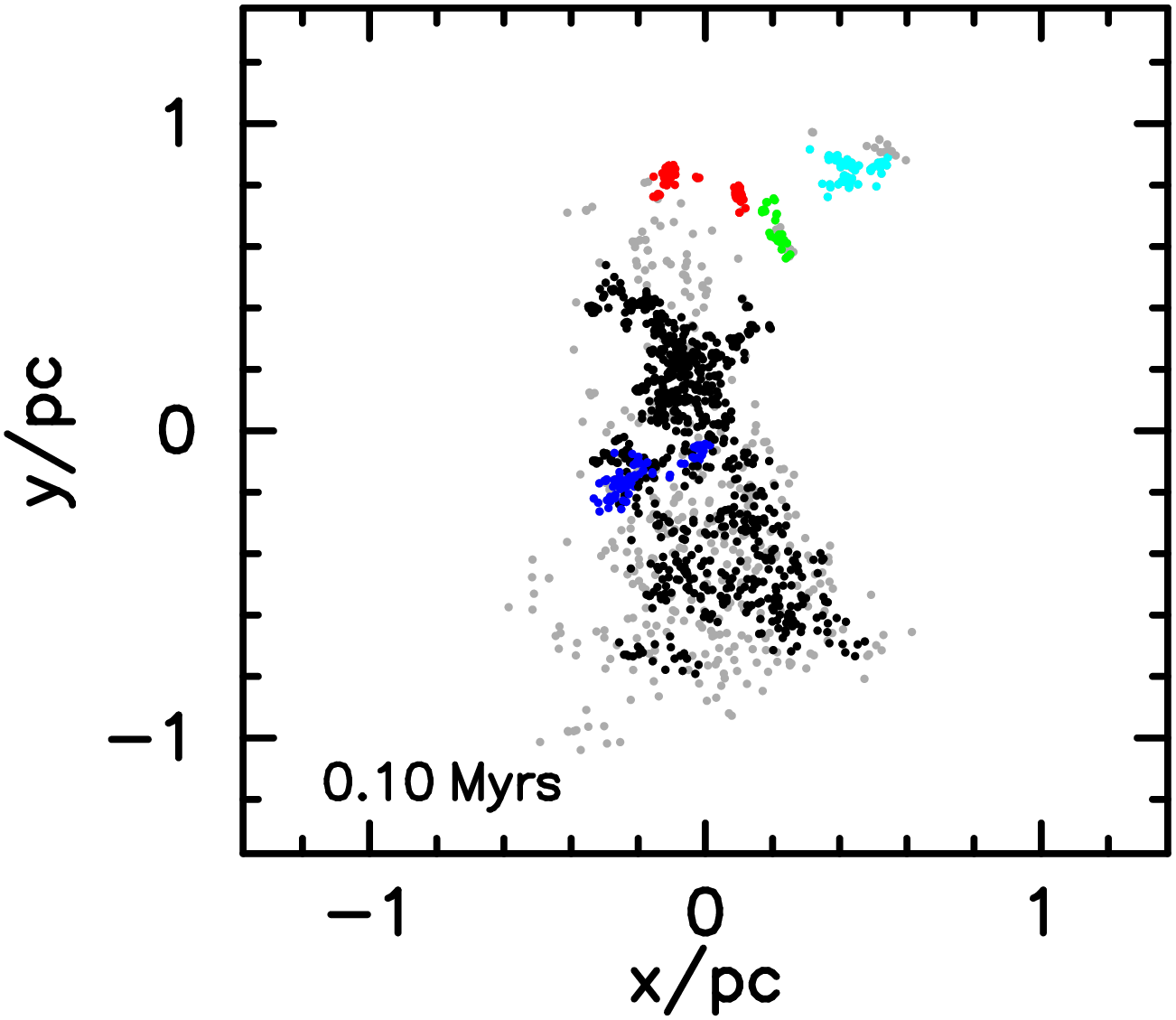}}}
\hspace*{0.3cm} 
\subfigure[0.1\,Myr]{\label{supervirial-e}\rotatebox{270}{\includegraphics[scale=0.20]{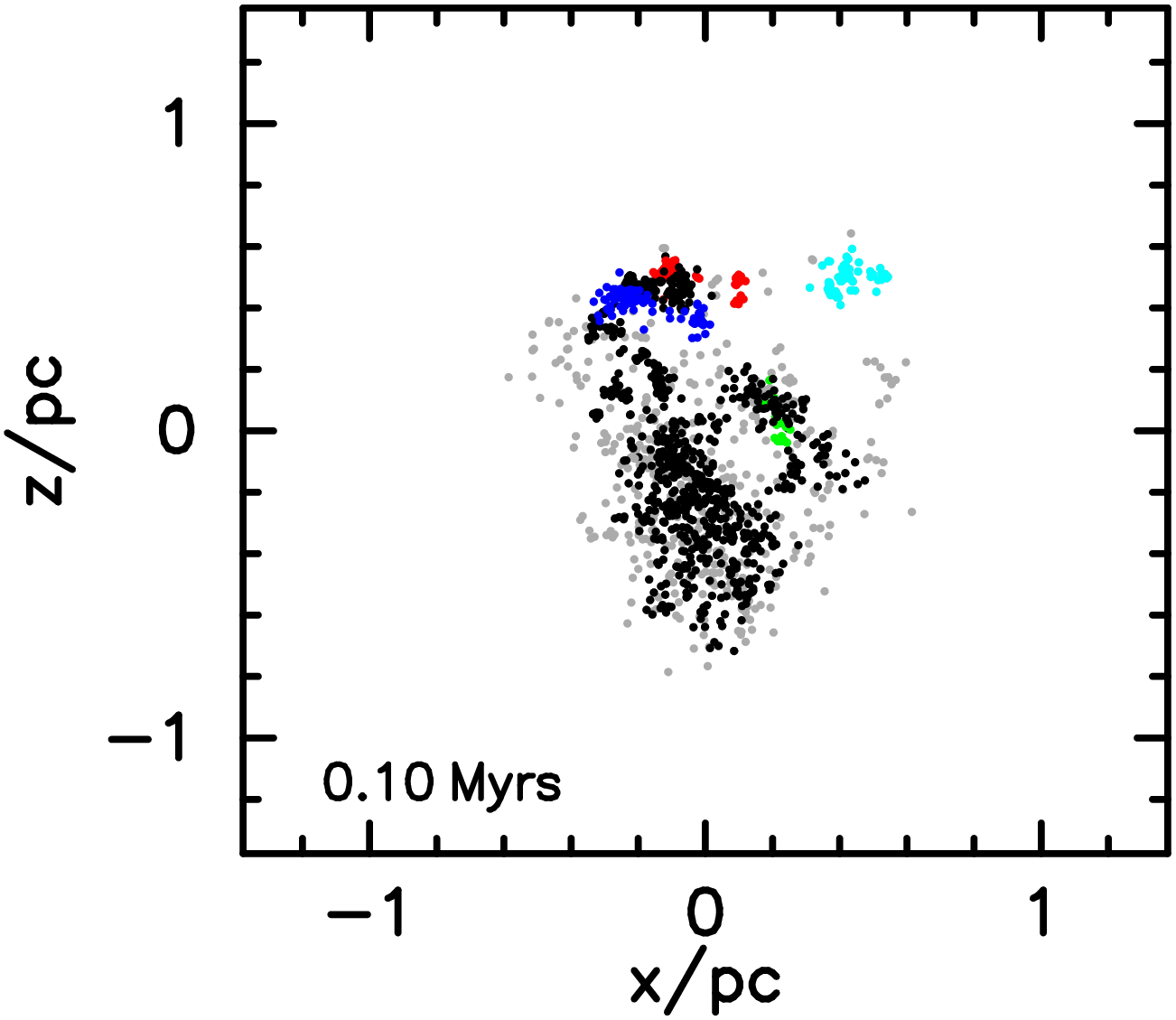}}} 
\hspace*{0.3cm}\subfigure[0.1\,Myr]{\label{supervirial-f}\rotatebox{270}{\includegraphics[scale=0.20]{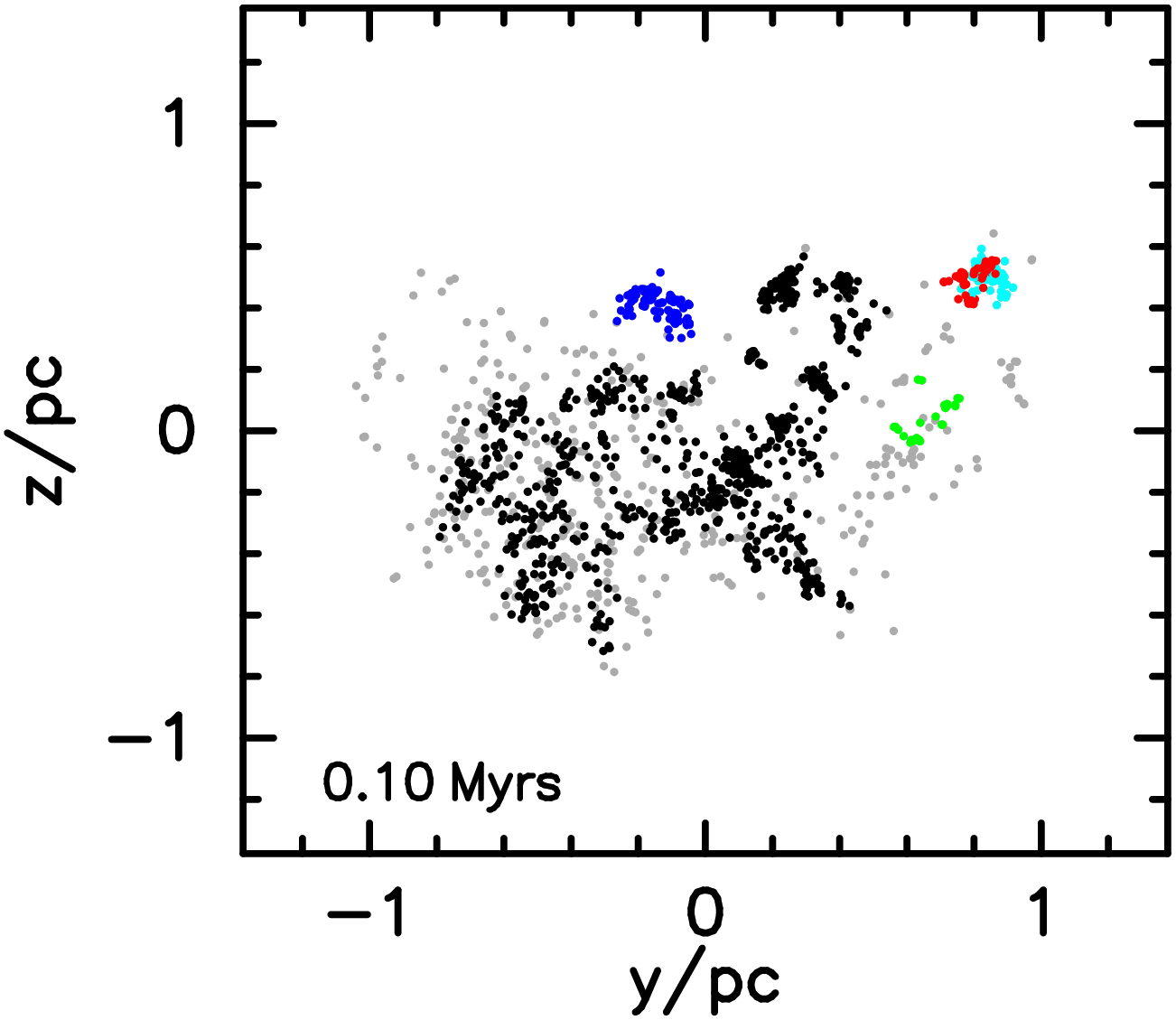}}}

\hspace*{-1.5cm}\subfigure[0.4\,Myr]{\label{supervirial-g}\rotatebox{270}{\includegraphics[scale=0.20]{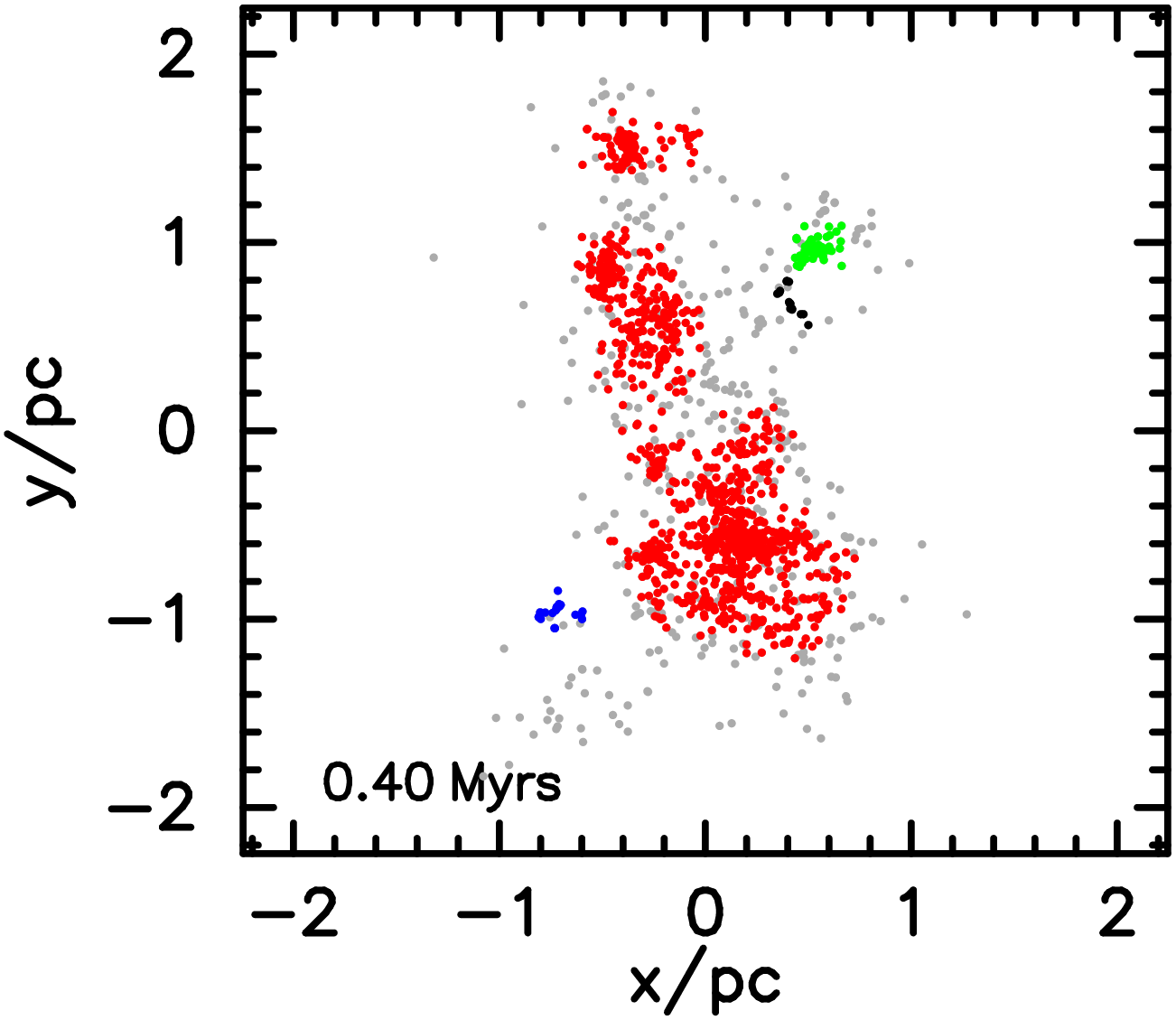}}}
\hspace*{0.3cm} 
\subfigure[0.4\,Myr]{\label{supervirial-h}\rotatebox{270}{\includegraphics[scale=0.20]{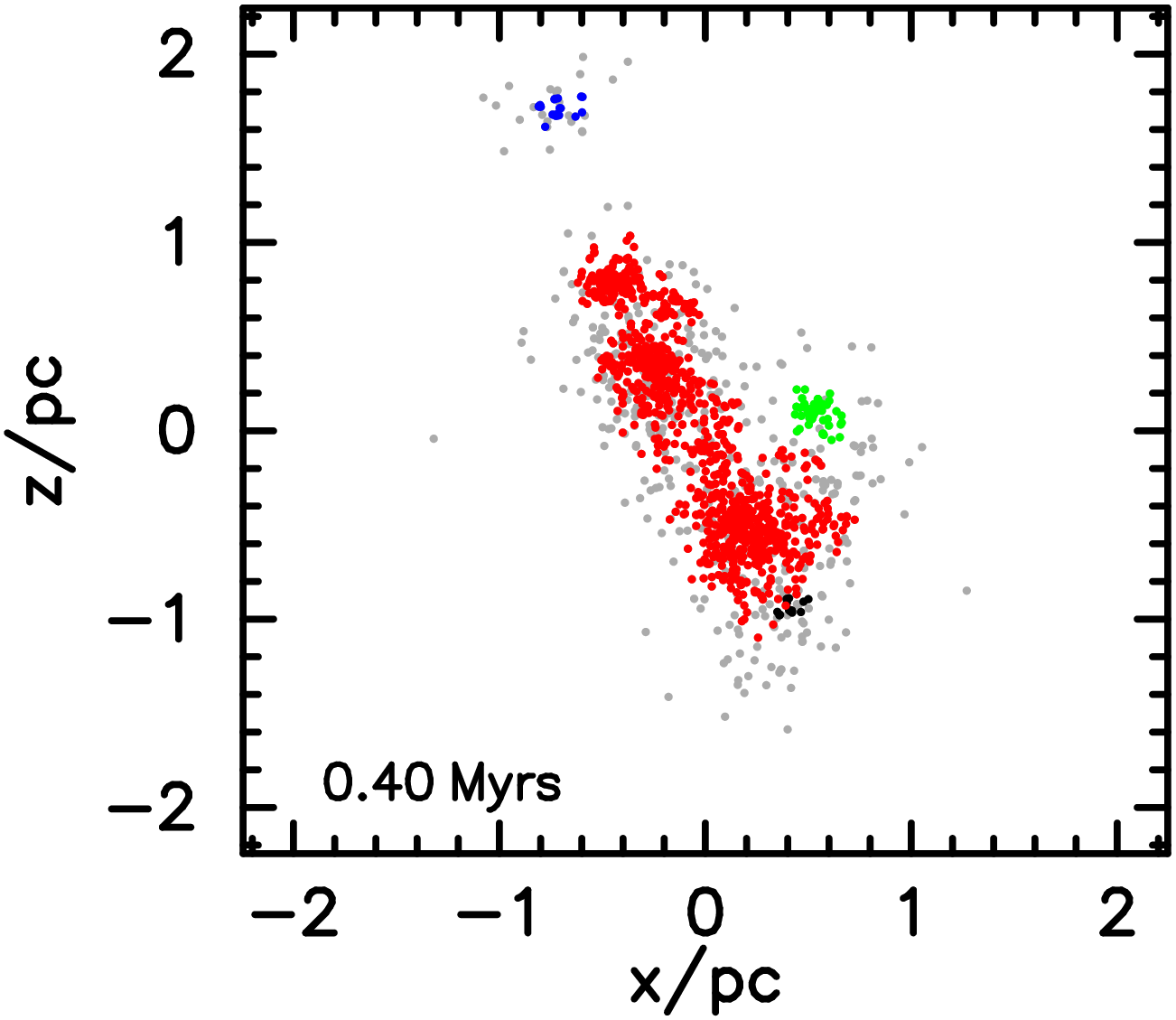}}} 
\hspace*{0.3cm}\subfigure[0.4\,Myr]{\label{supervirial-i}\rotatebox{270}{\includegraphics[scale=0.20]{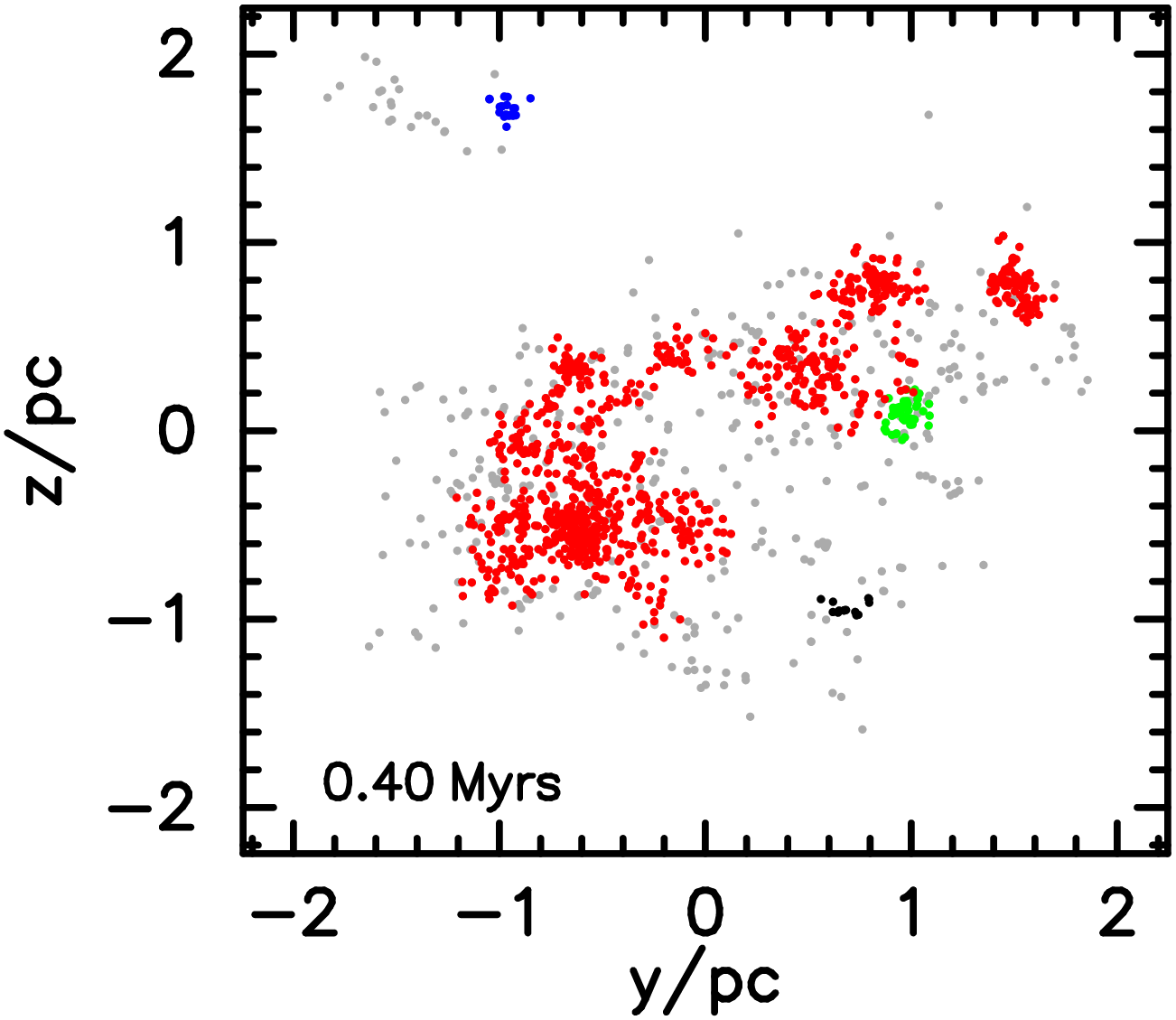}}}

\hspace*{-1.5cm}\subfigure[0.7\,Myr]{\label{supervirial-j}\rotatebox{270}{\includegraphics[scale=0.20]{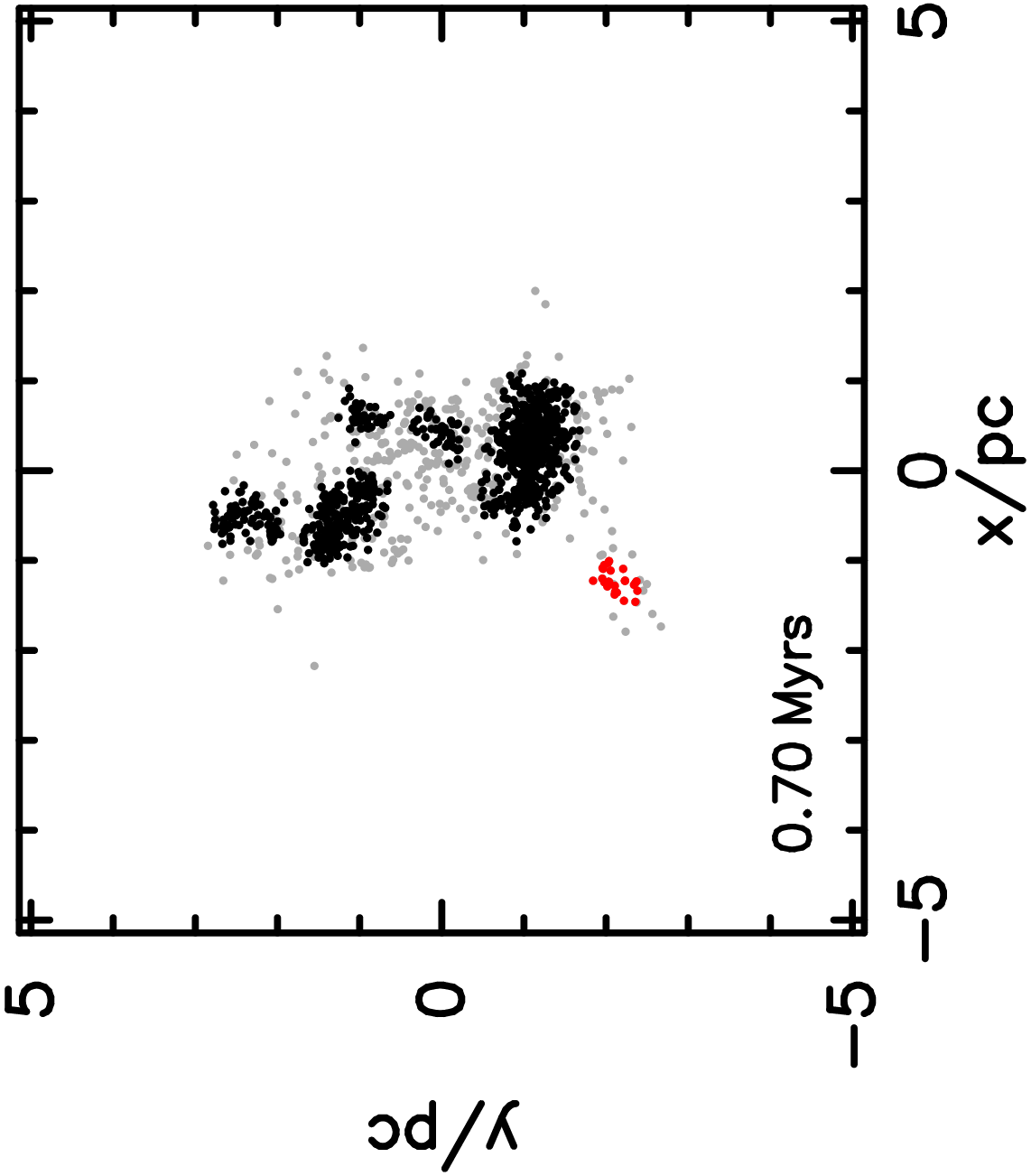}}}
\hspace*{0.3cm} 
\subfigure[0.7\,Myr]{\label{supervirial-k}\rotatebox{270}{\includegraphics[scale=0.20]{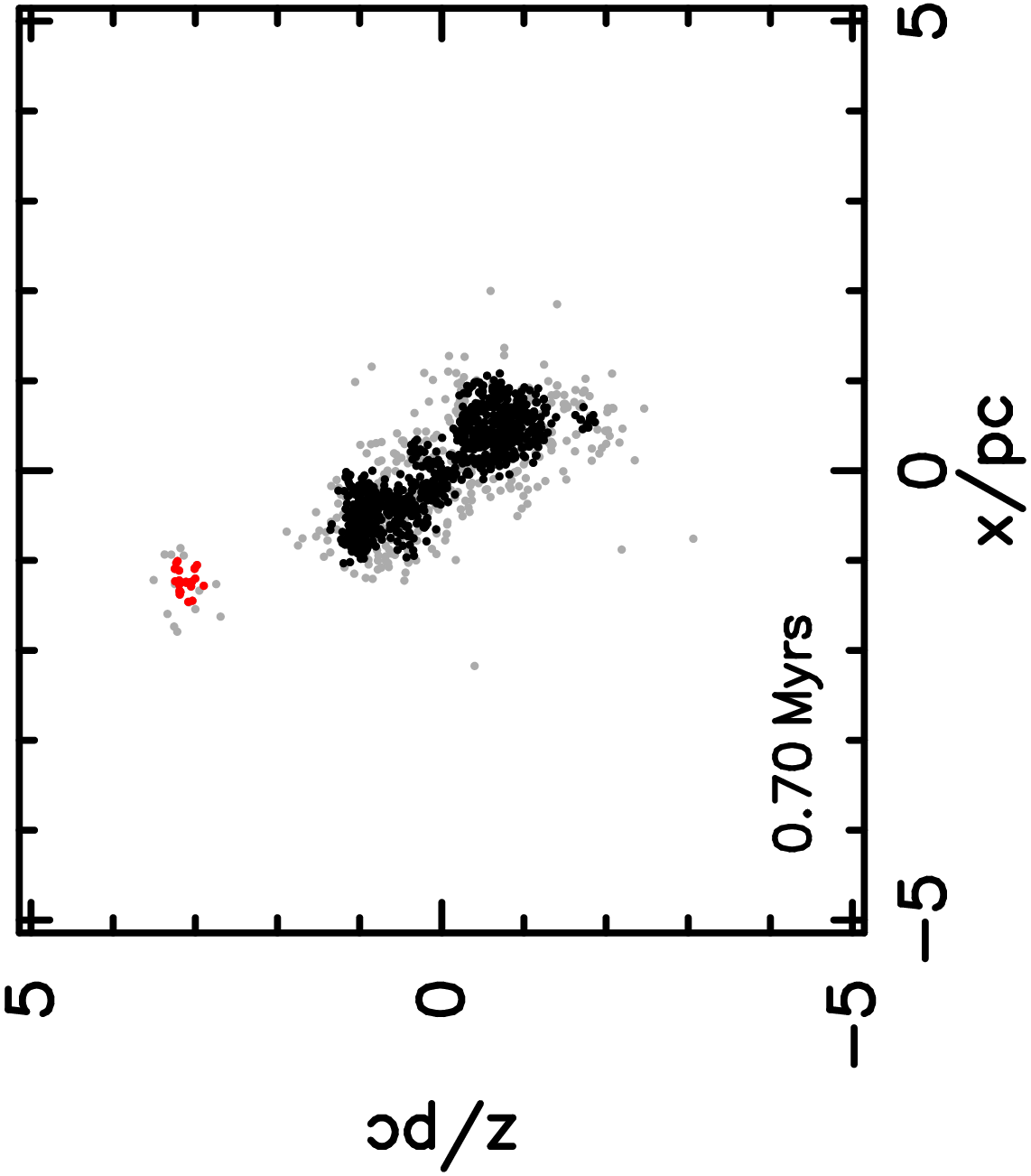}}} 
\hspace*{0.3cm}\subfigure[0.7\,Myr]{\label{supervirial-l}\rotatebox{270}{\includegraphics[scale=0.20]{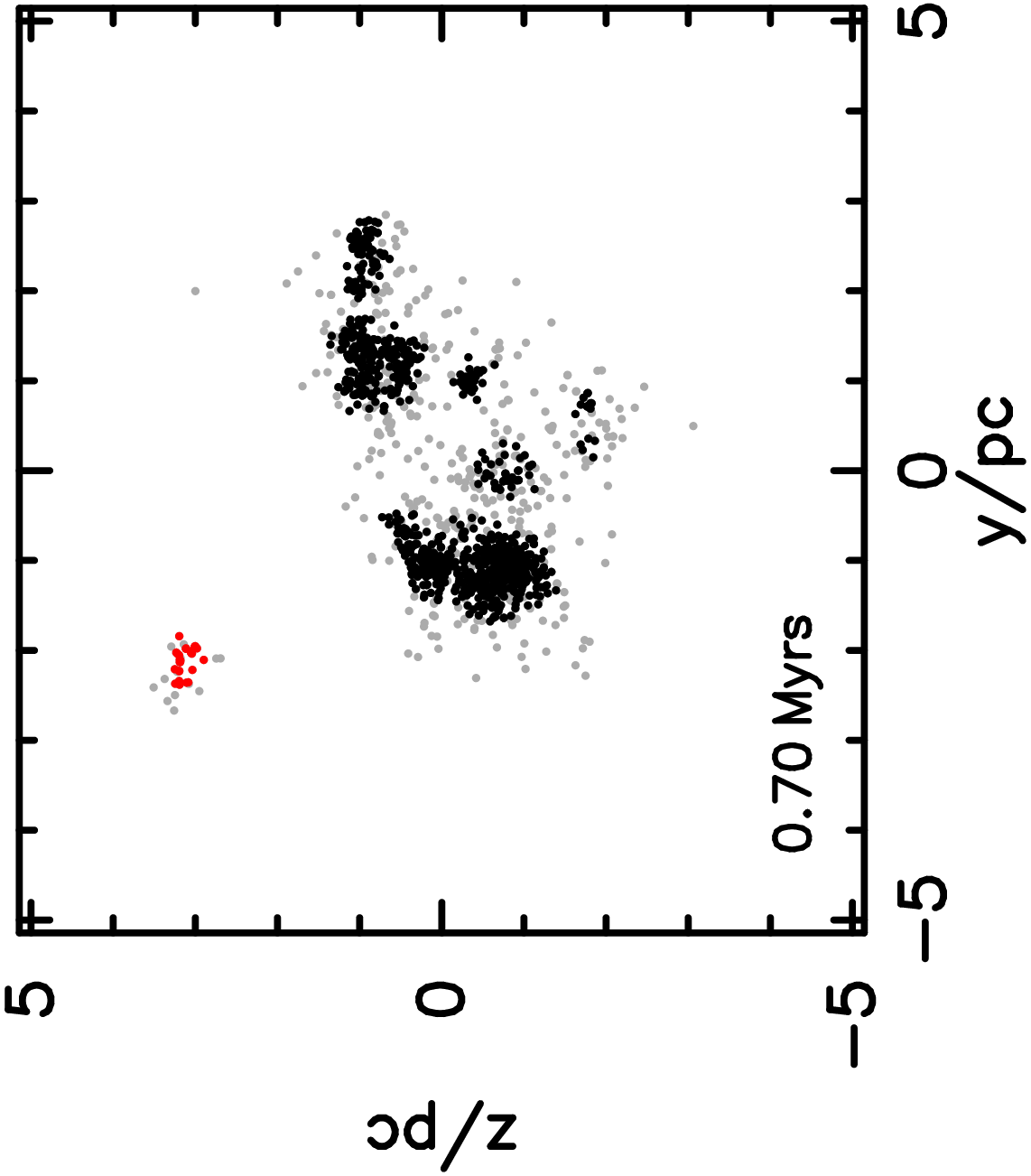}}}

\hspace*{-1.5cm}\subfigure[3\,Myr]{\label{supervirial-m}\rotatebox{270}{\includegraphics[scale=0.20]{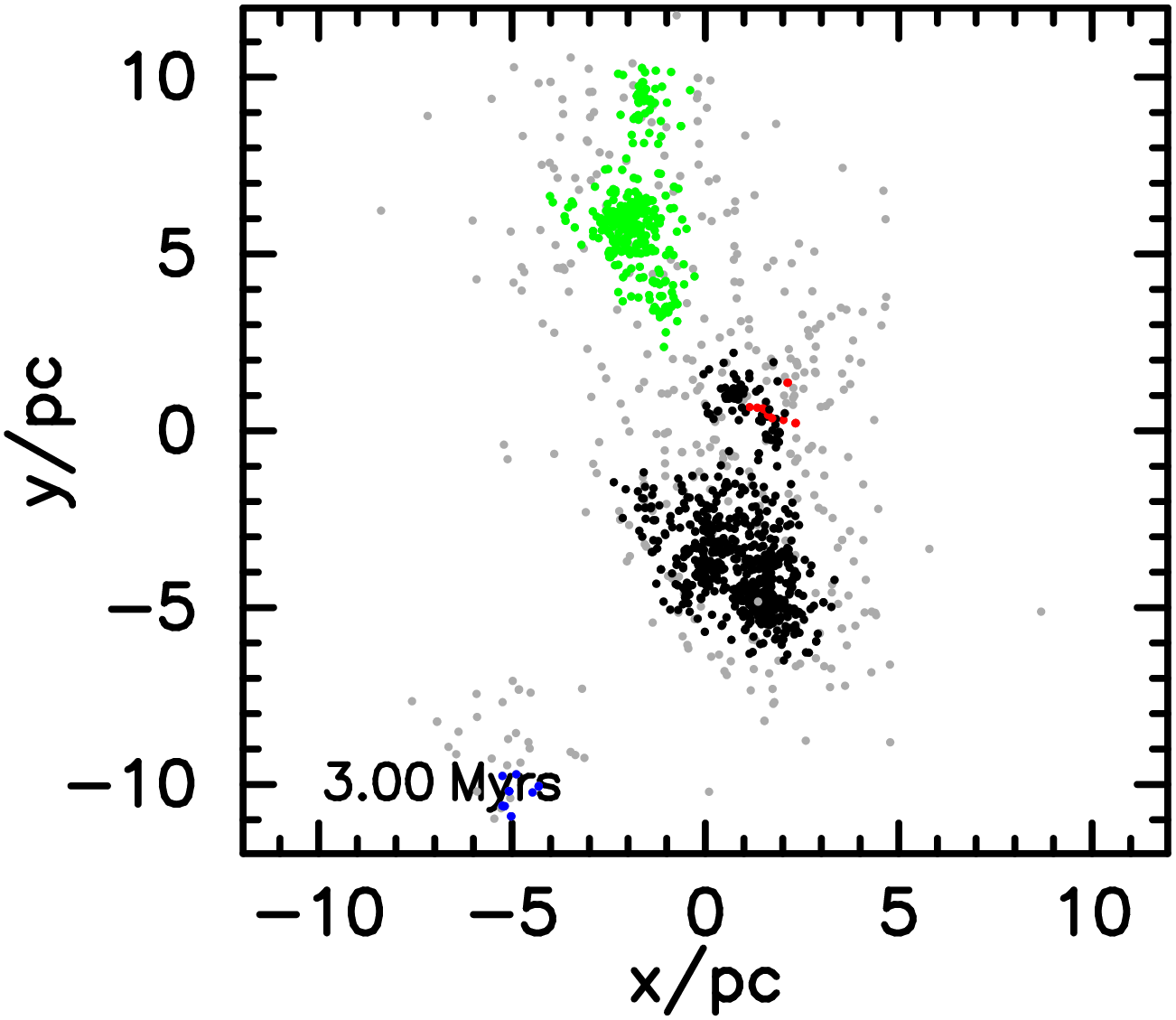}}}
\hspace*{0.3cm} 
\subfigure[3\,Myr]{\label{supervirial-n}\rotatebox{270}{\includegraphics[scale=0.20]{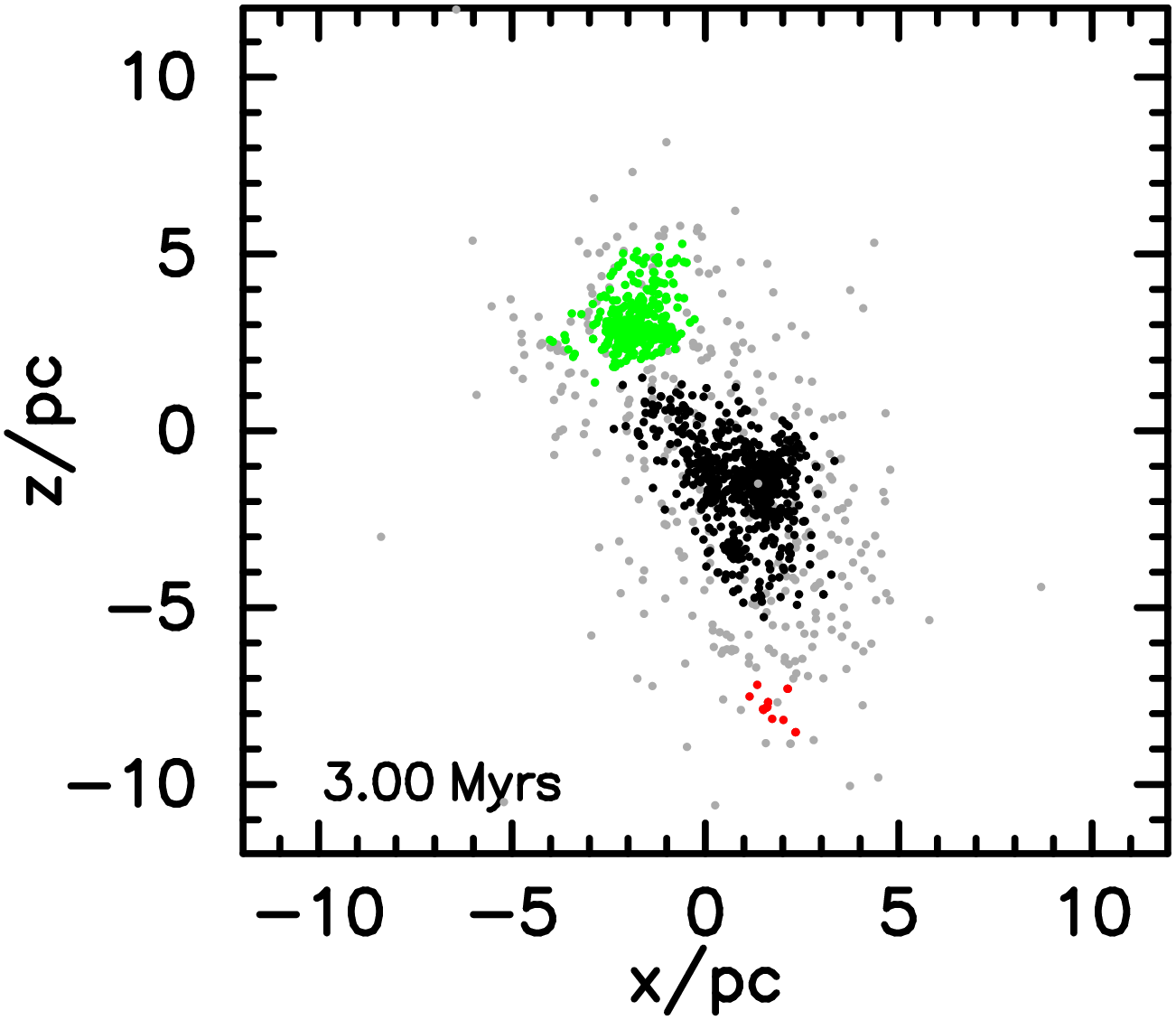}}} 
\hspace*{0.3cm}\subfigure[3\,Myr]{\label{supervirial-o}\rotatebox{270}{\includegraphics[scale=0.20]{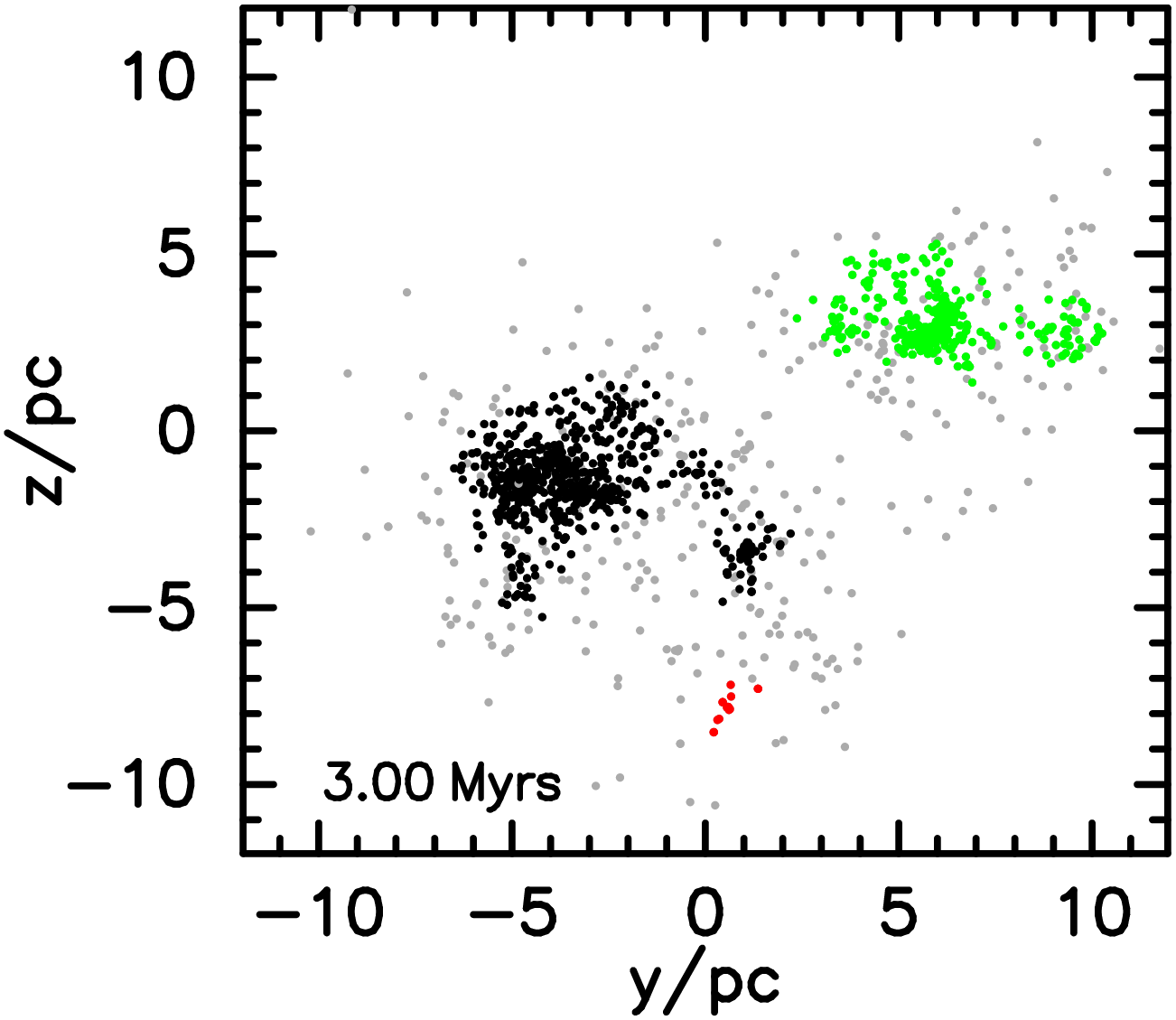}}}

\vspace*{0.25cm}
\caption[bf]{Evolution of groups defined by the Friends of Friends algorithm in a simulated supervirial ($\alpha_{\rm vir} = 1.5$) star-forming region with $D = 1.6$. Stars that have a local stellar density below the first quartile in the distribution are not assigned to a group and are coloured grey. The colours in the subsequent snapshots do not correspond to the colours in the first snapshot ($t = 0$\,Myr).}
  \label{supervirial}
  \end{center}
\end{figure*}

 \begin{figure}
 \begin{center}
\rotatebox{270}{\includegraphics[scale=0.39]{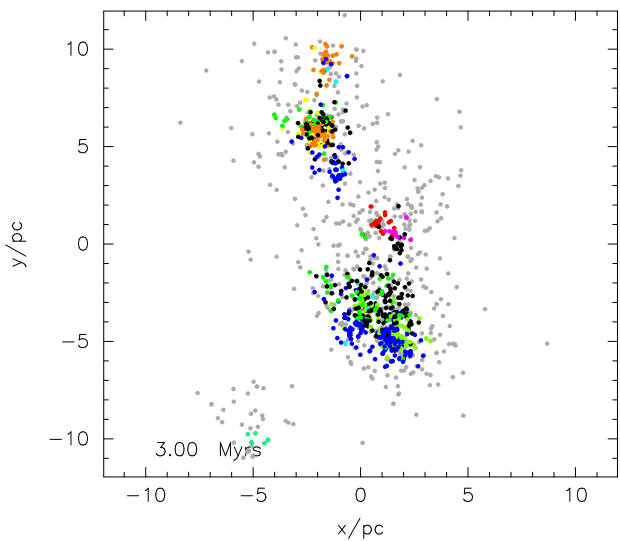}}
\caption[bf]{Same as Fig.~\ref{supervirial-m}, but showing the stars colour-coded according to their original groups in Fig.~\ref{supervirial-a}. Significant mixing of the groups has occurred, despite the region retaining spatial and kinematic substructure.}
\label{original_colours}
\end{center}
 \end{figure}

\begin{figure*}
  \begin{center}
\setlength{\subfigcapskip}{10pt}
\hspace*{-1.5cm}\subfigure[$\mathcal{Q}$ -- $\sigma$/IQR]{\label{supervirial_structure-a}\rotatebox{270}{\includegraphics[scale=0.25]{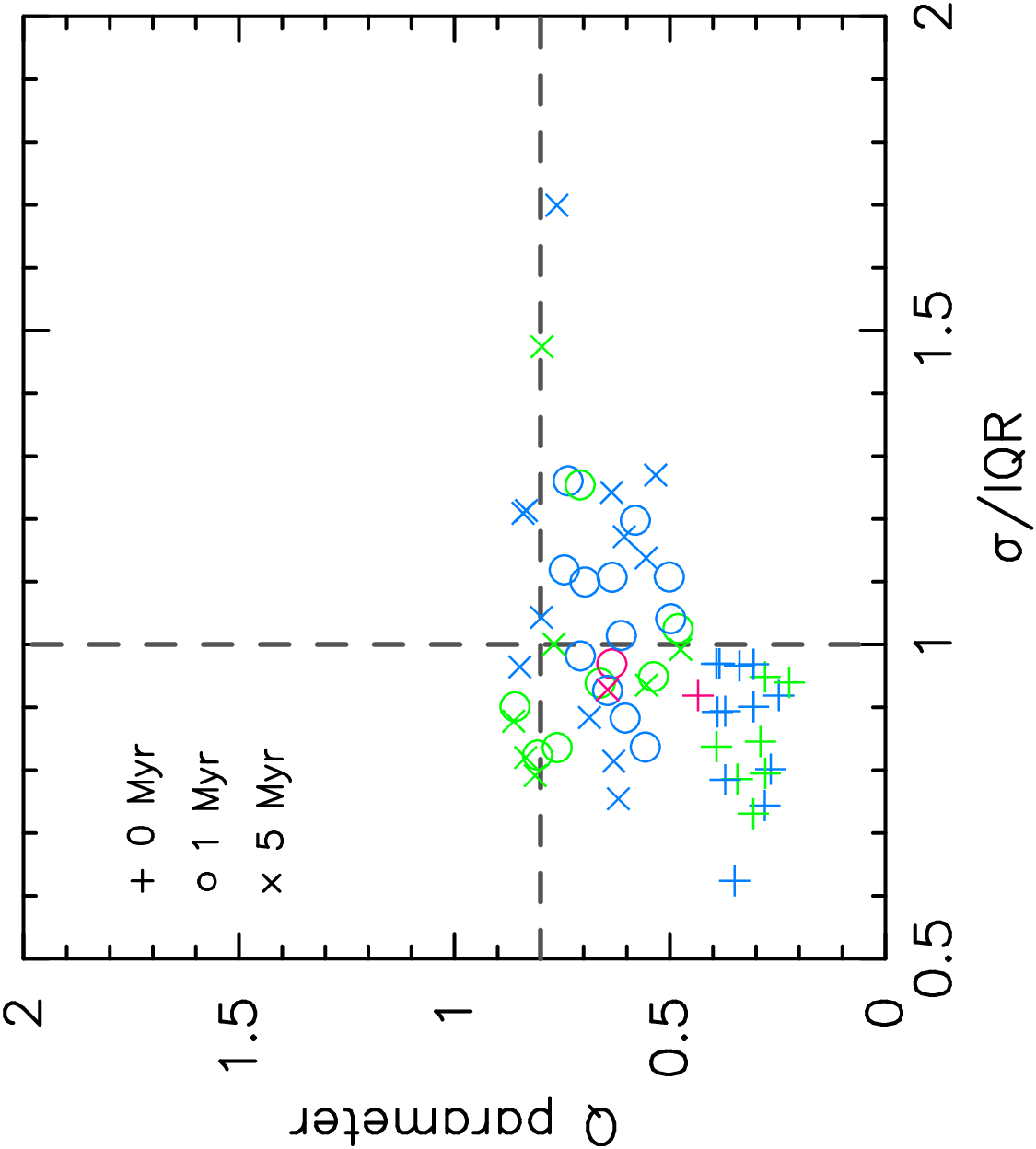}}}
\hspace*{0.3cm} 
\subfigure[$\mathcal{Q}$ -- $\Sigma_{\rm LDR}$]{\label{supervirial_structure-b}\rotatebox{270}{\includegraphics[scale=0.25]{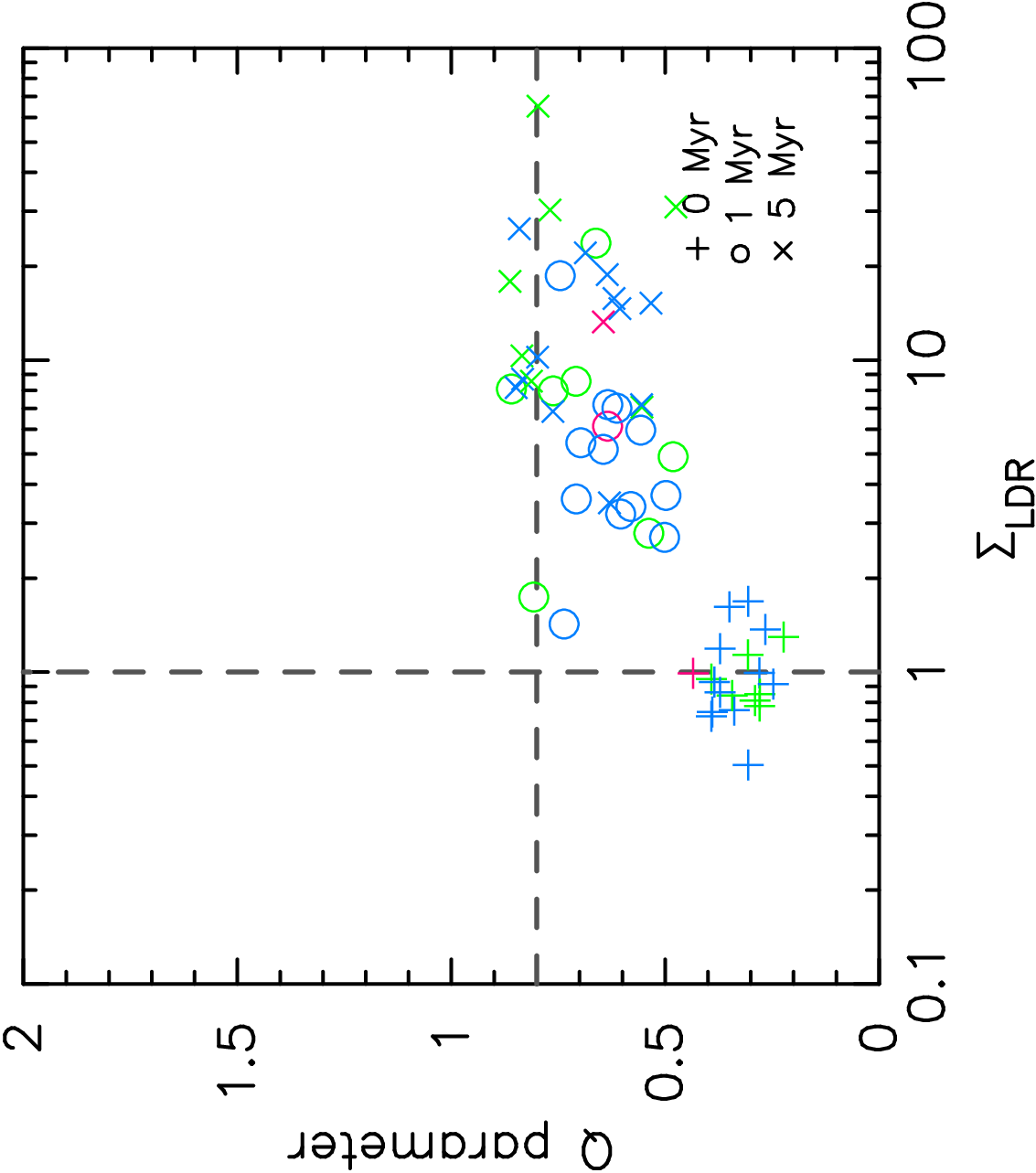}}} 
\hspace*{0.3cm}\subfigure[$\mathcal{Q}$ -- $\Lambda_{\rm MSR}$]{\label{supervirial_structure-c}\rotatebox{270}{\includegraphics[scale=0.25]{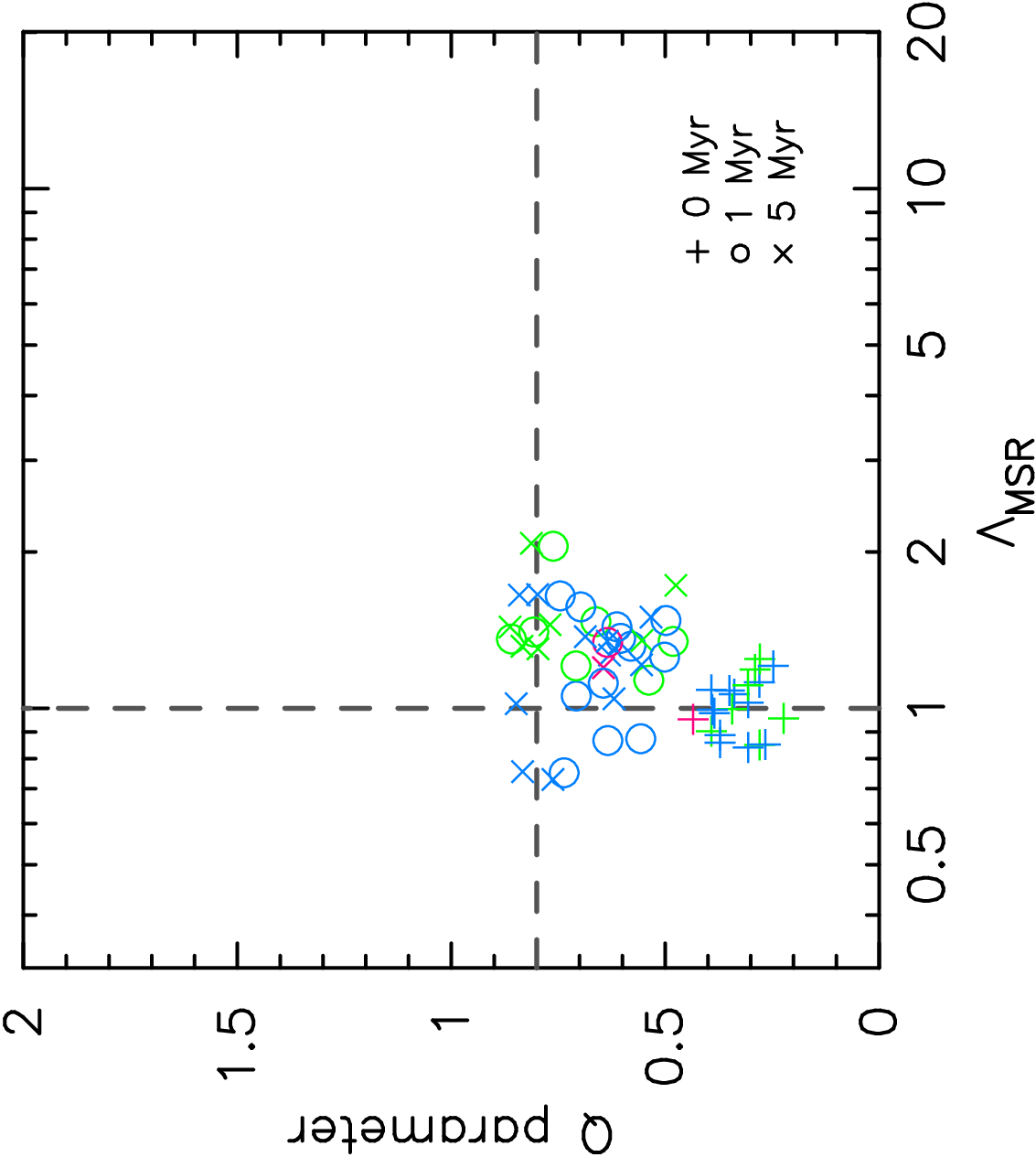}}}
\caption[bf]{Different measure of spatial and kinematic evolution in our supervirial ($\alpha_{\rm vir} = 1.5$) simulations. In panel (a) we show the \citet{Cartwright04} $\mathcal{Q}$-parameter against the radial velocity dispersion divided by the interquartile range of radial velocities \citep{Parker16b}. In panel (b) we show the $\mathcal{Q}$-parameter against the relative local surface density ratio of the ten most massive stars, $\Sigma_{\rm LDR}$ \citep{Kupper11,Parker14b}. In panel (c) we show the $\mathcal{Q}$-parameter against the mass segregation ratio $\Lambda_{\rm MSR}$ \citep{Allison09a}. The boundary between hierarchically substructured and centrally concentrated distributions is shown by the horizontal dashed lines and the vertical dashed lines correspond to unity for the other measures, indicating no special configuration for the massive stars. In all panels, the green symbols indicate simulations where the Friends of Friends algorithm picks out 5 groups or fewer at $t = 0$\,Myr; the blue symbols indicate between 5 and 15 groups, and the magenta symbols indicate that the Friends of Friends algorithm has picked out more than 15 groups. There is no strong dependence of the dynamical evolution on the initial number of groups.}
  \label{supervirial_structure}
  \end{center}
\end{figure*}

We show a typical example of a supervirial star-forming region in Fig.~\ref{supervirial}. As in Fig.~\ref{subvirial}, we show five snapshots in time and three viewing angles for each snapshot. Our first result is that the numbers of groups identified by the Friends of Friends algorithm at $t = 0$\,Myr in the suite of twenty simulations is almost identical to the numbers identified in the subvirial simulations, with a range between 1 and 20, a mode of 4 and a median of 8. However, the virial ratio in these supervirial simulations is $\alpha_{\rm vir} = 1.5$, meaning that the global velocity dispersion is significantly higher than in the subvirial simulations ($\alpha_{\rm vir} = 0.3$) presented above. This difference in velocity scaling between the two sets of simulations leads to very different dynamical evolution (expansion and preservation of some substructure versus collapse and erasing of all substructure), yet the numbers of groups identified by the Friends of Friends algorithm does not betray the future evolution of an individual star-forming region. This is because the threshold lengths are automatically calculated, and so any linear scaling of the velocities (such as changing the virial ratio) will not give statistically different results.

In these simulations a significant degree of substructure is retained, and so we might expect the Friends of Friends algorithm to identify multiple distinct groups of stars. Whilst this is the case, the number of groups identified after 0.1 - 0.5\,Myr of dynamical evolution is typically only 1--3, despite the region displaying rather obvious visual spatial substructure (e.g. panels j -- l in Fig.~\ref{supervirial}). At earlier stages of this simulation (0.4\,Myr -- panels g-i), bridges of stars are apparent between the (visual) groups of stars which would explain why the Friends of Friends algorithm classifies them as one large group (the red points in panels g--i). However, at later stages these bridges are not as apparent, and the physical distances between the visual groups are larger.

The reason the Friends of Friends algorithm does not detect many distinct groups in this simulation (and others), appears to be due to the large amount of spatial and kinematical mixing that occurs throughout the dynamical evolution of the star-forming region. In Fig.~\ref{original_colours} we show the $x - y$ plane of the snapshot at 3\,Myr from Fig.~\ref{supervirial-m}, but the stars are colour-coded according to their original groups at $t = 0$\,Myr (Fig.~\ref{supervirial-a}). Clearly, stars have migrated significantly, yet the star-forming region has preserved some spatial substructure. Interestingly, \citet{Arnold17} find a similar result when examining the formation of binary star clusters -- two clusters orbiting a common centre of mass -- from initially supervirial, substructured star-forming regions like those in this paper. \citet{Arnold17} find that the stars which constitute the components of the binary clusters do not originate in the same location as their fellow constituents, making it impossible to predict where a star will end up during the evolution of a supervirial star-forming region.

We now investigate whether the long-term evolution of the supervirial star-forming regions depends on the initial number of groups identified by the Friends of Friends algorithm. In Fig.~\ref{supervirial_structure} we plot the $\mathcal{Q}$-parameter against the kinematic and spatial diagnostics of dynamical evolution, $\sigma$/IQR, $\Sigma_{\rm LDR}$ and $\Lambda_{\rm MSR}$ and we colour code these plots according to the number of groups the Friends of Friends algorithm picks out. The orange points indicate simulations where the Friends of Friends algorithm picks out 5 groups or fewer at  $t = 0$\,Myr; the blue symbols indicate between 5 and 15 groups, and the magenta symbols indicate that the Friends of Friends algorithm has picked out more than 15 groups. As with the subvirial simulations, the initial number of groups picked out by the algorithm is not related to the subsequent magnitude of the $\sigma$/IQR ratio, or the spatial structure, $\mathcal{Q}$, the relative local surface density, $\Sigma_{\rm LDR}$, or the occurrence and amount of mass segregation, $\Lambda_{\rm MSR}$ (though the amount of mass segregation in supervirial star-forming regions is minimal because the massive stars rarely interact with each other).


\subsection{Varying the initial degree of substructure}

As discussed in Section~2.1, we might expect that the evolution of spatio-kinematic groups identified by the Friends of Friends algorithm to be correlated with the initial fractal dimension of the simulation. So far we have presented the results for the cases where $D = 1.6$, where we expect the timescale for structure erasure to be  $t(L) \propto L^{-0.4}$, which means that structure is erased more quickly on larger scales. Conversely, when the fractal dimension is higher ($D = 2.0$ or $D = 2.6$, then structure is erased more quickly on smaller scales (where $t(L) \propto L$  for $D = 2.0$ and for $D = 2.6$ $t(L) \propto L^{0.6}$). 

\begin{figure*}
  \begin{center}
\setlength{\subfigcapskip}{10pt}

\hspace*{-1.5cm}\subfigure[0\,Myr]{\label{F2p0-a}\rotatebox{270}{\includegraphics[scale=0.20]{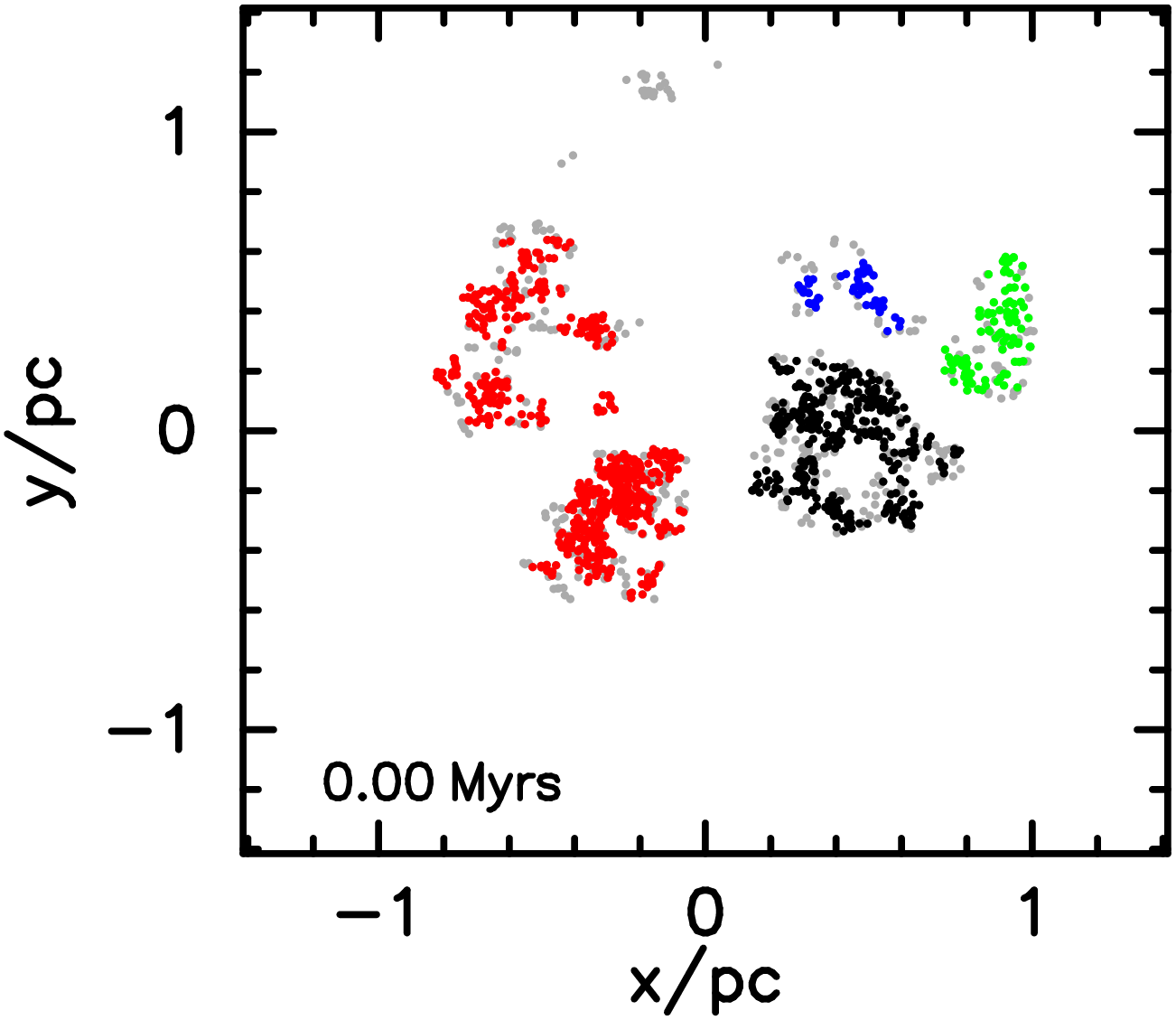}}}
\hspace*{0.3cm} 
\subfigure[0\,Myr]{\label{F2p0-b}\rotatebox{270}{\includegraphics[scale=0.20]{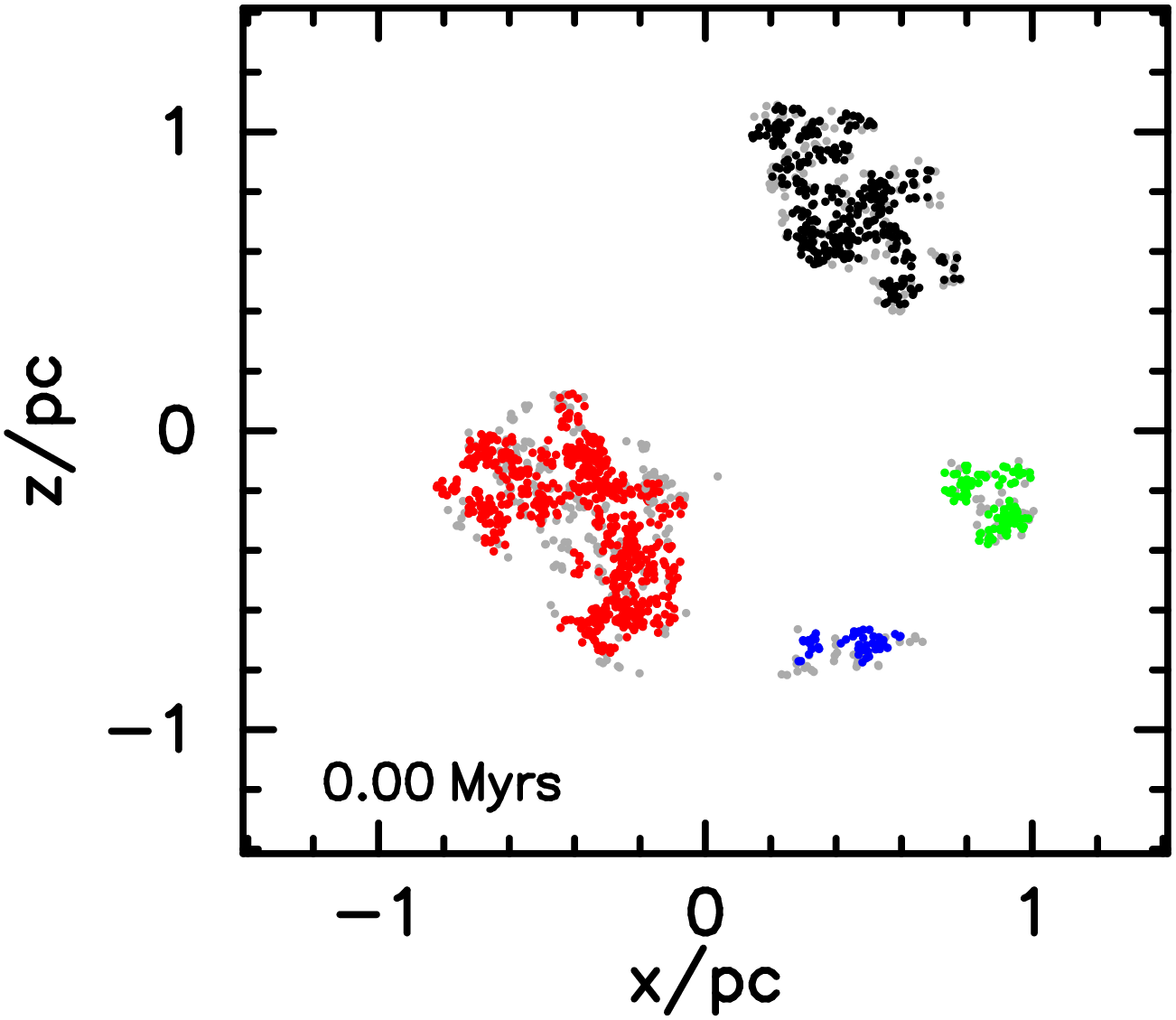}}} 
\hspace*{0.3cm}\subfigure[0\,Myr]{\label{F2p0-c}\rotatebox{270}{\includegraphics[scale=0.20]{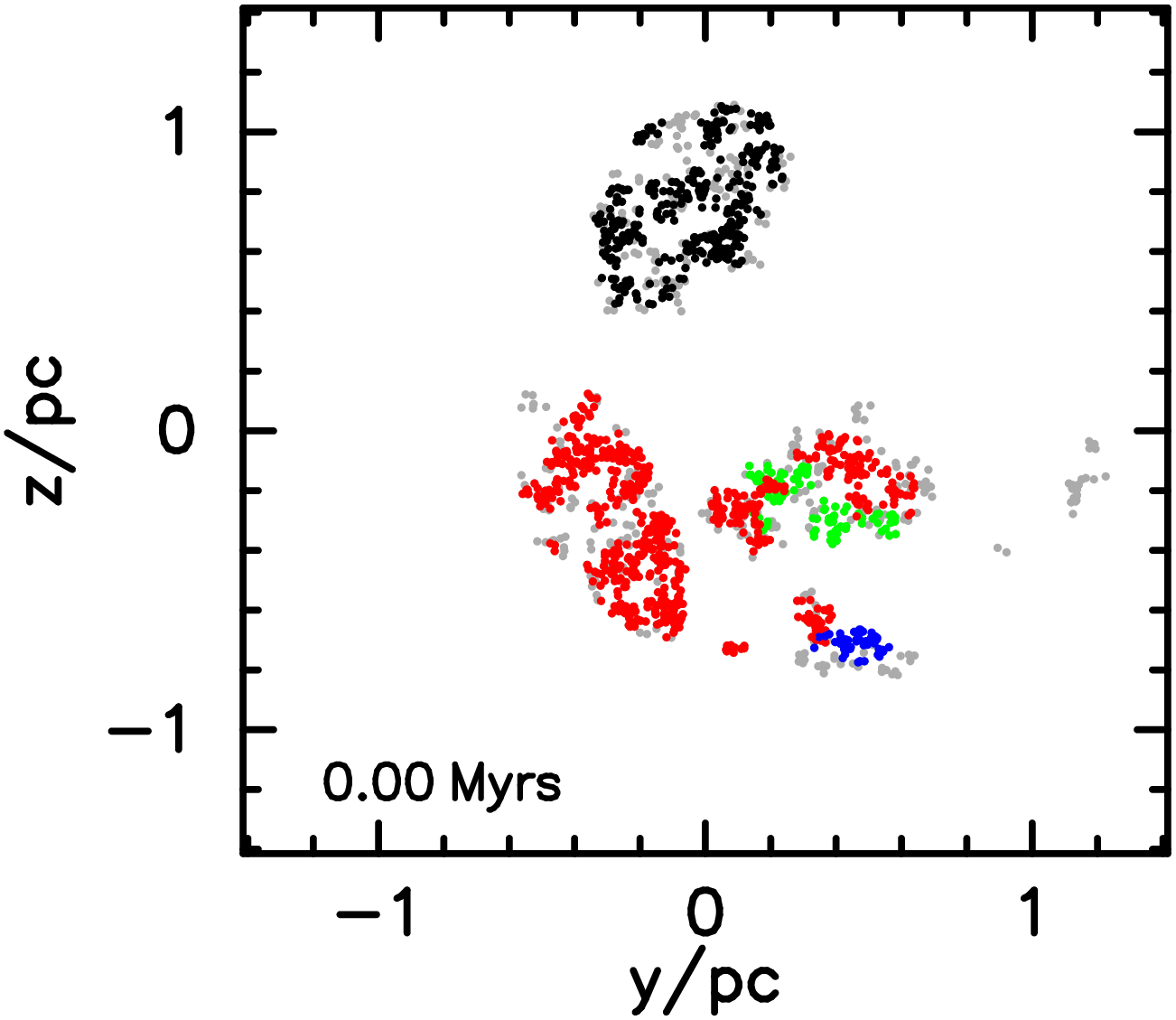}}}

\hspace*{-1.5cm}\subfigure[0.1\,Myr]{\label{F2p0-d}\rotatebox{270}{\includegraphics[scale=0.20]{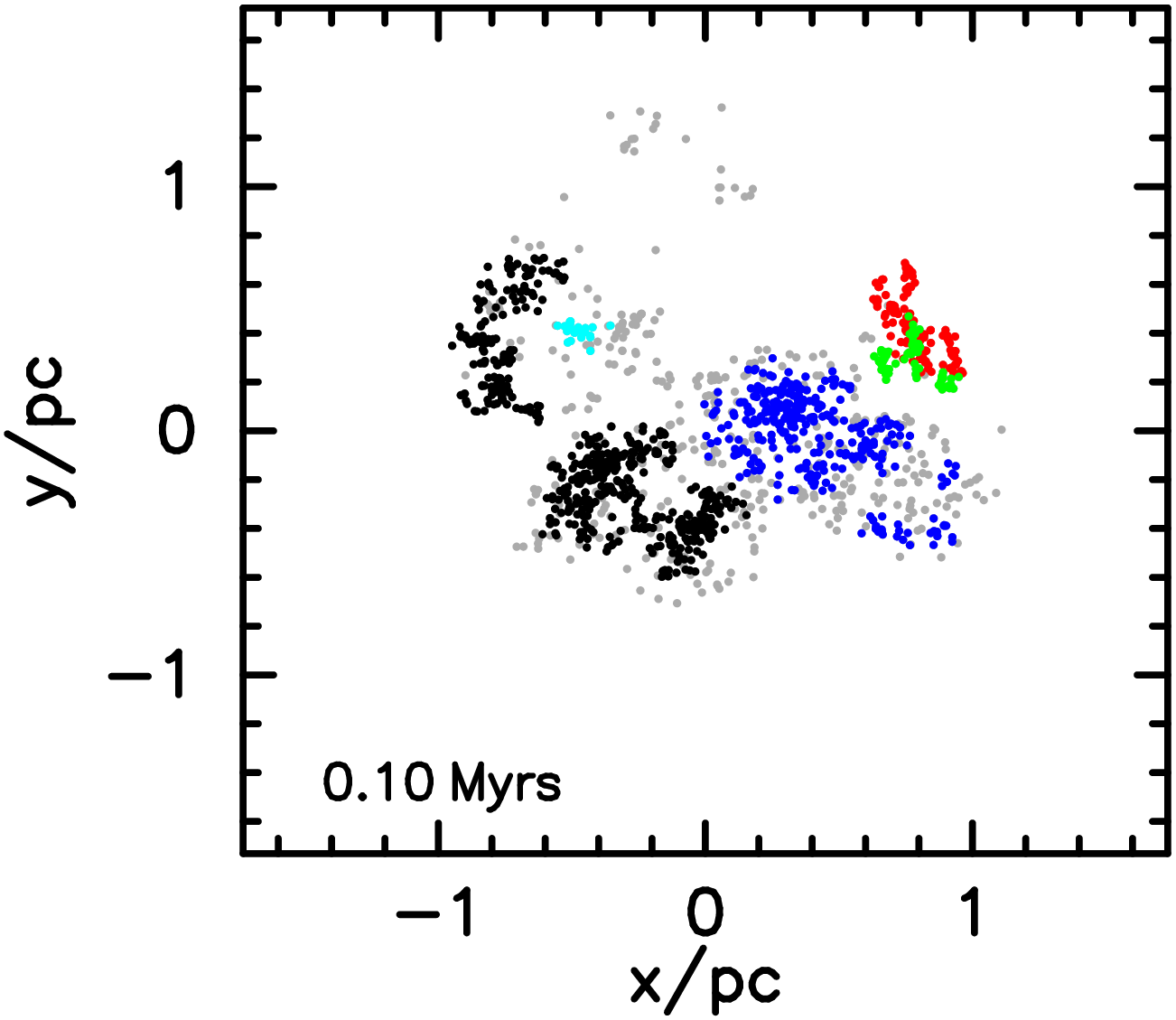}}}
\hspace*{0.3cm} 
\subfigure[0.1\,Myr]{\label{F2p0-e}\rotatebox{270}{\includegraphics[scale=0.20]{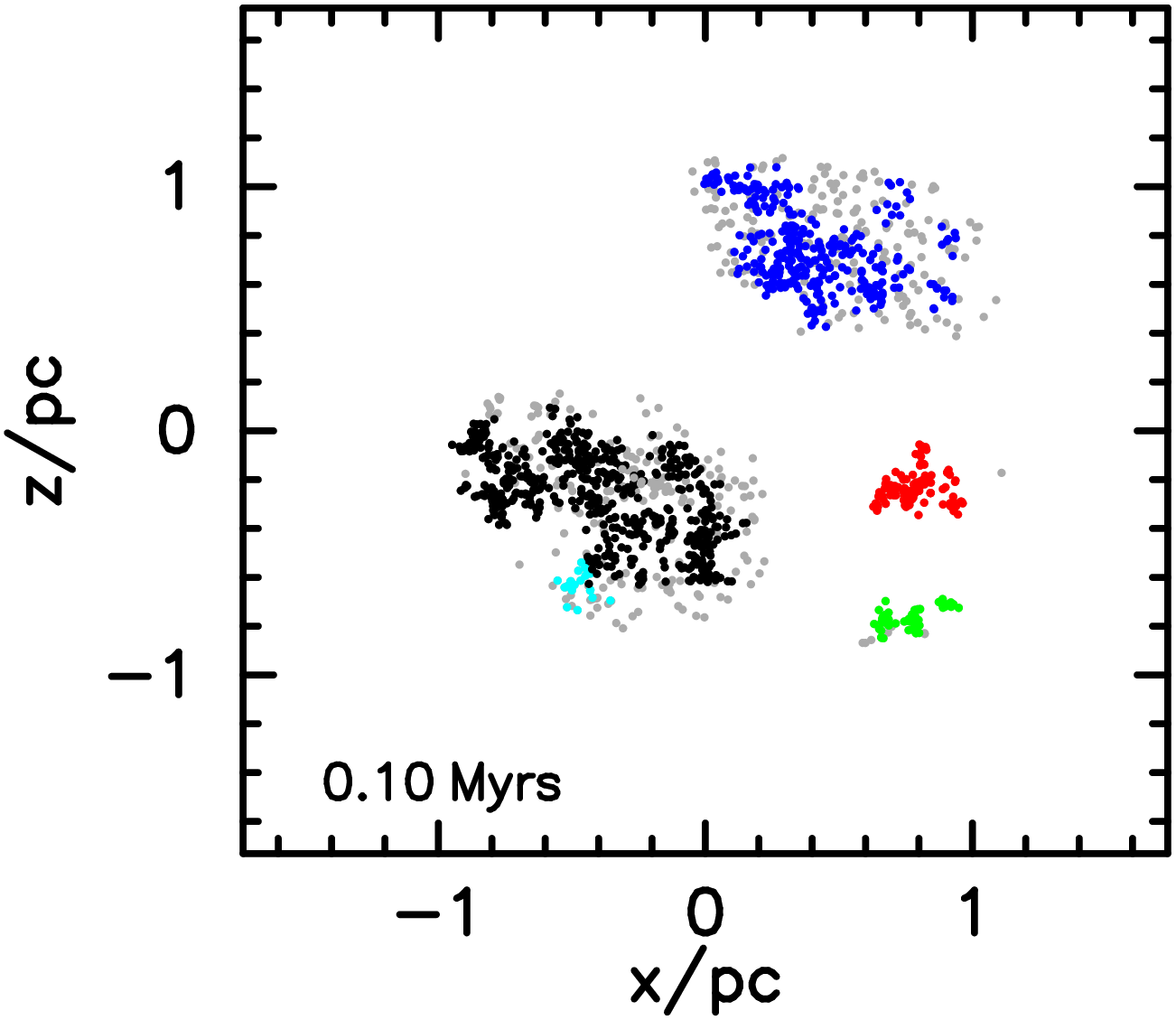}}} 
\hspace*{0.3cm}\subfigure[0.1\,Myr]{\label{F2p0-f}\rotatebox{270}{\includegraphics[scale=0.20]{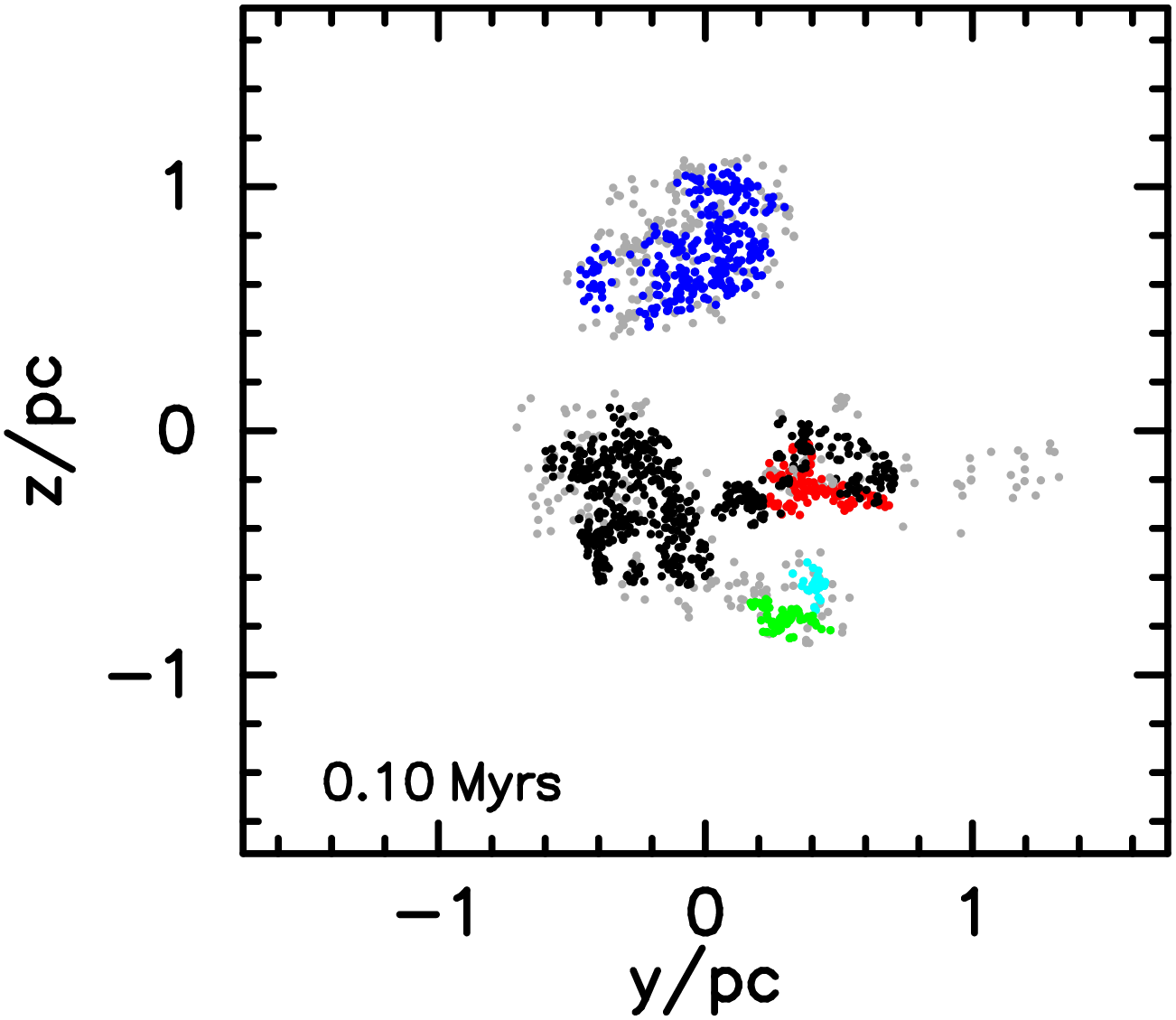}}}

\hspace*{-1.5cm}\subfigure[0.4\,Myr]{\label{F2p0-g}\rotatebox{270}{\includegraphics[scale=0.20]{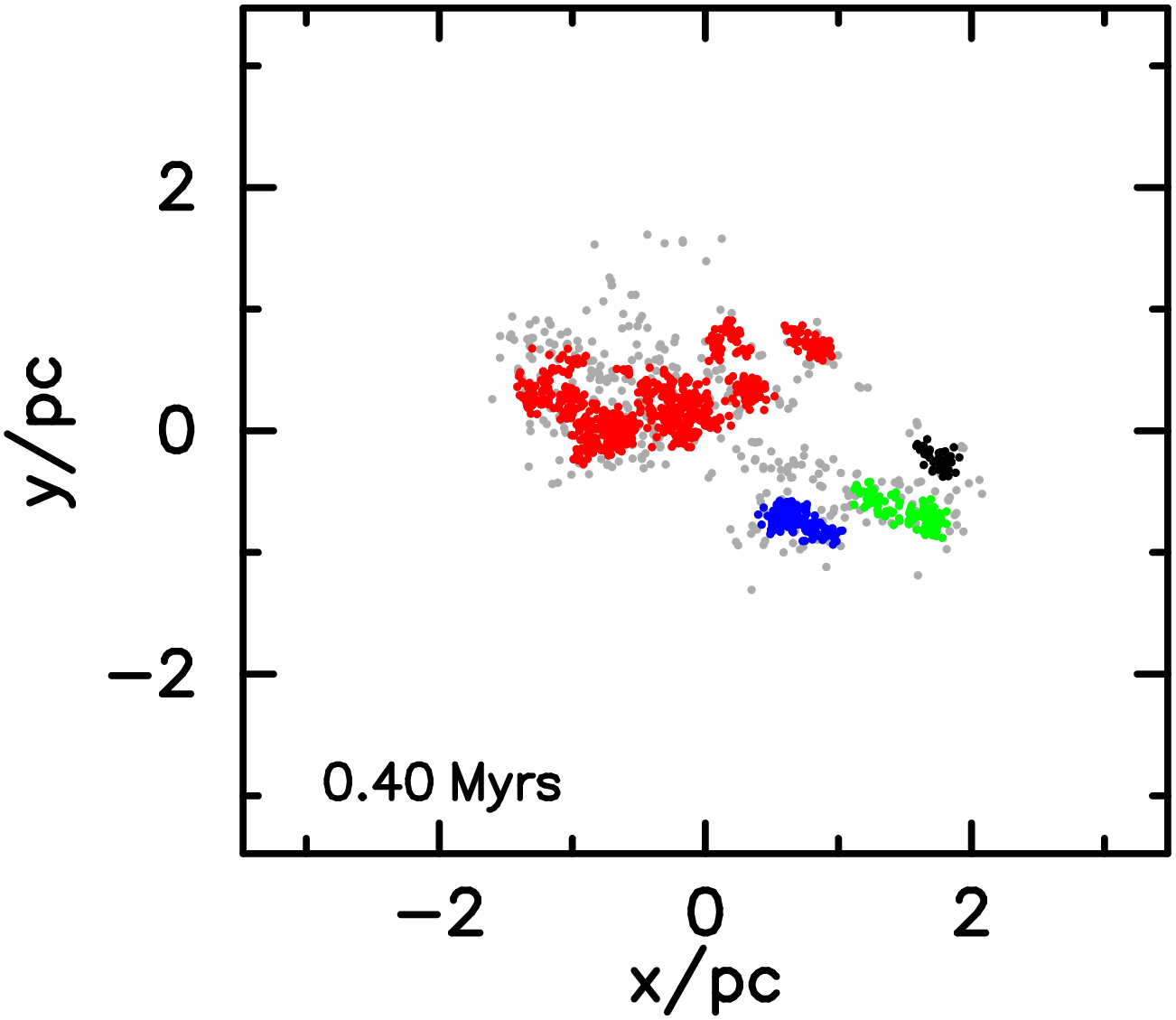}}}
\hspace*{0.3cm} 
\subfigure[0.4\,Myr]{\label{F2p0-h}\rotatebox{270}{\includegraphics[scale=0.20]{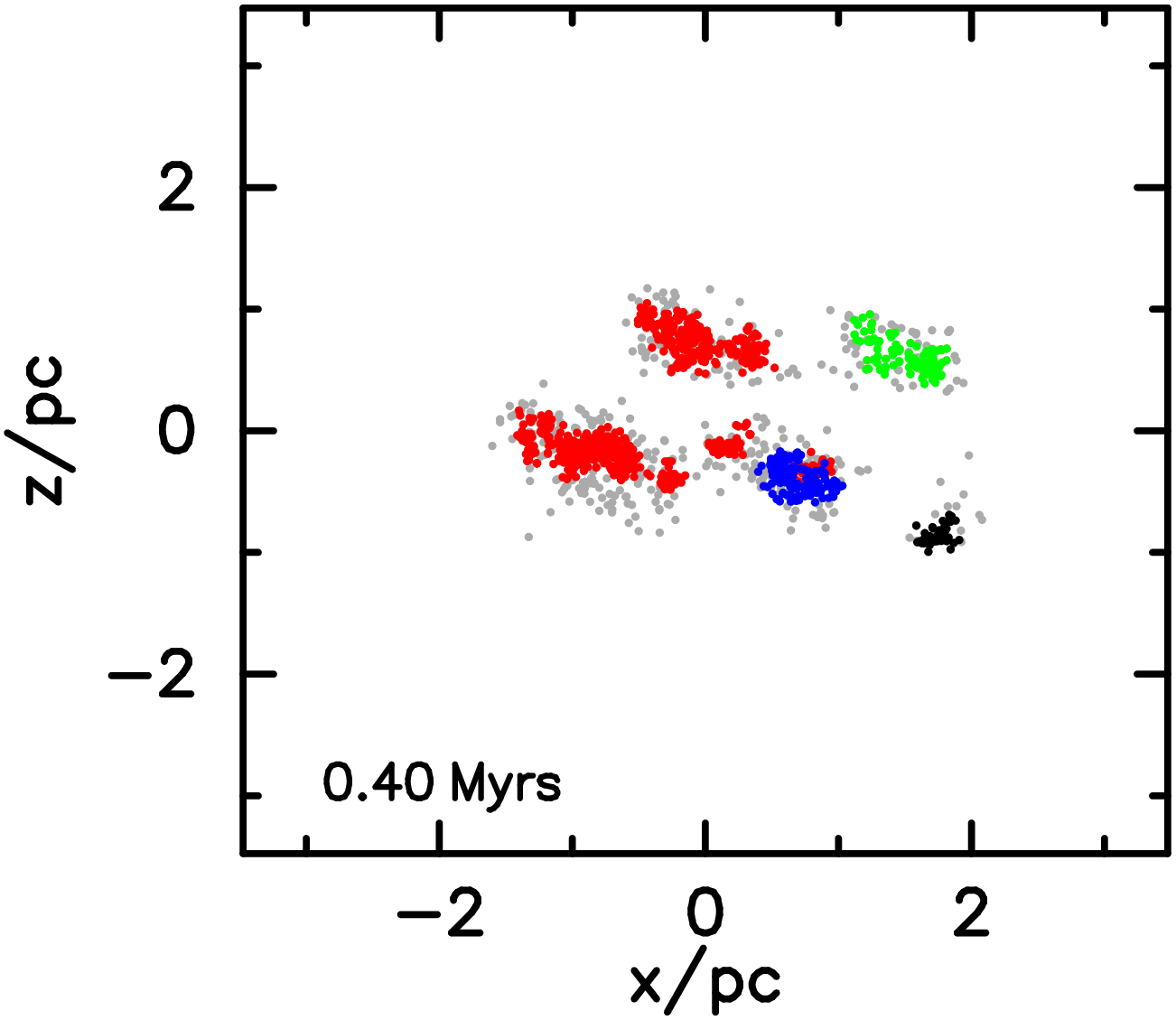}}} 
\hspace*{0.3cm}\subfigure[0.4\,Myr]{\label{F2p0-i}\rotatebox{270}{\includegraphics[scale=0.20]{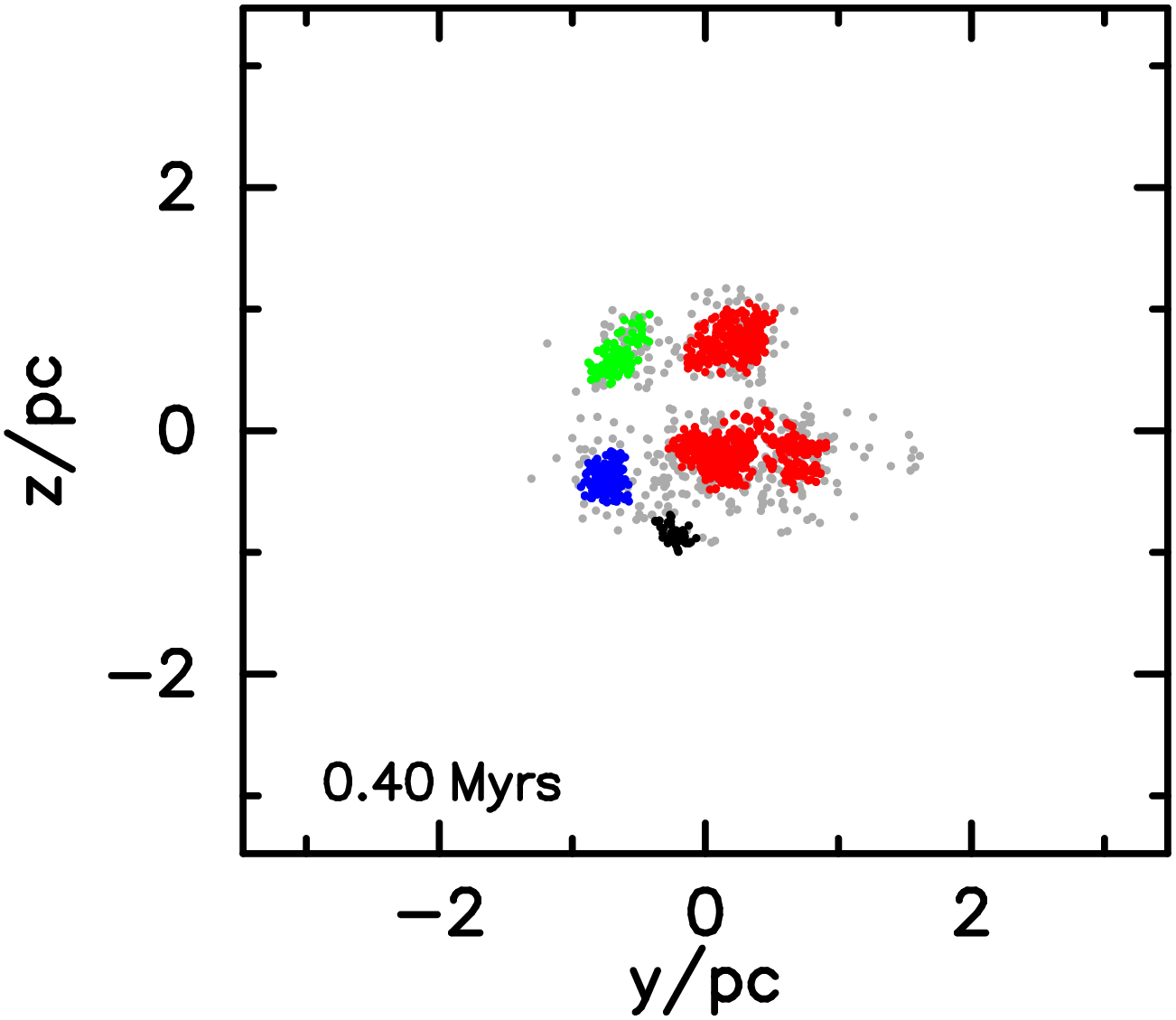}}}

\hspace*{-1.5cm}\subfigure[0.7\,Myr]{\label{F2p0-j}\rotatebox{270}{\includegraphics[scale=0.20]{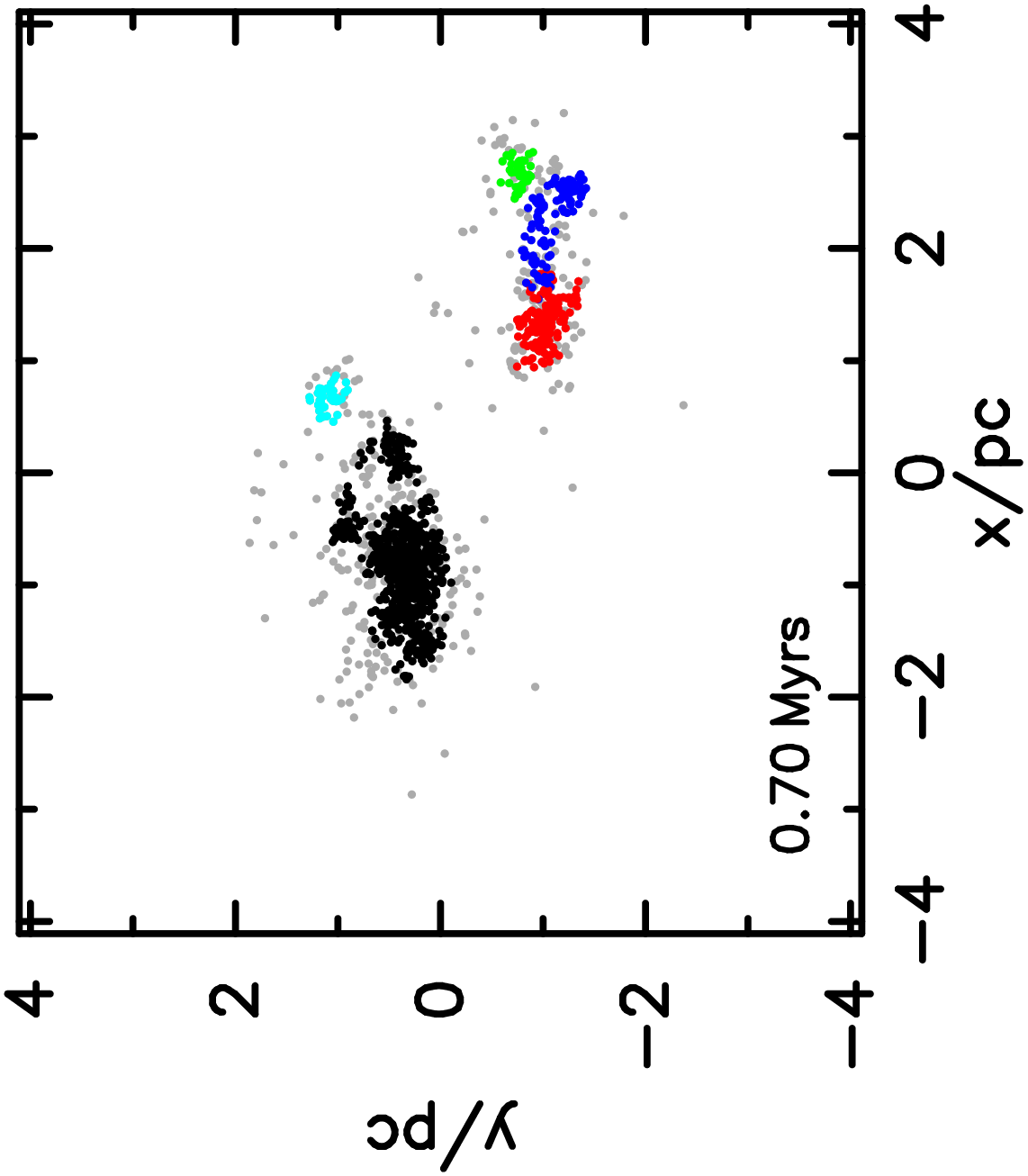}}}
\hspace*{0.3cm} 
\subfigure[0.7\,Myr]{\label{F2p0-k}\rotatebox{270}{\includegraphics[scale=0.20]{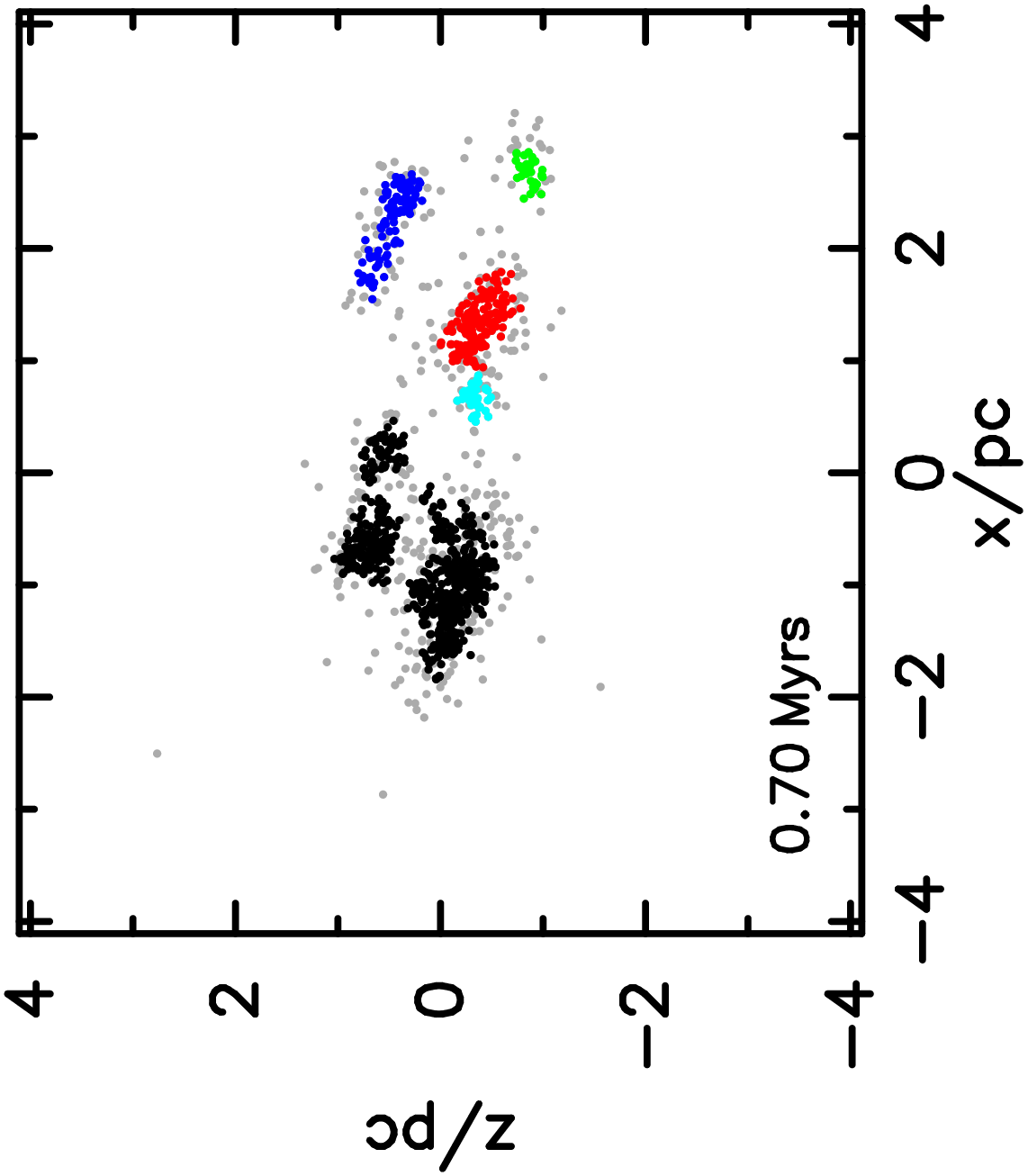}}} 
\hspace*{0.3cm}\subfigure[0.7\,Myr]{\label{F2p0-l}\rotatebox{270}{\includegraphics[scale=0.20]{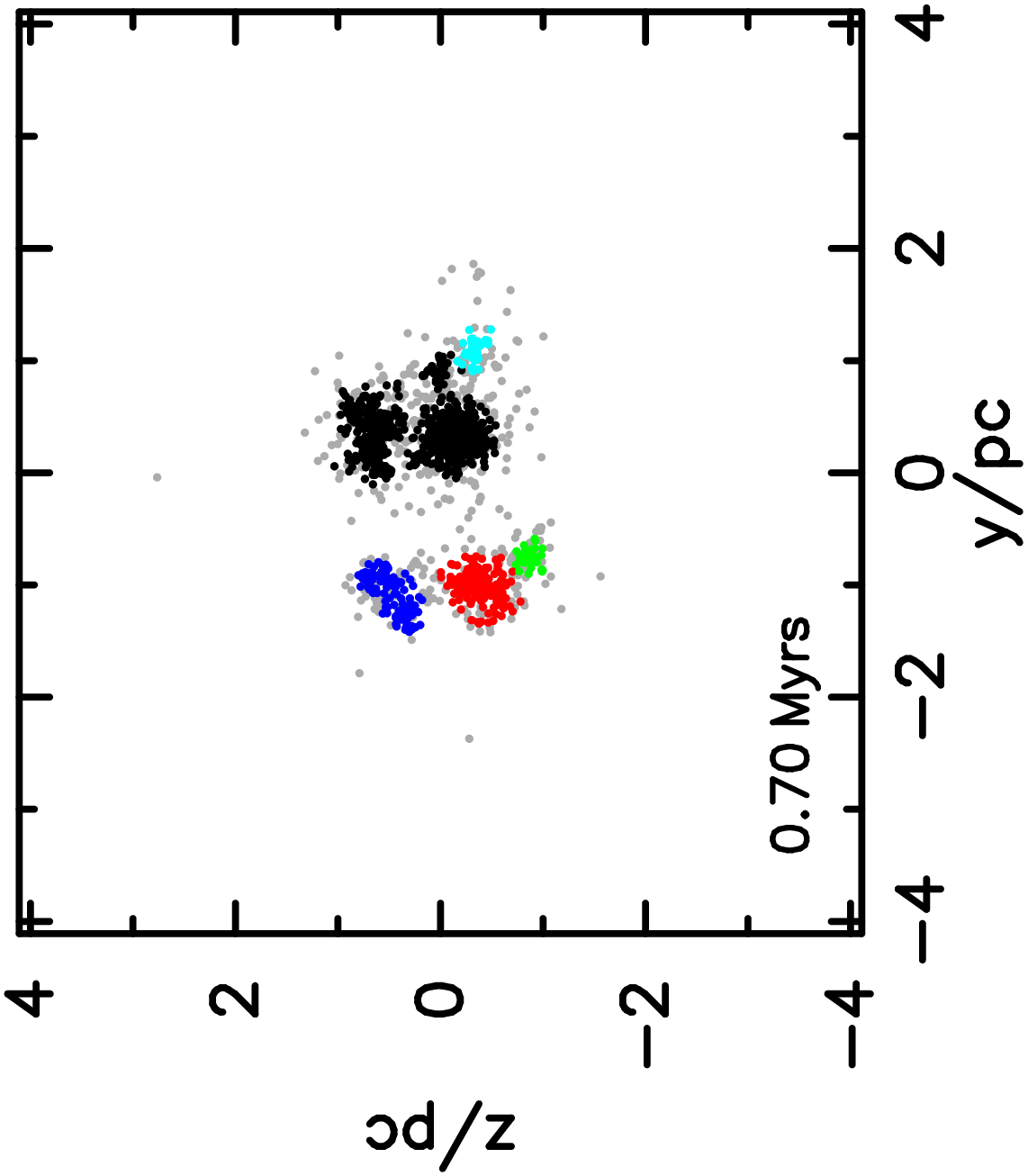}}}

\hspace*{-1.5cm}\subfigure[3\,Myr]{\label{F2p0-m}\rotatebox{270}{\includegraphics[scale=0.20]{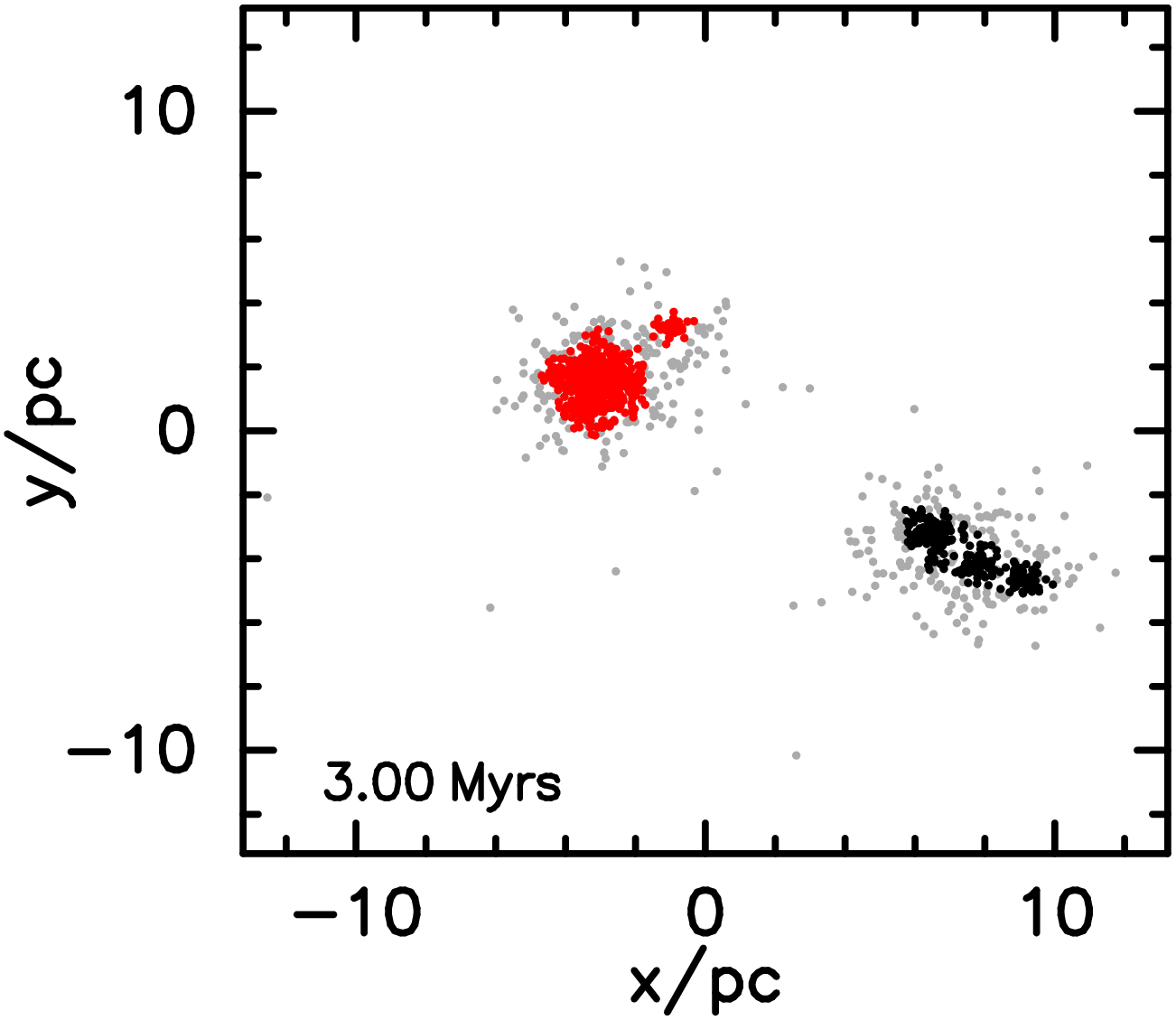}}}
\hspace*{0.3cm} 
\subfigure[3\,Myr]{\label{F2p0-n}\rotatebox{270}{\includegraphics[scale=0.20]{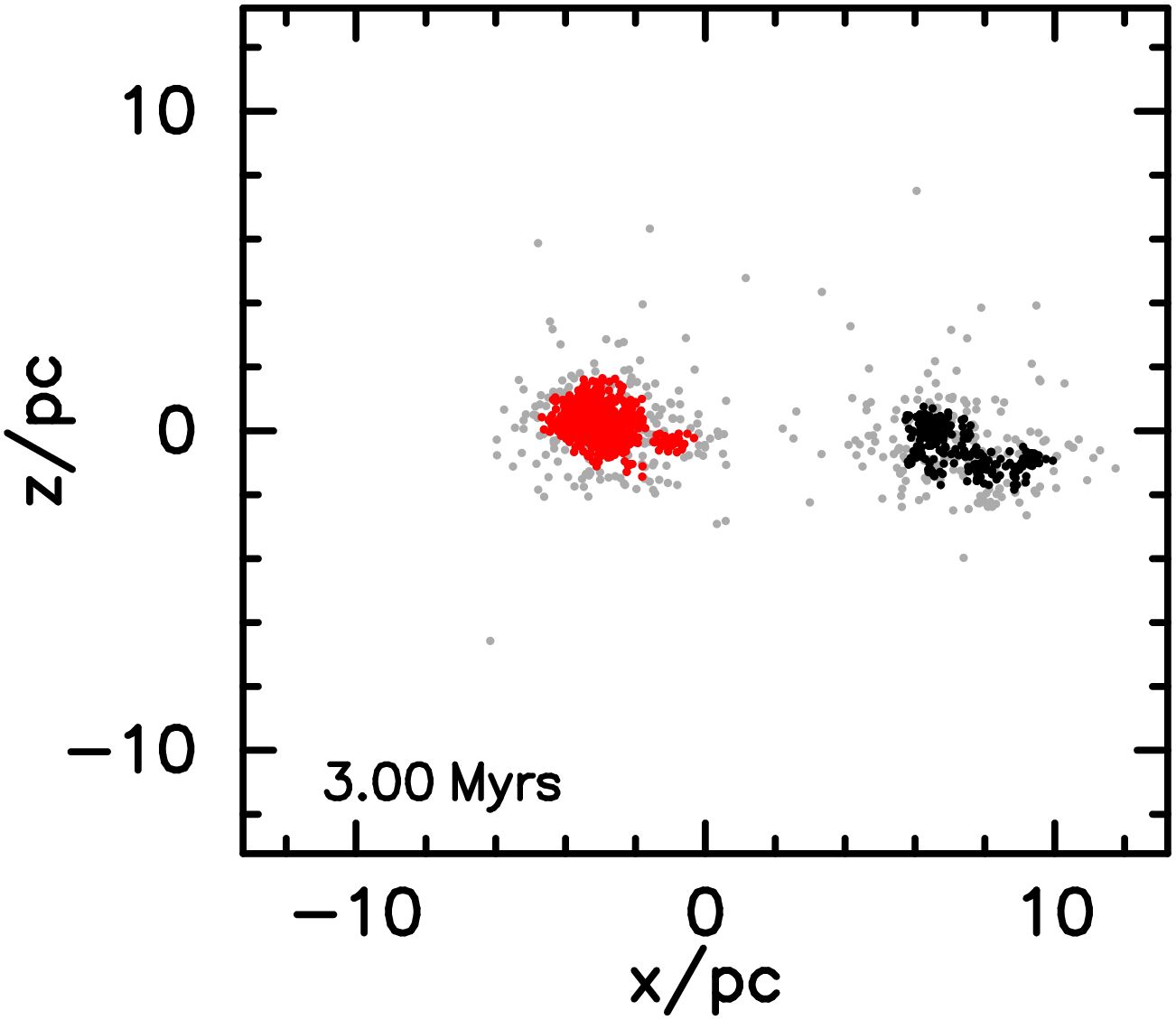}}} 
\hspace*{0.3cm}\subfigure[3\,Myr]{\label{F2p0-o}\rotatebox{270}{\includegraphics[scale=0.20]{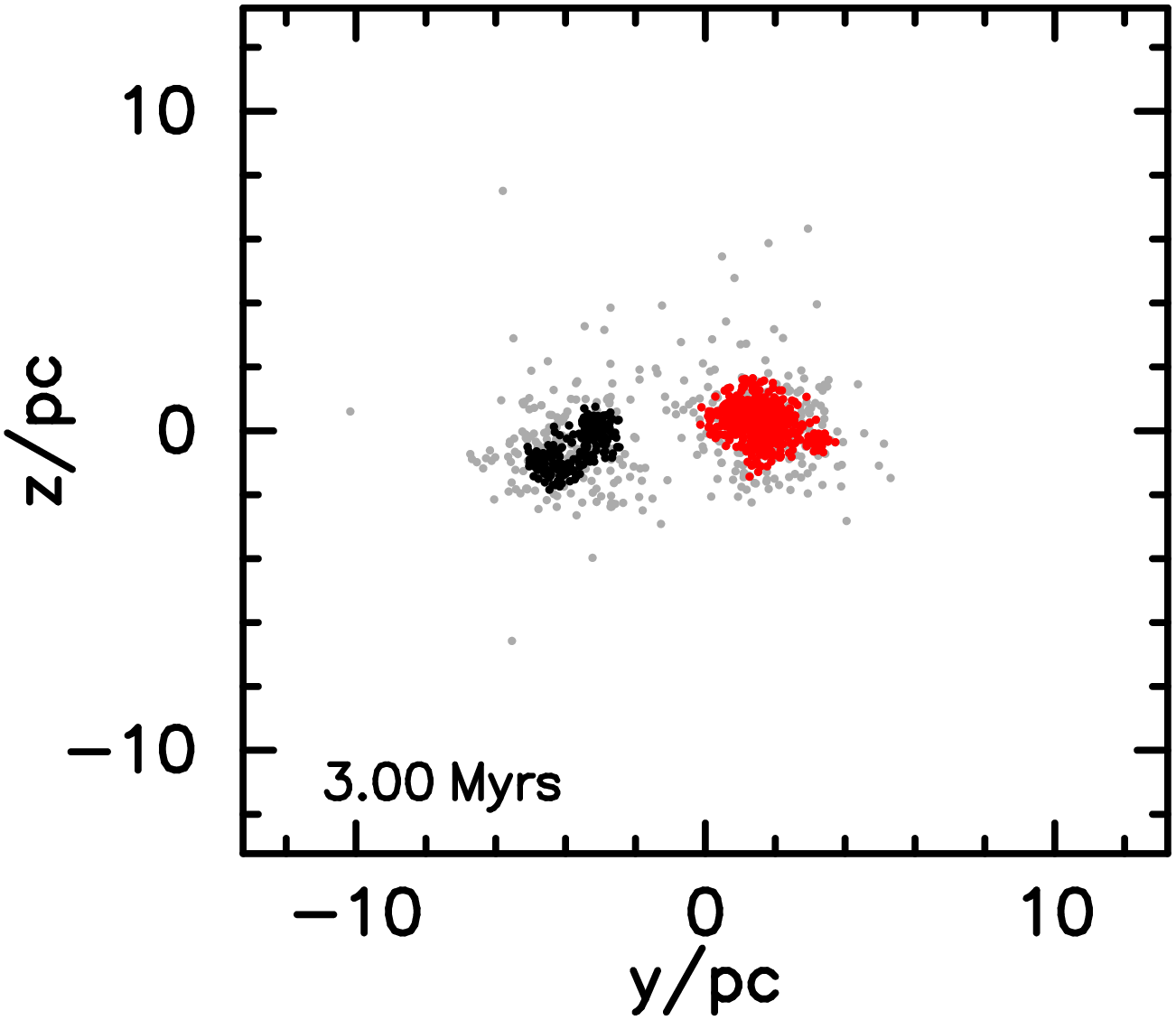}}}

\vspace*{0.2cm}
\caption[bf]{Evolution of groups defined by the Friends of Friends algorithm in a simulated  supervirial  ($\alpha_{\rm vir} = 1.5$) star-forming region with moderate levels of initial spatial and kinematic substructure (fractal dimension $D = 2.0$). Stars that have a local stellar density below the first quartile in the distribution are not assigned to a group and are coloured grey. The colours in the subsequent snapshots do not correspond to the colours in the first snapshot ($t = 0$\,Myr).}
  \label{F2p0}
  \end{center}
\end{figure*}

We show a typical example of a simulation where the initial fractal dimension is $D = 2.0$ in Fig.~\ref{F2p0}, with stars above the density threshold of the first quartile assigned to groups. Again, the colours of groups identified at later snapshots are not correlated with the colours of the groups identified at $t = 0$\,Myr. These initial conditions ($D = 2.0$ and $\alpha_{\rm vir} = 1.5$) often lead to the formation of a binary cluster \citep{Arnold17}, and this happens in more than 50\,per cent of the realisations of this set of initial conditions. 

We do not see any clear evidence for the timescale for substructure erasure to be different for the $D = 2.0$ simulations compared to the simulations with $D = 1.6$ (with similar results for the $D=2.6$ simulations that we do not show here for the sake of brevity). This may be because the groups identified by the Friends of Friends algorithm are not representative of the physical scales in our box fractal, or simply that the Friends of Friends algorithm cannot distinguish between multiple groups once dynamical evolution takes place.

As in the corresponding supervirial simulations with $D = 1.6$, the simulations with less substructure also expose the same problem with the Friends of Friends algorithm, namely that distinct clumps of stars are assigned to the same group due to bridging stars, and some of these groups are transient between snapshots (such as the cyan coloured group in Figs.~\ref{F2p0-j} -- \ref{F2p0-l}, which disappears in snapshots either side of 0.7\,Myr).

We identify fewer groups initially than in the $D = 1.6$ case (a mode of 2 and median of 5), but with  similarly large range in group number (1--15). We therefore conclude that the issues identified with the Friends of Friends technique are not unique to a particular set of initial conditions.

 \section{Discussion}

 Our analysis of $N$-body simulations using the Friends of Friends algorithm to identify spatio-kinematic groups exposes several issues with using this technique to quantify the initial conditions of star formation.

 Firstly, we have performed our analysis on fractal distributions. By definition, fractals are hierarchical and self-similar, with the only boundary conditions being the size scale of the distribution and the velocity dispersion. Therefore, any groups that are identified are necessarily artificial and arbitrary \citep[see also][]{Parker15b}, without any physical meaning. In an observed star-forming region, it is also usually extremely unclear whether the region can be broken down into constituent parts. This is especially relevant if star-formation is inherently hierarchical or self-similar \citep[e.g.][]{Elmegreen18}, which implies that there is no scale length for star-formation.

 Secondly, we find a wide range in the the number of groups identified by the Friends of Friends algorithm. Our simulations that are set up to be in subvirial collapse and have a high degree of initial spatial and kinematic substructure contain between 1 -- 25 groups, with a mode of 4 and a median of 8, for statistically similar fractal distributions, identical apart from the random number seed used to initialise the positions and velocities. Worryingly, we find very similar numbers of groups when the fractals are scaled to be supervirial (i.e. expanding), despite the regions having very different initial velocity dispersions. Therefore, the number of groups identified in a star-forming region \citep[or filaments -- c.f.][]{Hacar13,Hacar17} does not betray any information about the physical initial conditions and may be subject to line-of-sight confusion \citep{Clarke18}. 

 Third, during the subsequent dynamical evolution of our star-forming regions, stars move between groups, often creating bridges between groups so that two distinct groups become one larger group. Observational studies \citep{Hacar16,DaRio17} often attempt to mitigate for this by introducing a density threshold, so that lone stars (or gas parcels) do not unduly influence the number of groups (or filaments) identified. We also applied a density threshold to our $N$-body simulations, but find that bridges of stars still occur, and that the Friends of Friends algorithm cannot separate larger groups. In part, this is due to the fact that stars migrate significant distances during the dynamical evolution of the simulations (Fig.~\ref{original_colours}) and swap between groups \citep[see also][]{Arnold17}. The problem of `bridging' is likely ignored in observational studies if the thresholds for group definition are tuned to each specific region at a given time, rather than being automated as we have done here.

   Fourth, we have found no dependence on the later evolutionary state of the star-forming regions on the number of groups identified initially by the Friends of Friends algorithm. We measured the amount of spatial substructure, mass segregation, velocity dispersion and relative surface densities, and find no dependence on the number of initial groups.

   We note that some of our simulations have initially very high stellar densities ($\tilde{\rho} \sim 10^4$\,stars\,pc$^{-3}$). This facilitates rapid dynamical evolution, but does not affect the number of groups identified initially. We ran a set of low-density simulations   ($\tilde{\rho} \sim 10$\,stars\,pc$^{-3}$) and found similar behaviour, albeit on longer dynamical timescales. Similarly, changing the initial degree of spatial and kinematic substructure (which can invert the timescales on which we would expect structure to be erased) does not affect our conclusions.

   We have performed our analysis using the full six dimensional information (as the position and velocity vector of every star each contains three components). We also performed it in three dimensions ($x$, $y$ and $v_z$) to mimic the information available in (most) observational studies and found similar results. Indeed, one could argue that the confusion present when all of the spatio-kinematic information is available should preclude any further use of the technique in fewer dimensions.

\section{Conclusions}

We analyse $N$-body simulations of the dynamical evolution of subvirial (collapsing) and superviral (expanding) star-forming regions and apply an automated Friends of Friends algorithm to pick out groups or stars that have similar spatial and kinematic properties. Our conclusions are the following:

(i) The Friends of Friends technique picks out wide-ranging numbers of groups in statistically identical spatio-kinematic fractal distributions, despite the threshold lengths for distance and velocity varying little between individual simulations. The mode is 4, the median is 8 but the number of groups identified across twenty identical simulations ranges from 1 -- 25.

(ii) We do not see any difference in the number of groups identified in subvirial and supervirial simulations. The mode and median numbers of groups are identical, and the range is almost identical. This is because although the initial velocity scalings are very different (0.3\,km\,s$^{-1}$ for the subvirial simulations versus 1.5\,km\,s$^{-1}$ for the supervirial simulations), the scaling is linear and our Friends of Friends algorithm automatically calculates distance and velocity thresholds. In practice, this means that any automated analysis would not be able to distinguish between very different initial star formation conditions.

(iii) The dynamical evolution of the star-forming regions causes the groups to merge together in Friends of Friends space, even if (as in the case of the supervirial simulations), there are still distinct spatial substructures. This occurs because stars migrate between groups as the simulation progresses, but the groups do not dynamically mix with each other. Therefore, at a given point in the evolution of a star-forming region, a spatio-kinematic group is not retaining any information on the initial properties of that group. 

(iv) Furthermore, there is no dependence of the later spatial and kinematic evolution of the star-forming regions on the number of groups identified by the Friends of Friends algorithm. The amount of mass segregation, overall spatial structure and velocity dispersion that develops with time are unrelated to the initial number of groups, implying that the global dynamical evolution of the star-forming region cannot be related to any group structure defined by the Friends of Friends method.

Taken together, our results suggest that the Friends of Friends algorithm may not be particularly useful for quantifying the initial conditions of star-forming regions, and we urge users to include tests on synthetic datasets in any future analyses \citep[see also][]{Clarke18}. 








\section*{Acknowledgements}

 We are grateful to the anonymous referee for a helpful report. RJP acknowledges support from the Royal Society in the form of a Dorothy Hodgkin Fellowship. NJW acknowledges an STFC Ernest Rutherford Fellowship (grant number ST/M005569/1). 

\bibliographystyle{mnras}  
\bibliography{general_ref}

\begin{thebibliography}{}
\makeatletter
\relax
\def\mn@urlcharsother{\let\do\@makeother \do\$\do\&\do\#\do\^\do\_\do\%\do\~}
\def\mn@doi{\begingroup\mn@urlcharsother \@ifnextchar [ {\mn@doi@}
  {\mn@doi@[]}}
\def\mn@doi@[#1]#2{\def\@tempa{#1}\ifx\@tempa\@empty \href
  {http://dx.doi.org/#2} {doi:#2}\else \href {http://dx.doi.org/#2} {#1}\fi
  \endgroup}
\def\mn@eprint#1#2{\mn@eprint@#1:#2::\@nil}
\def\mn@eprint@arXiv#1{\href {http://arxiv.org/abs/#1} {{\tt arXiv:#1}}}
\def\mn@eprint@dblp#1{\href {http://dblp.uni-trier.de/rec/bibtex/#1.xml}
  {dblp:#1}}
\def\mn@eprint@#1:#2:#3:#4\@nil{\def\@tempa {#1}\def\@tempb {#2}\def\@tempc
  {#3}\ifx \@tempc \@empty \let \@tempc \@tempb \let \@tempb \@tempa \fi \ifx
  \@tempb \@empty \def\@tempb {arXiv}\fi \@ifundefined
  {mn@eprint@\@tempb}{\@tempb:\@tempc}{\expandafter \expandafter \csname
  mn@eprint@\@tempb\endcsname \expandafter{\@tempc}}}

\bibitem[\protect\citeauthoryear{Adams, Hollenbach, Laughlin  \& Gorti}{Adams
  et~al.}{2004}]{Adams04}
Adams F.~C.,  Hollenbach D.,  Laughlin G.,   Gorti U.,  2004, ApJ, 611, 360

\bibitem[\protect\citeauthoryear{Allison \& Goodwin}{Allison \&
  Goodwin}{2011}]{Allison11}
Allison R.~J.,  Goodwin S.~P.,  2011, MNRAS, 415, 1967

\bibitem[\protect\citeauthoryear{Allison, Goodwin, Parker, {Portegies Zwart},
  de Grijs  \& Kouwenhoven}{Allison et~al.}{2009}]{Allison09a}
Allison R.~J.,  Goodwin S.~P.,  Parker R.~J.,  {Portegies Zwart} S.~F.,  de
  Grijs R.,   Kouwenhoven M. B.~N.,  2009, MNRAS, 395, 1449

\bibitem[\protect\citeauthoryear{Allison, Goodwin, Parker, {Portegies Zwart}
  \& de Grijs}{Allison et~al.}{2010}]{Allison10}
Allison R.~J.,  Goodwin S.~P.,  Parker R.~J.,  {Portegies Zwart} S.~F.,   de
  Grijs R.,  2010, MNRAS, 407, 1098

\bibitem[\protect\citeauthoryear{{Arnold}, {Goodwin}, {Griffiths}  \&
  {Parker}}{{Arnold} et~al.}{2017}]{Arnold17}
{Arnold} B.,  {Goodwin} S.~P.,  {Griffiths} D.~W.,   {Parker} R.~J.,  2017,
  \mn@doi [\mnras] {10.1093/mnras/stx1719}, \href
  {http://adsabs.harvard.edu/abs/2017MNRAS.471.2498A} {471, 2498}

\bibitem[\protect\citeauthoryear{Bastian, Covey  \& Meyer}{Bastian
  et~al.}{2010}]{Bastian10}
Bastian N.,  Covey K.~R.,   Meyer M.~R.,  2010, ARA\&A, 48, 339

\bibitem[\protect\citeauthoryear{Bressert et~al.,}{Bressert
  et~al.}{2010}]{Bressert10}
Bressert E.,  et~al., 2010, MNRAS, 409, L54

\bibitem[\protect\citeauthoryear{{Cai}, {Kouwenhoven}, {Portegies Zwart}  \&
  {Spurzem}}{{Cai} et~al.}{2017}]{Cai17a}
{Cai} M.~X.,  {Kouwenhoven} M.~B.~N.,  {Portegies Zwart} S.~F.,   {Spurzem} R.,
   2017, \mn@doi [\mnras] {10.1093/mnras/stx1464}, \href
  {http://adsabs.harvard.edu/abs/2017MNRAS.470.4337C} {470, 4337}

\bibitem[\protect\citeauthoryear{Cartwright}{Cartwright}{2009}]{Cartwright09b}
Cartwright A.,  2009, MNRAS, 400, 1427

\bibitem[\protect\citeauthoryear{Cartwright \& Whitworth}{Cartwright \&
  Whitworth}{2004}]{Cartwright04}
Cartwright A.,  Whitworth A.~P.,  2004, MNRAS, 348, 589

\bibitem[\protect\citeauthoryear{{Cartwright}, {Whitworth}  \&
  {Nutter}}{{Cartwright} et~al.}{2006}]{Cartwright06}
{Cartwright} A.,  {Whitworth} A.~P.,   {Nutter} D.,  2006, \mn@doi [MNRAS]
  {10.1111/j.1365-2966.2006.10389.x}, \href
  {http://adsabs.harvard.edu/abs/2006MNRAS.369.1411C} {369, 1411}

\bibitem[\protect\citeauthoryear{{Clarke}, {Whitworth}, {Spowage},
  {Duarte-Cabral}, {Suri}, {Jaffa}, {Walch}  \& {Clark}}{{Clarke}
  et~al.}{2018}]{Clarke18}
{Clarke} S.~D.,  {Whitworth} A.~P.,  {Spowage} R.~L.,  {Duarte-Cabral} A.,
  {Suri} S.~T.,  {Jaffa} S.~E.,  {Walch} S.,   {Clark} P.~C.,  2018, \mn@doi
  [\mnras] {10.1093/mnras/sty1675}, \href
  {http://adsabs.harvard.edu/abs/2018MNRAS.479.1722C} {479, 1722}

\bibitem[\protect\citeauthoryear{{Da Rio} et~al.,}{{Da Rio}
  et~al.}{2017}]{DaRio17}
{Da Rio} N.,  et~al., 2017, \mn@doi [\apj] {10.3847/1538-4357/aa7a5b}, \href
  {http://adsabs.harvard.edu/abs/2017ApJ...845..105D} {845, 105}

\bibitem[\protect\citeauthoryear{{Elmegreen}}{{Elmegreen}}{2018}]{Elmegreen18}
{Elmegreen} B.~G.,  2018, \mn@doi [\apj] {10.3847/1538-4357/aaa252}, \href
  {http://adsabs.harvard.edu/abs/2018ApJ...853...88E} {853, 88}

\bibitem[\protect\citeauthoryear{{Foster} et~al.,}{{Foster}
  et~al.}{2015}]{Foster15}
{Foster} J.~B.,  et~al., 2015, \mn@doi [ApJ] {10.1088/0004-637X/799/2/136},
  \href {http://adsabs.harvard.edu/abs/2015ApJ...799..136F} {799, 136}

\bibitem[\protect\citeauthoryear{Goodwin \& Whitworth}{Goodwin \&
  Whitworth}{2004}]{Goodwin04a}
Goodwin S.~P.,  Whitworth A.~P.,  2004, A\&A, 413, 929

\bibitem[\protect\citeauthoryear{{Gouliermis}, {Hony}  \&
  {Klessen}}{{Gouliermis} et~al.}{2014}]{Gouliermis14}
{Gouliermis} D.~A.,  {Hony} S.,   {Klessen} R.~S.,  2014, \mn@doi [MNRAS]
  {10.1093/mnras/stu228}, \href
  {http://cdsads.u-strasbg.fr/abs/2014MNRAS.439.3775G} {439, 3775}

\bibitem[\protect\citeauthoryear{Gutermuth, Megeath, Myers, Allen  \&
  Fazio}{Gutermuth et~al.}{2009}]{Gutermuth09}
Gutermuth R.~A.,  Megeath S.~T.,  Myers P.~C.,  Allen L.~E.,   Fazio J. L. P.
  G.~G.,  2009, ApJS, 184, 18

\bibitem[\protect\citeauthoryear{{Hacar}, {Tafalla}, {Kauffmann}  \&
  {Kov{\'a}cs}}{{Hacar} et~al.}{2013}]{Hacar13}
{Hacar} A.,  {Tafalla} M.,  {Kauffmann} J.,   {Kov{\'a}cs} A.,  2013, A\&A,
  554, A55

\bibitem[\protect\citeauthoryear{{Hacar}, {Alves}, {Forbrich}, {Meingast},
  {Kubiak}  \& {Gro{\ss}schedl}}{{Hacar} et~al.}{2016}]{Hacar16}
{Hacar} A.,  {Alves} J.,  {Forbrich} J.,  {Meingast} S.,  {Kubiak} K.,
  {Gro{\ss}schedl} J.,  2016, \mn@doi [\aap] {10.1051/0004-6361/201527805},
  589, A80

\bibitem[\protect\citeauthoryear{{Hacar}, {Tafalla}  \& {Alves}}{{Hacar}
  et~al.}{2017}]{Hacar17}
{Hacar} A.,  {Tafalla} M.,   {Alves} J.,  2017, \mn@doi [\aap]
  {10.1051/0004-6361/201630348}, \href
  {http://adsabs.harvard.edu/abs/2017A%26A...606A.123H} {606, A123}

\bibitem[\protect\citeauthoryear{{Hacar}, {Tafalla}, {Forbrich}, {Alves},
  {Meingast}, {Grossschedl}  \& {Teixeira}}{{Hacar} et~al.}{2018}]{Hacar18}
{Hacar} A.,  {Tafalla} M.,  {Forbrich} J.,  {Alves} J.,  {Meingast} S.,
  {Grossschedl} J.,   {Teixeira} P.~S.,  2018, \mn@doi [\aap]
  {10.1051/0004-6361/201731894}, 610, A77

\bibitem[\protect\citeauthoryear{{Henshaw} et~al.,}{{Henshaw}
  et~al.}{2016a}]{Henshaw16a}
{Henshaw} J.~D.,  et~al., 2016a, \mn@doi [\mnras] {10.1093/mnras/stw121}, \href
  {http://adsabs.harvard.edu/abs/2016MNRAS.457.2675H} {457, 2675}

\bibitem[\protect\citeauthoryear{{Henshaw} et~al.,}{{Henshaw}
  et~al.}{2016b}]{Henshaw16}
{Henshaw} J.~D.,  et~al., 2016b, \mn@doi [\mnras] {10.1093/mnras/stw1794},
  \href {http://adsabs.harvard.edu/abs/2016MNRAS.463..146H} {463, 146}

\bibitem[\protect\citeauthoryear{{Huchra} \& {Geller}}{{Huchra} \&
  {Geller}}{1982}]{Huchra82}
{Huchra} J.~P.,  {Geller} M.~J.,  1982, \mn@doi [\apj] {10.1086/160000}, \href
  {http://adsabs.harvard.edu/abs/1982ApJ...257..423H} {257, 423}

\bibitem[\protect\citeauthoryear{{Jaehnig}, {Da Rio}  \& {Tan}}{{Jaehnig}
  et~al.}{2015}]{Jaehnig15}
{Jaehnig} K.~O.,  {Da Rio} N.,   {Tan} J.~C.,  2015, \mn@doi [\apj]
  {10.1088/0004-637X/798/2/126}, \href
  {http://adsabs.harvard.edu/abs/2015ApJ...798..126J} {798, 126}

\bibitem[\protect\citeauthoryear{{Jaffa}, {Whitworth}  \& {Lomax}}{{Jaffa}
  et~al.}{2017}]{Jaffa17}
{Jaffa} S.~E.,  {Whitworth} A.~P.,   {Lomax} O.,  2017, \mn@doi [MNRAS]
  {10.1093/mnras/stw3140}, \href
  {http://adsabs.harvard.edu/abs/2017MNRAS.466.1082J} {466, 1082}

\bibitem[\protect\citeauthoryear{{Jaffa}, {Whitworth}, {Clarke}  \&
  {Howard}}{{Jaffa} et~al.}{2018}]{Jaffa18}
{Jaffa} S.~E.,  {Whitworth} A.~P.,  {Clarke} S.~D.,   {Howard} A.~D.~P.,  2018,
  \mn@doi [\mnras] {10.1093/mnras/sty696}, \href
  {http://adsabs.harvard.edu/abs/2018MNRAS.tmp..667J} {}

\bibitem[\protect\citeauthoryear{{Jeffries} et~al.,}{{Jeffries}
  et~al.}{2014}]{Jeffries14}
{Jeffries} R.~D.,  et~al., 2014, \mn@doi [\aap] {10.1051/0004-6361/201323288},
  563, A94

\bibitem[\protect\citeauthoryear{{Kere{\v s}}, {Katz}, {Dav{\'e}}, {Fardal}  \&
  {Weinberg}}{{Kere{\v s}} et~al.}{2009}]{Keres09}
{Kere{\v s}} D.,  {Katz} N.,  {Dav{\'e}} R.,  {Fardal} M.,   {Weinberg} D.~H.,
  2009, \mn@doi [\mnras] {10.1111/j.1365-2966.2009.14924.x}, \href
  {http://adsabs.harvard.edu/abs/2009MNRAS.396.2332K} {396, 2332}

\bibitem[\protect\citeauthoryear{{Kuhn} et~al.,}{{Kuhn} et~al.}{2014}]{Kuhn14}
{Kuhn} M.~A.,  et~al., 2014, \mn@doi [\apj] {10.1088/0004-637X/787/2/107},
  \href {http://adsabs.harvard.edu/abs/2014ApJ...787..107K} {787, 107}

\bibitem[\protect\citeauthoryear{{K{\"u}pper}, {Maschberger}, {Kroupa}  \&
  {Baumgardt}}{{K{\"u}pper} et~al.}{2011}]{Kupper11}
{K{\"u}pper} A.~H.~W.,  {Maschberger} T.,  {Kroupa} P.,   {Baumgardt} H.,
  2011, \mn@doi [MNRAS] {10.1111/j.1365-2966.2011.19412.x}, \href
  {http://adsabs.harvard.edu/abs/2011MNRAS.417.2300K} {417, 2300}

\bibitem[\protect\citeauthoryear{{Kuznetsova}, {Hartmann}  \&
  {Ballesteros-Paredes}}{{Kuznetsova} et~al.}{2018}]{Kuznetsova18}
{Kuznetsova} A.,  {Hartmann} L.,   {Ballesteros-Paredes} J.,  2018, \mn@doi
  [\mnras] {10.1093/mnras/stx2480}, \href
  {http://adsabs.harvard.edu/abs/2018MNRAS.473.2372K} {473, 2372}

\bibitem[\protect\citeauthoryear{Lada \& Lada}{Lada \& Lada}{2003}]{Lada03}
Lada C.~J.,  Lada E.~A.,  2003, ARA\&A, 41, 57

\bibitem[\protect\citeauthoryear{Larson}{Larson}{1981}]{Larson81}
Larson R.~B.,  1981, MNRAS, 194, 809

\bibitem[\protect\citeauthoryear{Larson}{Larson}{1995}]{Larson95}
Larson R.~B.,  1995, MNRAS, 272, 213

\bibitem[\protect\citeauthoryear{{Lomax}, {Whitworth}  \& {Cartwright}}{{Lomax}
  et~al.}{2011}]{Lomax11}
{Lomax} O.,  {Whitworth} A.~P.,   {Cartwright} A.,  2011, \mn@doi [MNRAS]
  {10.1111/j.1365-2966.2010.17935.x}, \href
  {http://adsabs.harvard.edu/abs/2011MNRAS.412..627L} {412, 627}

\bibitem[\protect\citeauthoryear{{Lynden-Bell}}{{Lynden-Bell}}{1967}]{LyndenBell67}
{Lynden-Bell} D.,  1967, \mn@doi [\mnras] {10.1093/mnras/136.1.101}, \href
  {http://adsabs.harvard.edu/abs/1967MNRAS.136..101L} {136, 101}

\bibitem[\protect\citeauthoryear{{Marks} \& {Kroupa}}{{Marks} \&
  {Kroupa}}{2012}]{Marks12}
{Marks} M.,  {Kroupa} P.,  2012, \mn@doi [A\&A] {10.1051/0004-6361/201118231},
  543, A8

\bibitem[\protect\citeauthoryear{Maschberger}{Maschberger}{2013}]{Maschberger13}
Maschberger T.,  2013, MNRAS, 429, 1725

\bibitem[\protect\citeauthoryear{Maschberger \& Clarke}{Maschberger \&
  Clarke}{2011}]{Maschberger11}
Maschberger T.,  Clarke C.~J.,  2011, MNRAS, 416, 541

\bibitem[\protect\citeauthoryear{Parker}{Parker}{2014}]{Parker14e}
Parker R.~J.,  2014, MNRAS, 445, 4037

\bibitem[\protect\citeauthoryear{Parker \& {Alves de Oliveira}}{Parker \&
  {Alves de Oliveira}}{2017}]{Parker17a}
Parker R.~J.,  {Alves de Oliveira} C.,  2017, MNRAS, 468, 4340

\bibitem[\protect\citeauthoryear{Parker \& Dale}{Parker \&
  Dale}{2015}]{Parker15c}
Parker R.~J.,  Dale J.~E.,  2015, MNRAS, 451, 3664

\bibitem[\protect\citeauthoryear{Parker \& Goodwin}{Parker \&
  Goodwin}{2015}]{Parker15b}
Parker R.~J.,  Goodwin S.~P.,  2015, MNRAS, 449, 3381

\bibitem[\protect\citeauthoryear{Parker \& Meyer}{Parker \&
  Meyer}{2012}]{Parker12d}
Parker R.~J.,  Meyer M.~R.,  2012, MNRAS, 427, 637

\bibitem[\protect\citeauthoryear{Parker \& Quanz}{Parker \&
  Quanz}{2012}]{Parker12a}
Parker R.~J.,  Quanz S.~P.,  2012, MNRAS, 419, 2448

\bibitem[\protect\citeauthoryear{Parker \& Wright}{Parker \&
  Wright}{2016}]{Parker16b}
Parker R.~J.,  Wright N.~J.,  2016, MNRAS, 457, 3430

\bibitem[\protect\citeauthoryear{Parker, Wright, Goodwin  \& Meyer}{Parker
  et~al.}{2014}]{Parker14b}
Parker R.~J.,  Wright N.~J.,  Goodwin S.~P.,   Meyer M.~R.,  2014, MNRAS, 438,
  620

\bibitem[\protect\citeauthoryear{{Peretto}, {Andr{\'e}}  \&
  {Belloche}}{{Peretto} et~al.}{2006}]{Peretto06}
{Peretto} N.,  {Andr{\'e}} P.,   {Belloche} A.,  2006, \mn@doi [A\&A]
  {10.1051/0004-6361:20053324}, \href
  {http://adsabs.harvard.edu/abs/2006A\%26A...445..979P} {445, 979}

\bibitem[\protect\citeauthoryear{{Portegies Zwart}}{{Portegies
  Zwart}}{2016}]{Zwart16}
{Portegies Zwart} S.~F.,  2016, \mn@doi [MNRAS] {10.1093/mnras/stv2831}, \href
  {http://adsabs.harvard.edu/abs/2016MNRAS.457..313P} {457, 313}

\bibitem[\protect\citeauthoryear{{Portegies Zwart}, Makino, McMillan  \&
  Hut}{{Portegies Zwart} et~al.}{1999}]{Zwart99}
{Portegies Zwart} S.~F.,  Makino J.,  McMillan S. L.~W.,   Hut P.,  1999, A\&A,
  348, 117

\bibitem[\protect\citeauthoryear{{Portegies Zwart}, McMillan, Hut  \&
  Makino}{{Portegies Zwart} et~al.}{2001}]{Zwart01}
{Portegies Zwart} S.~F.,  McMillan S. L.~W.,  Hut P.,   Makino J.,  2001,
  MNRAS, 321, 199

\bibitem[\protect\citeauthoryear{Salpeter}{Salpeter}{1955}]{Salpeter55}
Salpeter E.~E.,  1955, ApJ, 121, 161

\bibitem[\protect\citeauthoryear{Scally \& Clarke}{Scally \&
  Clarke}{2001}]{Scally01}
Scally A.,  Clarke C.,  2001, MNRAS, 325, 449

\bibitem[\protect\citeauthoryear{{Traficante} et~al.,}{{Traficante}
  et~al.}{2018}]{Traficante18}
{Traficante} A.,  et~al., 2018, \mn@doi [\mnras] {10.1093/mnras/sty798}, \href
  {http://adsabs.harvard.edu/abs/2018MNRAS.477.2220T} {477, 2220}

\bibitem[\protect\citeauthoryear{{V{\'a}zquez-Semadeni},
  {Gonz{\'a}lez-Samaniego}  \& {Col{\'{\i}}n}}{{V{\'a}zquez-Semadeni}
  et~al.}{2017}]{Vazquez17}
{V{\'a}zquez-Semadeni} E.,  {Gonz{\'a}lez-Samaniego} A.,   {Col{\'{\i}}n} P.,
  2017, \mn@doi [MNRAS] {10.1093/mnras/stw3229}, \href
  {http://adsabs.harvard.edu/abs/2017MNRAS.467.1313V} {467, 1313}

\bibitem[\protect\citeauthoryear{{Vincke}, {Breslau}  \& {Pfalzner}}{{Vincke}
  et~al.}{2015}]{Vincke15}
{Vincke} K.,  {Breslau} A.,   {Pfalzner} S.,  2015, \mn@doi [A\&A]
  {10.1051/0004-6361/201425552}, 577, A115

\bibitem[\protect\citeauthoryear{{Williams}, {Peretto}, {Avison},
  {Duarte-Cabral}  \& {Fuller}}{{Williams} et~al.}{2018}]{Williams18}
{Williams} G.~M.,  {Peretto} N.,  {Avison} A.,  {Duarte-Cabral} A.,   {Fuller}
  G.~A.,  2018, arXiv: 1801.07253, \href
  {http://adsabs.harvard.edu/abs/2018arXiv180107253W} {}

\bibitem[\protect\citeauthoryear{{Wright} \& {Mamajek}}{{Wright} \&
  {Mamajek}}{2018}]{Wright18}
{Wright} N.~J.,  {Mamajek} E.~E.,  2018, \mn@doi [\mnras]
  {10.1093/mnras/sty207}, \href
  {http://adsabs.harvard.edu/abs/2018MNRAS.476..381W} {476, 381}

\bibitem[\protect\citeauthoryear{Wright, Parker, Goodwin  \& Drake}{Wright
  et~al.}{2014}]{Wright14}
Wright N.~J.,  Parker R.~J.,  Goodwin S.~P.,   Drake J.~J.,  2014, MNRAS, 438,
  639

\bibitem[\protect\citeauthoryear{{Wright}, {Bouy}, {Drew}, {Sarro}, {Bertin},
  {Cuillandre}  \& {Barrado}}{{Wright} et~al.}{2016}]{Wright16}
{Wright} N.~J.,  {Bouy} H.,  {Drew} J.~E.,  {Sarro} L.~M.,  {Bertin} E.,
  {Cuillandre} J.-C.,   {Barrado} D.,  2016, \mn@doi [MNRAS]
  {10.1093/mnras/stw1148}, \href
  {http://adsabs.harvard.edu/abs/2016MNRAS.460.2593W} {460, 2593}

\makeatother
\end{thebibliography}

\label{lastpage}

\end{document}